\documentclass[11pt]{article}

\newcommand{\be}{\begin{equation}}
\newcommand{\ee}{\end{equation}}
\usepackage{amsmath}
\usepackage{amssymb}
\usepackage{latexsym}
\usepackage{epsfig}

\newcommand{\half}{{\textstyle {1\over 2}}}

\newcommand{\bra}[1]{\langle #1|}
\newcommand{\ket}[1]{|#1\rangle}

\newcommand{\Z}{\mathbb{Z}}
\newcommand{\R}{\mathbb{R}}

\newcommand{\X}{\mathbb{X}}

\newcommand{\bb}{{\mathrm{b}}}
\newcommand{\Gg}{{\mathrm{g}}}
\newcommand{\ddd}{{\mathrm{d}}}

\newcommand{\HH}{\mathcal{H}}
 \newcommand{\GG}{\mathcal{G}}
  \newcommand{\EE}{{\cal{E}}}

\newcommand{\fluc}{\check}

\newcommand{\ti}{\tilde}
\newcommand{\pa}{\partial}

\newcommand{\RRR}{{\hbox{\rm R\kern-2.35mm R}}}

\def\ZZZ{{\hbox{ Z\kern-1.6mm Z}}}

\newcommand{\sectiono}[1]{\section{#1}\setcounter{equation}{0}}

\begin{document}

\begin{titlepage}
\rightline{April 2009} 
\rightline{\tt arXiv:0904.4664}
\rightline{\tt  Imperial-TP-2009-CH-02}
\rightline{\tt MIT-CTP-4031}    
\begin{center}
\vskip 2.5cm
{\Large \bf {Double Field Theory}}\\
\vskip 1.0cm
{\large {Chris Hull${}^1$ and Barton Zwiebach${}^2$}}
\vskip 1.0cm
{\it {${}^1$The Blackett Laboratory}}\\
{\it {Imperial College London}}\\
{\it {Prince Consort Road, London SW7 @AZ, U.K.}}\\
c.hull@imperial.ac.uk
\vskip 0.5cm
{\it {${}^2$Center for Theoretical Physics}}\\
{\it {Massachusetts Institute of Technology}}\\
{\it {Cambridge, MA 02139, USA}}\\
zwiebach@mit.edu
\vskip 1.0cm
{\bf Abstract}
\end{center}

\noindent
\begin{narrower}

The zero modes of closed strings on a torus --the   
torus coordinates plus dual coordinates
conjugate to winding number--
parameterize a doubled torus. 
In closed  string field theory, the string field depends 
on all zero-modes and so can be expanded to give 
 an infinite set of fields on the doubled torus. 
We use  string field theory
to  construct a theory of massless fields on 
the doubled torus.  Key to the consistency is a constraint on fields
and gauge parameters that arises from the $L_0 - \bar L_0=0$ 
condition in closed string theory.  
 The symmetry of this double field theory  
includes usual and \lq dual diffeomorphisms', together with
a  T-duality  acting on fields that have explicit dependence on the torus coordinates and the dual coordinates.
We find that, along with gravity, a Kalb-Ramond field
and a dilaton must be added to support 
both usual and dual diffeomorphisms.
 We construct a fully consistent and gauge invariant action on 
 the doubled torus to cubic order in the fields.  
 We discuss the challenges involved in the construction of the full 
nonlinear theory.  
We emphasize that the doubled geometry is physical 
and the dual dimensions should not
 be viewed as an auxiliary structure or a gauge artifact.

\end{narrower}

\end{titlepage}

\newpage

\tableofcontents
\baselineskip=16pt
\section{Introduction and summary}

T-duality is a striking property of string 
theory.\footnote{See \cite{Giveon:1994fu}
for a review of T-duality and references.}
Closed strings can wrap around
non-contractible cycles in spacetime, giving
  winding states that have no analogue for particle theories.   
 The existence of both momentum and winding states is the key property
of strings that allows T-duality: the complete physical equivalence
of string theories on dual backgrounds that have very
  different geometries.

String field theory provides a complete gauge-invariant
formulation of string dynamics around any consistent background, and we will use it here to study  T-duality.
A   closed string field theory for a 
flat spacetime  with some  spatial directions   
curled up   into a torus was examined 
long ago by Kugo and 
 Zwiebach~\cite{Kugo:1992md}\footnote{While this work used a 
 covariantised light-cone formulation of the
string field theory, the results are largely
applicable to the  covariant closed string field theory~\cite{Zwiebach:1992ie} which we use here.}, following earlier work 
in~\cite{Hata:1986mz,Maeno:1989uc}.  
In particular, \cite{Kugo:1992md} showed how T-duality is realised as a symmetry of the string field theory.
 The string field theory  treats momenta and winding rather   
symmetrically and, as a consequence,
expanding the string field gives component fields that
depend on {\em both} 
momentum and winding number.
Fourier transforming to position space then gives component fields that depend on both the spacetime
coordinates conjugate to momentum 
 and on new periodic coordinates conjugate to winding number.
For a spacetime which is a product of a Minkowski space $M$ with  a 
$d$-dimensional torus $T^d$, the component fields are  then fields on
$M\times  T^{2d}$ 
where the  {\em doubled torus}   
$T^{2d}$  contains the original spacetime torus $T^d$ together 
with another torus $T^d$ parameterised by  
 the winding coordinates. 
 In fact, the doubled torus contains the original torus $T^d$ as well
as  the tori related to it by T-duality.  Then T-duality can be viewed as changing which $T^d$ subspace of the doubled torus is to be regarded as part of the spacetime~\cite{Hull:2004in}.

The complete  closed string 
field theory on a torus is exotic and complicated.  
To our knowledge, it has not been examined
in detail at the component level to try to uncover how spacetime fields
realise the magic of T-duality. 
 This is one of the main purposes of the present
paper.  As a simplification, we restrict
ourselves to the \lq massless' 
sector or, more precisely, to the set of
fields that would be massless in the uncompactified theory.  We thus
focus on the 
gravity,
 antisymmetric tensor (Kalb-Ramond),  and dilaton fields.  
We include all momenta and winding excitations of these fields 
by keeping their full dependence  on
the coordinates of the doubled torus. 
 T-duality   exchanges 
momentum and winding excitations,
so that we expect T-duality to be a symmetry of this massless theory. 
A T-duality symmetric field theory on the doubled torus that can incorporate all T-dual geometries is likely
to be novel and perhaps even exotic.  Our hope is that this massless theory exists and it is not so complicated as to defy construction.
Our results 
so far are encouraging:
we have constructed the theory to cubic order in the fields. No higher derivatives are needed: each term 
has two  derivatives, as in Einstein gravity.

Previous work on double field theory   
includes that of Tseytlin~\cite{Tseytlin:1990nb} who 
used a first-quantized approach with 
non-covariant actions for left and right-moving 
string coordinates on the torus. 
He calculated amplitudes for vertex operators depending on both coordinates,  finding partially gauge-fixed cubic interactions  
 for metric perturbations  that are consistent with our action.
It would be interesting to develop the first-quantised approach further, perhaps using the covariant formulation of~\cite{Hull:2004in}.
Siegel~\cite{Siegel:1993th} considered the field theory for the massless sector of closed strings without winding modes, but this restriction is implemented in an
$O(d,d,\mathbb{Z})$ covariant fashion through an intriguing formulation of  T-duality.
An effective field theory on a doubled torus also arose
in the study of open strings on a torus with Êspace-fillingÊ
and point-likeÊ D-branes~\cite{VanRaamsdonk:2003gj}.

The gauge symmetry of the theory we build   should  include diffeomorphisms for each
$T^d$ subspace of the doubled torus that can arise as a possible spacetime.
We find that this is the case, and the linearised transformations include
linearised diffeomorphisms on the doubled torus as well as a doubled
version of the antisymmetric tensor field gauge symmetry. 
The non-linear structure is rather 
intricate and a simple characterization remains to be found.
We find that the Jacobi identities are not satisfied, so the symmetry appears not to be diffeomorphisms on the doubled torus or even a Lie algebra.
Gauge invariance requires that the fields and 
gauge parameters  
satisfy a constraint that arises from the $L_0 - \bar L_0=0$ constraint
of closed string field theory.

\medskip
The doubled torus $T^{2d}$  
arises naturally in the 
first-quantized approach to strings on a torus, leading to a number of approaches involving sigma models whose target  is the space  
with doubled torus 
fibres~\cite{Hull:2004in,Tseytlin:1990nb,Duff:1990,Hull:1988dp,Hull:1988si,Maharana:1992my,Hull:2006va,Berman:2007yf,Berman:2007xn,HackettJones:2006bp}.  
T-duality extends to spacetimes that have a torus fibration  
if the fields  are independent of the coordinates of the torus fibres.
The Buscher rules~\cite{buscher}  for $d=1$, 
and their extension to $d>1$~\cite{GivRoc},
encode the transformation of such a   background under  T-duality.
In the doubled torus formalism 
of Refs.~\cite{Hull:2004in,Hull:2006va},    
the $T^d$ fibres of such a background are replaced with doubled torus fibres $T^{2d}$. 
A key feature of this formalism  is that T-duality is a manifest geometric symmetry,  as the 
T-duality group acts through diffeomorphisms on the doubled torus fibres.
Moreover,    the target space   with  doubled torus fibres 
    incorporates all possible T-dual geometries.
The conventional picture emerges only on choosing a $T^d$ subspace of each $T^{2d}$ fibre to be the spacetime torus, and T-duality acts to change which $T^d$ subspace is chosen \cite{Hull:2004in}. The fact that T-duality is a symmetry means that the physics is the same in each case.

If fields have explicit dependence on the torus coordinates, the situation is not well understood.
It is expected that fields that depend on the spacetime torus coordinates $x$ should transform into fields that depend on the dual coordinates $\ti x$. Dependence on the dual coordinates is puzzling, but one would expect that while $x$-dependence affects particles, $\ti x$-dependence should affect  winding modes, so that particles and winding modes could experience different backgrounds; see e.g. \cite{Gregory:1997te}. 
Dependence on $\ti x$ has been associated with world-sheet instanton effects~\cite{Gregory:1997te}, and a number of calculations have supported this 
view~\cite{Tong:2002rq,Harvey:2005ab,Okuyama:2005gx,Witten:2009xu}.
  General string backgrounds,  
  however,  should involve fields depending on both $x$ and $\ti x$, and it is to be expected that there should be an extension of the T-duality transformation rules to this general case~\cite{Dabholkar:2005ve,Hull:2009sg}. 
 We find the T-duality transformations
 that are a symmetry of 
 the double action for
 fields that depend   on both $x$ and $\ti x $.
The fields in this action arise naturally from string field theory.
 In the case with no dependence on the dual coordinates $\ti x$, we use the non-linear relation between these fields   and the familiar metric and $B$-field to find a generalisation of the Buscher rules to the case of fields with general dependence on the torus coordinates $x$ (or any set of coordinates related to these by a duality).
The form of these transformations then suggest a natural  further generalisation to the case in which the fields have full dependence on $x$ and $\ti x$.

We would like to emphasize
 that the inclusion of dual coordinates
in double field theory is not a gauge redundancy or a reformulation
of an underlying non-doubled geometry.  The dual coordinates are needed
to represent physical degrees of freedom; one cannot eliminate the
dependence of fields on the additional coordinates 
using gauge conditions
or solving constraints.
This is perhaps less obvious in first quantization than in second
quantization.  In first quantization the familiar sigma
 model  for closed strings on tori 
 defines a conformal field theory.  Using a
doubled torus or other additional structures for the sigma model gives a 
better and more useful description of the {\it same} conformal field theory.
It allows, for example,  a natural construction of vertex operators  for states with 
 both momentum and winding. The physics, however, is in the conformal field theory, which includes
momentum and winding, however they are described.  
In the string field theory a non-doubled formulation is not even
an option.  The string field, always defined by the conformal field theory state
space, necessarily depends on coordinates conjugate to momentum and dual coordinates conjugate to winding.  This dependence is nontrivial.  While
string field theory is now known to have 
nonperturbative information (at least in the open sector), 
our use of closed string field theory here has
been more limited.  String field theory was useful  
in the construction of
a nontrivial action and gauge transformations that 
would have been hard to guess
or construct directly.

\bigskip 
 Let us now discuss in some detail the setup and results in
 the present paper.
We shall be interested in closed string theory 
  in $D$-dimensional flat space with $d$ compactified directions, $\R^{n-1,1}\times T^d$ where $n+d=D$. We shall present our discussion for the critical
  $D=26$  bosonic closed string, 
  but much of this applies to   
  closed superstring theories.
 We use coordinates $x^i=(x^\mu, x^a)$ with $i=0,...,D-1$ which split into coordinates   
  $x^\mu$ on the $n$-dimensional Minkoswski space $\R^{n-1,1}$  and  coordinates  $x^a$ on the $d$-torus $T^d$.
States are 
labelled by the momentum   $p_i=(k_\mu,p_a)$  and the string 
windings 
 $w^a$.
For coordinates with periodicity $x^a\sim x^a +2\pi$, the operators
$p_a$  and $w^a$ have 
integer  eigenvalues -- these are the momentum
and winding quantum numbers. 
Perturbative states are  of the form  
\be
\label{cft-states} \sum_I\int dk \sum_{p_a, w^a}  \phi_I( k_\mu , p_a, w^a) 
\, \mathcal{O}^I\vert k_\mu, p_a, w^a \rangle \,,\ee
where $\mathcal{O}^I$ are operators built from  matter and ghost oscillators 
and   $\phi_I(k,p_a, w^a)$ are   
   momentum-space fields which also depend on the winding numbers.
 Fourier transforming, dependence on the momenta $k_\mu,p_a$ becomes dependence on the spacetime coordinates 
$x^\mu, x^a$ as usual, while dependence on $w^a$ is replaced by dependence on a new periodic 
coordinate $\tilde x_a$ conjugate to  winding numbers $w^a$. 
Thus 
  the fields $\phi_I$ above  give us coordinate-space fields 
  \be
  \label{coor-fields}
   \phi_I (x^\mu, \, x^a\,, \tilde x_a )\,. \ee
Then $(x^a, \tilde x_a)$ are periodic coordinates for the doubled 
torus $T^{2d}$.
All physical string states must 
satisfy the level matching condition, {\it i.e.}, they must
be annihilated by $ L_0-\bar L_0$:
\be
\label{lmexp}  
L_0-\bar L_0=  N - \bar N - p_a w^a = 0 \,. 
\ee
This   
constraint will play a central role in our work.
The free string on-shell condition   
$L_0 + \bar L_0 -2=0$ 
takes a simple form 
when the   background antisymmetric tensor vanishes:
\be M^2 \equiv   
  -(k^2+p^2 +w^2) = {2\over \alpha'} (N+\bar N-2)\,.
  \ee 
Here $\alpha' p^2=\hat G^{ab}p_ap_b$   
 and $\alpha' w^2=
\hat G_{ab}w^aw^b$ where $\hat G_{ab}$ is the torus metric and $N,\bar N$ are the number operators for the  left and right moving oscillators. 
We can view  $M^2$   
as the $D$-dimensional mass-squared and the associated massless states ($M^2=0$)  
satisfy  $N+\bar N=2$.

The mass $M$ in 
$D$-dimensions should not be confused with the mass ${\cal M}$ in the $n$-dimensional Minkoswki  space obtained after compactification:   
 \be
 {\cal M}^2\equiv   
 -k^2=   p^2 +w^2 + {2\over \alpha'} (N+\bar N-2)\,.
 \ee
For a rectangular torus 
the metric is
$\hat G_{ab}=\delta _{ab} R^2_a /\alpha '$,  
where $R_a$ is the radius of the circle along~$x^a$. 
If all the circles   are 
sufficiently
large compared with the string length ($R_a^2 \gg \alpha '$), then $w^2=\sum _a w_a^2 R_a^2/\alpha '$ is large and  $p^2=\sum _a p_a^2 \alpha '/R_a^2$ is small, so that the states 
that are light compared to the string scale 
include those which
have $w^a=0$ and $N+\bar N=2$. This is the Kaluza-Klein tower of states obtained by compactifying the theory of massless states in $D$ dimensions.
A conventional effective field theory
in the $n$-dimensional Minkowski space
 would keep  
 states for which   
 ${\cal M}^2$ is zero or small, and would give the leading terms in a systematic expansion in 
${\cal M}^2$.  Instead, here we focus on $M^2=0$ states and in so doing,
we are keeping certain states that, from the lower-dimensional point of view, are heavy  while neglecting some which 
are lighter.\footnote{We thank 
David Gross for emphasizing this point to us. }  It is possible that the
theory we are trying to build should be considered as an effective
theory in which we keep a set of massless fields, including all of their
large-energy excitations, and integrate out everything else.  
At special points in the torus moduli space   
 there are extra states   
 with ${\cal M}^2=0$ giving enhanced gauge symmetry, 
 while near these special points these states 
will have small ${\cal M}^2$.  
  These have $(N,\bar N) = (1,0)$ or $(N,\bar N) = (0,1)$ and so have $M^2=-2/\alpha '$; we will not include these here.

T-duality is an $O(d,d; \Z)$  symmetry of the string theory acting linearly on the torus coordinates
$x^a,\tilde x_a$ and preserving their boundary conditions. 
This includes a $\Z_2$ symmetry for   each direction $a$  that
interchanges $x^a$ with $\tilde x_a$. For a rectangular torus in which $x^a$ is a coordinate for a circle of radius $R_a$, 
 $\tilde x_a$ is the coordinate for a T-dual circle of radius $\alpha '/ R_a$.  Performing  a $\Z_2$ on each of the toroidal dimensions takes a theory on the original spacetime $\R^{n-1,1}\times T^d$ with coordinates $ x^\mu, x^a$ to a theory in the dual 
spacetime $\R^{n-1,1}\times \widetilde T^d$ with coordinates $ x^\mu, \tilde x_a$.

\smallskip
In the closed string field theory for  
this toroidal background
   the string field $\ket{\Psi}$ is a general state of the form (\ref{cft-states}),
  and so can be viewed as a collection of 
  component fields $\phi_I (x^\mu, x^a, \tilde x_a)$.
   It should be emphasized 
  that the  
    difference between the toroidally compactified theory  
    and  the $D$-dimensional Minkowski space theory is that the
  toroidal zero modes are doubled; no new oscillators are added.
Two off-shell constraints must be satisfied by both the string field and
the gauge parameter $\ket{\Lambda}$.  We must have
\be 
(b_0 - \bar b_0) \ket{\Psi} = 0, \qquad
 (b_0 - \bar b_0) \ket{\Lambda} = 0\,,
\ee
 and the associated level-matching conditions
\be
\label{lmcondroj}
(L_0 - \bar L_0) \ket{\Psi} = 0 \,, ~~ (L_0 - \bar L_0) \ket{\Lambda} = 0\,.
\ee
The free field equation is
$Q\ket{\Psi}=0$,
where $Q$ is the BRST operator, and it is 
invariant under gauge transformations 
$\delta \ket{\Psi} = Q \ket{\Lambda}$.
 The ket $\ket{\Lambda}$ 
 gives rise to an infinite set of gauge parameters
that depend on $x^\mu, x^a,$ and $\tilde x_a$.
On account of (\ref{lmexp}) and (\ref{lmcondroj})  the string field satisfies
\be
(N - \bar N) \ket{\Psi} = p_a w^a  \ket{\Psi} \,,
\ee
 and for a 
component field $ \phi _I(x^\mu, x^a, \tilde x_a )$ we have 
\be   
\label{constraint-fields}
({N}_I - \bar {{N}}_I)\,  \phi _I= {1\over 2}   \, \alpha'\Delta  \phi _I \,, \qquad 
\hbox{with} \quad \Delta\equiv -{2\over \alpha'} {\partial \over \partial x^a}{\partial \over \partial \tilde  x_a}  \,.
\ee
Here the $ {N}_I$ and $\bar{ {N}}_I$ are the eigenvalues of
$N$ and $\bar N$ on the CFT state for which $\phi_I$ is the expansion
coefficient.  Thus string field theory is a theory of constrained fields, but the constraint still allows fields with non-trivial dependence on 
both $x^a$ and $\tilde x_a$
if $d>1$.

For $N= \bar  N=1$ 
we have the following fields, all with $M^2=0$:\footnote{There are additional auxiliary 
fields and gauge trivial fields 
that do not contribute propagating degrees of freedom.}
\be
\label{mssfds}
h_{ij}(  x^\mu, x^a, \tilde x_a), \qquad b_{ij}(  x^\mu, x^a, \tilde x_a),
 \qquad d\,(  x^\mu, x^a, \tilde x_a)\,.
\ee
The
constraint requires that 
these fields are all 
annihilated by the differential operator $\Delta$.
The solutions independent of $\tilde x$ give the gravity field  $h_{ij}(  x^\mu, x^a)$, the antisymmetric tensor field $b_{ij}(x^\mu, x^a )$, and the dilaton $ d(  x^\mu, x^a )$ in $D$ dimensions. The solutions independent of $x^a$ give dual versions of these fields, while again the general case depends on both $x^a$ and $\tilde x_a$
(for $d>1$).
Note that e.g. $h_{ij}$ decomposes as usual into $h_{\mu \nu }$, $h_{\mu a}$, $h_{ab}$ and there is no doubling of the tensor indices.
At higher levels
 the fields have the  same index structure as for the uncompactified  string theory, but now depend on  $\tilde x$ as well as $x^\mu, x^a$
and are subject to the constraint  (\ref{constraint-fields}).

\medskip
In this paper we focus on  the 
$M^2=0$ fields
 in (\ref{mssfds}).  The relevant
gauge parameters are a pair of vector fields
$\epsilon_i (  x^\mu, x^a, \tilde x_a)$ and  $\tilde\epsilon_i (  x^\mu, x^a, \tilde x_a)$, both of which are annihilated by $\Delta$.  
Our analysis of the quadratic theory shows that the linearised gauge
transformations take the form
\be
\label{diffeo-tilda}
\begin{split}
\delta h_{ij} &= \phantom{-}\,\partial_i \epsilon_j \,+ \partial_j \epsilon_i \,\,+\,\,\tilde \partial_i \tilde\epsilon_j + \tilde\partial_j \tilde\epsilon_i \,,
  \\[1.0ex]
\delta b_{ij} &=
- (\tilde \partial_i \epsilon_j - \tilde\partial_j \epsilon_i) -(\partial_i \tilde\epsilon_j - \partial_j \tilde\epsilon_i) \,,
~~\\[1.0ex]
\delta d   \,\,    & =   -\half \,\partial\cdot \epsilon   \,+ \half\,
\tilde \partial \cdot \tilde\epsilon \,.
\end{split}
\ee
We use the notation $\ti x_i =( \ti x_a, 0)$  
 and $\tilde \partial_i =( \partial/  \partial \ti x_a\,,0, )$ which makes it clear that only the coordinates on the
torus are doubled.  
The above gauge  structure is rather intricate and novel.  
For parameters and fields that are independent of $\tilde x$, these are the standard linearised diffeomorphisms (acting on $x^i$) with parameter $\epsilon_i $ and antisymmetric tensor gauge transformations
with parameter $\tilde \epsilon_i $.
 A dilaton $\phi$   
which is a scalar (invariant under these linearised transformations) can be defined  by
$\phi = d + \frac 1 4 \eta ^{ij} h_{ij}\,.$  
Parameters and fields that are independent of $  x^a$ live on the dual space with coordinates  $x^\mu, \tilde x_a$.  These are again linearised diffeomorphisms, now acting on 
$x^\mu, \tilde x_a$,
 and antisymmetric tensor gauge transformations, but the roles of the parameters 
 $\epsilon_i $ and  $\tilde \epsilon_i $ have been interchanged. 
 Now  $\tilde \epsilon_i $ is   the diffeomorphism parameter and  $\epsilon_i $ the antisymmetric tensor gauge parameter.
In this case, the scalar dilaton would be 
$\tilde \phi = d - \frac 1 4 \eta ^{ij} h_{ij}\,.$
While $\phi$ is invariant under $\epsilon$
transformations and $\tilde\phi$ is invariant under $\tilde\epsilon$
transformations, there is no combination 
of $d$ and $\eta ^{ij} h_{ij}$ that is invariant under both.
In the full non-linear theory  there is no dilaton that is a scalar under both diffeomorphisms and dual diffeomorphisms, and $d$ is the natural field to use.
 Nonlinearly,  one has a relation of the form  $e^{-2d}= e^{-2 \phi}\sqrt {- g}$; the dilaton $d$ is invariant under T-duality and its expectation value provides the duality-invariant string coupling constant \cite{Kugo:1992md,Alvarez:1996vt,Hull:2006va}.

In the general case with dependence on both $x^a$ and $\tilde x_a$
one has 
both diffeomorphisms and dual diffeomorphisms, giving an intriguing structure of \lq doubled diffeomorphisms'. 
Moreover, we will show the diffeomorphisms and 
antisymmetric tensor gauge transformations become closely linked, with the roles of the parameters interchanged by T-duality.  
The consistency of this free theory
hinges crucially on the constraint $\Delta =0$ satisfied by the fields
and gauge parameters.  
Given the general interest in theories on doubled tori, we 
analyze the free theory further 
 and  find that  linearised double diffeomorphisms 
{\em cannot} be realised with 
the $h_{ij}$ field 
alone:  the Kalb-Ramond and dilaton fields  must be added.  
While diffeomorphism
symmetry does not fix the field content of the massless sector of 
closed string theory,  `double diffeomorphisms' does!   

 We are guided by string field theory to build  a remarkable interacting generalisation of the linearised massless theory 
described above.  In doing so we obtain a two-derivative theory
with a gauge invariance that is the nonlinear version of the
doubled diffeomorphisms found in the quadratic theory.  
The constraint
$\Delta =0$ remains unmodified and the theory remains a theory of constrained 
fields.  The  action is given in (\ref{redef-action})  and  
 the gauge transformations are given 
in (\ref{gaugetrans}).  The theory also has a discrete $\mathbb{Z}_2$
symmetry~(\ref{disc-sym-final}) that arises from the orientation invariance of the underlying closed string theory.  It should be emphasized that
the quadratic part of the action that we write is exactly that of the
string field theory, but the cubic part of the action is not.  In constructing
this cubic part we
drop all terms with more than two derivatives.
We also  drop the momentum-dependent
sign factors due to cocycles that enforce the mutual locality of
 vertex operators~\cite{Frenkel,Gross:1985rr,Hata:1986mz,Maeno:1989uc,Hellerman:2006tx}.  
Gauge invariance works to this order   
without the inclusion of such terms, 
although some may be needed 
to achieve a complete nonlinear construction. 
The role of  sign factors is discussed in
Section~\ref{constandcoc}.

The symmetry
algebra of closed string field theory is not a Lie algebra (the Jacobi identitites do not hold) as in familiar theories, but rather 
  a homotopy Lie algebra~\cite{Zwiebach:1992ie}.  
The structure of the interactions we find in our double field theory leads to a symmetry algebra that appears not to be a Lie algebra, 
suggesting that some of the homotopy structure of the string field theory
survives in the massless theory. 
 As we discuss in 
 Section~\ref{constandcoc},
  an explicit projector is needed 
 so that the product of two fields in the kernel of $\Delta$
 is also in the kernel of $\Delta$.
 The presence of  this projection is part of the
 reason the brackets that define the composition of gauge parameters
 do not satisfy a Jacobi identity.   Understanding the full symmetry of
 the theory 
  is a central open problem.  Further discussion of 
 open problems and directions for further research can be found 
 in~\S\ref{coandopenque}.

\medskip
In closing this introduction we note that the work here furnishes some new
results in closed string field theory.  The cubic 
theory of the massless fields, required to see the full structure of diffeomorphisms, was not worked out before.
The formulation of gravity
in string theory uses auxiliary fields that must be eliminated
using their equations of motion as well as a gauge
trivial scalar field that must be carefully gauged away.  Field redefinitions are needed to obtain a simple
form of the gauge transformations.  
In  the end, the formulation
of gravity plus antisymmetric field and a dilaton in string theory
is extremely efficient; it uses $e_{ij}   
= h_{ij} + b_{ij}$ and 
a duality-invariant scalar $d$ (related to linearized 
order to
the usual dilaton $\phi $ by $d= \phi- \frac 1 4 h$).
 The 
 cubic action we present is much  
 simpler than the cubic action
obtained by direct expansion of the familiar action for gravity, antisymmetric tensor, and dilaton.  
 The results in this paper suffice
to find the field redefinitions that connect the string field theory and sigma
model fields for the massless sector of the closed string to quadratic
order in the fields and without derivatives.   
Earlier work in this direction includes that of~\cite{Ghoshal:1991pu},
which discussed general coordinate invariance in closed string field
theory and~\cite{Michishita:2006dr}, 
which studied the constraints that
T-duality imposes on the relation between closed string fields and sigma
model fields.

\sectiono{The Free Theory}

In this section we begin by giving an
argument supporting our claim that linearised
double diffeomorphism invariance 
requires the massless multiplet of closed string theory.  
We then 
review
 closed string theory on toroidal backgrounds,
setting the notation and giving the basic results 
used in this paper.  We then use the
free closed string field theory to
construct the free double field theory.   We study the symmetries in detail and
emphasize the differences with the conventional free theory
of gravity, antisymmetric field, and dilaton.

\subsection{Linearised double diffeomorphism  symmetry}\label{amultfordoub}

 In the introduction we introduced $M^2=0$ fields depending on  
 ($x^\mu, x^a, \tilde x_a$)  with 
   linearised transformations (\ref{diffeo-tilda}). 
   These included a  field
       $h_{ij}( x^\mu, x^a, \tilde x_a)$
   transforming under 
   linearised diffeomorphisms~as
\be
\label{lindiff}
\delta h_{ij} = \partial_i \epsilon_j + \partial_j \epsilon_i \,.
\ee
and under  
linearised \lq dual diffeomorphisms' as
\be
\label{lindiffdual}
\tilde\delta h_{ij} = \tilde\partial_i \tilde \epsilon_j + \tilde\partial_j 
\tilde\epsilon_i \,.
\ee
We will now show why we cannot have a theory of $h_{ij}$ alone that is invariant under 
such
 \lq double diffeomorphisms'. We will find that introducing a Kalb-Ramond field and a dilaton is essential, and that the constraint $\Delta =0$ must
 be satisfied for invariance.

For Einstein's gravity  $S= {1\over 2\kappa^2} \int \sqrt{-g} R$, and
to quadratic order in the fluctuation field $h_{ij}(x) \equiv g_{ij}(x) - \eta_{ij}$ one 
has
\be
\label{linein}
(2\kappa^2)\,S_0=  \int dx \Bigl[~ {1\over 4}  h^{ij} \partial^2 h_{ij} 
 - {1\over 4}  \, h \partial^2 h  
+ {1\over 2} (\partial^i h_{ij} )^2  
+ {1\over 2}  h \,\partial_i \partial_j \, h^{ij}~ \Bigr]\,.
\ee
This action, of course, is invariant under (\ref{lindiff}), but we wish to
 implement also the dual 
diffeomorphisms
(\ref{lindiffdual}). 
For a field $h_{ij} (\tilde x, x)$ depending on both $x$ and $\ti x$, the action is an integral over   the full $n+2d$ dimensional doubled space. 
We will denote this integral as $\int [dx d\tilde x]$.   
The natural action is
\be
\label{eoi983hg}
\begin{split}
(2\kappa^2)\,S=  \int [dx d\tilde x] \Bigl[~& {1\over 4}  h^{ij} \partial^2 h_{ij} 
 - {1\over 4}  \, h \partial^2 h  
+ {1\over 2} (\partial^i h_{ij} )^2  
+ {1\over 2}  h \,\partial_i \partial_j \, h^{ij} \\
+&{1\over 4}  h^{ij} \tilde\partial^2 h_{ij} 
 - {1\over 4}  \, h \tilde \partial^2 h  
+ {1\over 2} (\tilde\partial^i h_{ij} )^2  
+ {1\over 2}  h \,\tilde\partial_i \tilde\partial_j \, h^{ij} ~ \Bigr]\,.   
\end{split}
\ee
For a gravity field  
 $h_{ij}(x^i)$
independent of $\ti x_a$ the action reduces to the linearised Einstein action 
on the space with coordinates $x^i$.  
 For a gravity field $h_{ij}(x^\mu, \ti x_a)$
independent of $x^a$ the action reduces
 to the linearised Einstein action on the dual space with coordinates $x^\mu , \ti x_a$. 
The first line  in (\ref{eoi983hg})   
is invariant under the $\delta$ transformations (\ref{lindiff}), the second is invariant
under the $\tilde \delta$ transformations (\ref{lindiffdual}). 

Let us vary  
the double action $S$ under $\tilde\delta$.  The second line is invariant and varying the first   
gives
\be   
\begin{split}
(2\kappa^2) \,\tilde\delta S = \int  [dx d\tilde x] \, \Bigl[~ ~& 
h^{ij} \partial^2 \tilde\partial_i \tilde\epsilon_j   
+\partial_i h^{ij}  \, (\partial^k  \tilde  \partial_k) \tilde \epsilon_j\\
 -&  \, h\, \partial^2  \, \tilde \partial \cdot \tilde\epsilon 
~+ h\,(\partial_i  \tilde\partial^i) \partial_j\tilde \epsilon^j  \\[0.4ex]
 + &
 \,\partial_i h^{ij}  \, \partial^k  \, \tilde\partial_j \tilde\epsilon_k   
 +    \,( \partial_i \partial_j  h^{ij} ) \tilde\partial\cdot \tilde\epsilon   
  ~\Bigr] \,.
\end{split}
\ee
We have organised the right-hand side so that the terms on
each line would cancel if the tilde derivatives were replaced
by ordinary derivatives.  As we can see, no cancellation 
whatsoever takes place!
Grouping related terms we have 
\be
\begin{split}
(2\kappa^2)\, \tilde\delta S = \int  [dx d\tilde x] \, \Bigl[~& 
h^{ij} \partial^2 \tilde\partial_i \tilde\epsilon_j   
-h^{ij} \partial_i \partial^k \tilde \partial_j \tilde \epsilon_k
+ (\partial_i \partial_j  h^{ij} -  \partial^2h\,)  \, \tilde \partial \cdot \tilde\epsilon 
\\
+&\,(\partial^i h_{ij}   - \partial_jh  ) \, (\partial \cdot \tilde  \partial) \tilde \epsilon^j~\Bigr] \,.
\end{split}
\ee
The terms on the second line    vanish when the gauge
parameter $\tilde \epsilon$ satisfies the constraint $\partial \cdot \tilde\partial =0$. Relabeling the indices on the first two terms, we 
get 
\be
\label{jhdfsjka}
\begin{split}
(2\kappa^2) \, \tilde\delta S = \int  [dx d\tilde x] \, \Bigl[~ &
 (\tilde\partial_j h^{ij}) \partial^k ( 
  \partial_i \tilde \epsilon_k - \partial_k \tilde\epsilon_i   )
+ (\partial_i \partial_j  h^{ij} -  \partial^2h\,)  \, \tilde \partial \cdot \tilde\epsilon
\\ 
+&(\partial^i h_{ij}   - \partial_jh  ) \, (\partial \cdot \tilde  \partial) \tilde \epsilon^j~\Bigr] \,.
\end{split}
\ee
In order to   cancel this variation  we need new
fields with new gauge transformations.  To cancel the first term
we can use a Kalb-Ramond field $b_{ij}$ and a new term $S_1$ in the
action:
\be
(2\kappa^2) \, S_1 = \int [dx d\tilde x] ~   (\tilde\partial_j h^{ij}) \partial^k b_{ik} \,,
\qquad \hbox{with} \quad  \tilde \delta b_{ij} = - (  \partial_i \tilde \epsilon_j - \partial_j \tilde\epsilon_i)\,.
\ee
To cancel the second term in (\ref{jhdfsjka}) we introduce  a 
dilaton $\phi$ 
and a  new 
term $S_2$ given by
\be
(2\kappa^2)  S_2 = \int [dx d\tilde x] (-2) (\partial_i \partial_j  h^{ij} -  \partial^2h\,)  \phi \,,\quad \hbox{with} \qquad \tilde \delta \phi = {1\over 2} \, \tilde
 \partial\cdot \tilde \epsilon\,.
\ee
The above are the first steps in the construction of a consistent
quadratic theory.  More terms are
needed, and we will find the full, invariant quadratic action from the closed string field theory in~\S\ref{quaactfro}.  
The lessons are clear, however. 
Implementation of 
linearised doubled diffeomorphisms for 
$h_{ij}$  requires the addition of further
fields, most naturally, a Kalb-Ramond gauge field and a dilaton.
Moreover, a second-order differential constraint is required: fields and
gauge parameters must be annihilated by $\partial\cdot \tilde \partial$.
In fact, to this order, it suffices
for the gauge parameters to satisfy the constraint.

It is natural to ask if by adding further fields one can find
an action that is invariant without the constraint.  
The offending term on the second line of (\ref{jhdfsjka}) can be cancelled in this way, but then further terms are needed. 
We have not been able to find a non-trivial theory
that is invariant under both $\delta $ and $\ti \delta $ transformations 
 without use of the constraint.

\subsection{General toroidal backgrounds}\label{gentorback}

An explicit discussion of closed string field theory in
toroidal backgrounds was given in  the work of Kugo and Zwiebach~\cite{Kugo:1992md}. 
 Following this work, we  review the basic results that will be needed
 here.
 We begin with the string action, given by\footnote{Our formulae will
 keep explicit factors of $\alpha'$.  In the worldsheet action
 (\ref{wsaction}) $G_{ij}, B_{ij},$ and the $X^i$ are all
dimensionless.} 
\be
\label{wsaction}
S = -{1\over 4\pi}  \int_0^{2\pi} d\sigma \int d\tau \bigl(  \sqrt{\gamma} \gamma^{\alpha\beta} \partial_\alpha X^i \partial_\beta X^j
  G_{ij}  
 + \epsilon^{\alpha\beta} \partial_\alpha X^i \partial_\beta X^j 
 B_{ij} \bigr)\,.
\ee
The string coordinates
\be
X^i  =  \{  X^a \,,  X^\mu\} \,,   
\ee
split into string  coordinates $X^\mu$ for $n$-dimensional Minkowski space 
and periodic string coordinates $ X^a$ for the 
internal $d$-dimensional torus:
\be
\label{identif}
X^a \sim X^a + 2\pi \,.
\ee
In the above action $G_{ij}$ and $B_{ij}$ are the constant
background metric and antisymmetric tensor, respectively.
As usual, we define the inverse metric with upper indices:
\be
G^{ij} G_{jk} =
\delta^i_k\,.
\ee    
The background fields are taken to be
    \be
 G_{ij} = \begin{pmatrix} \hat G_{ab}  & 0 \\[0.5ex] 0 & \eta_{\mu\nu} \end{pmatrix}\,, \qquad
   B_{ij} = \begin{pmatrix} \hat B_{ab} & 0 \\[0.5ex] 0 & 0
   \end{pmatrix}\,.
 \ee 
and we define
\be
 E_{ij}   \equiv \,  G_{ij} + B_{ij}
= \begin{pmatrix} \hat E_{ab}  &0 \\ 0 &\eta_{\mu\nu} \end{pmatrix}\,,
~~ \hat E_{ab} \equiv  \hat G_{ab} + \hat B_{ab}\,.
\ee
The Hamiltonian $H$   
for this theory takes the form
\be
4\pi H =  (X'\,, 2\pi P)  \,\HH(E)  \begin{pmatrix} X'\\[1.0ex] 2\pi P\end{pmatrix}
\,, \ee
where
   the derivatives of the coordinates ${X^i}'= \partial _\sigma X^i$ and the momenta $P_i$  are combined into a $2D$ dimensional column
vector
and 
the $2D\times 2D $ matrix $\HH$ is given by
\be
\label{genmet}
 \HH(E) = \begin{pmatrix}  ~G - B G^{-1}
B~  &  B G^{-1}  \\[2.0ex]  -G^{-1} B  &  G^{-1} \end{pmatrix}\,.   
\ee
The matrix $\HH(E)$ satisfies the constraint 
$\HH^{-1} =  \eta  \HH \eta$.   

The mode expansions for $X^i, P_i$ and the dual coordinates  
 $\tilde X_i$ take the form
\be
\begin{split}
X^i(\tau, \sigma) &=  x^i  + w^i \sigma +  \tau \, G^{ij} (p_j - B_{jk} w^k)  + 
{i\over \sqrt{2}} \sum_{n\not=0}  {1\over n} \bigl[ \alpha_n^i e^{in\sigma}
+\bar \alpha_n^i e^{-in\sigma} \bigr] e^{-in\tau}\,,
\\
2\pi P_i (\tau, \sigma) &= p_i  + {i\over \sqrt{2}} \sum_{n\not=0}  \bigl[
E_{ij}^t \,\alpha_n^j e^{in\sigma}
+E_{ij} \,\bar \alpha_n^j  e^{-in\sigma} \bigr] e^{-in\tau}\,,\\
\tilde X_i (\tau, \sigma) &= \tilde x_i + p_i \sigma  +\hskip-1pt \tau \bigl[ (G\hskip-1pt -\hskip-1pt BG^{-1}\hskip-1pt B)_{ij} w^j 
\hskip-1pt +\hskip-1pt (B G^{-1}\hskip-1pt)_i^{\, j}p_j \bigr] \hskip-1pt  + \hskip-1pt
{i\over \sqrt{2}} \sum_{n\not=0}  {1\over n} \bigl[ -E_{ij}^t \alpha_n^j e^{in\sigma}
\hskip-1pt+\hskip-2pt E_{ij} \bar \alpha_n^j e^{-in\sigma} \bigr] e^{-in\tau}.
\end{split}
\ee
Given (\ref{identif}), $x^a \sim x^a + 2\pi$ and 
$w^a$ and $p_a$ take integer  values.  Conjugate to the winding charges $w^a$, there are periodic coordinates   $\tilde x_a$ satisfying $\tilde x_a \sim \tilde x_a + 2\pi$.
In the above expansions we use 
\be  
\begin{split}
w^i &= \{ w^a,  w^\mu\} =  \{  w^a, 0 \}   \,,\\
\tilde x_i &= \{  \tilde x_a,  \tilde x_\mu\} =  \{  \tilde x_a, 0 \}   \,,
\end{split}
\ee 
which state that there are no windings nor dual coordinates
along the Minkowski directions.  
We have the  commutation relations:
\be
[\, x^i , p_j\, ]  = i \,\delta^i_j \,, ~~ [\, \tilde x_i,  w^j\, ]  =  i \,\hat\delta^j_i\,,
\ee
where $\hat \delta^j_i = \hbox{diag} \{ \hat\delta^a_b, 0\}$  
so that the second relation is just
$[\, \tilde x_a,  w^b\, ]  =  i \,\hat\delta^b_a
$.
Moreover  
\be
\label{erocgkjer}
[\, \alpha_m^i, \,\alpha_n^j\, ] = [\, \bar\alpha_m^i, \,
\bar\alpha_n^j\, ]
= \,  m \,  
G^{ij}  \, \delta_{m+n,0}\,.
\ee
Finally, we have the zero-modes   
given by
\be
\begin{split}
\alpha_0^i &=~ {1\over \sqrt{2}} \, G^{ij}\, \bigl(p_j - E_{jk} w^k\bigr)\,,\\[1.0ex]
\alpha_0^i &= ~{1\over \sqrt{2}} \, G^{ij}\, \bigl(p_j + {E}^t_{jk} w^k\bigr)\,.
\end{split}
\ee
Lowering the indices and writing  in 
terms of the dimensionless coordinates $x^i$ and $\tilde x_i$ gives
\be
\label{derorign}
\begin{split}
\alpha_{0i} &=~- {i\over \sqrt{2}} \,  \Bigl({\partial\over \partial x^i} 
- E_{ik} {\partial \over \partial\tilde x_k}\Bigr)    
=- i \sqrt{\alpha'\over 2} \, D_i \,, \\[1.0ex]
\bar \alpha_{0i} &= ~-{i\over \sqrt{2}} \,  \Bigl({\partial\over \partial x^i} + {E}^t_{ik} {\partial \over \partial\tilde x_k}\Bigr)   
=- i \sqrt{\alpha'\over 2} \, \bar D_i \,,
\end{split}
\ee
where we introduced derivatives $D_i$ and $\bar D_i$ with the dimensions of inverse length and used $p_j = {1\over i} \partial_j$ as well as
$w^k = {1\over i} \tilde \partial^k$. 
 The derivatives $D$ and $\bar D$ can then be written as
\be
\label{groihfgruu8774}
\boxed{
\begin{split}
\phantom{\Biggl(}~D_i &=~{ 1\over \sqrt{\alpha'}} \,\Bigl(\,{\partial\over \partial x^i} - E_{ik} \,{\partial \over \partial\tilde x_k}\Bigr)\,,  \\[1.0ex]
\bar D_i &= ~{ 1\over \sqrt{\alpha'}} \,  \Bigl(\,{\partial\over \partial x^i} + {E}^t_{ik}\, {\partial \over \partial\tilde x_k}\Bigr)\,.
\phantom{\Biggr)}~~
\end{split}}
\ee
We work in Lorentzian signature (both for the worldsheet and 
spacetime) and  $D$ and $\bar D$ are independent real derivatives 
with respect 
to right- and left-moving coordinates  
$\tilde x_i - E_{ij} x^j$ and $ \tilde x_i + E^t_{ij}  x^j$, respectively. 
Indeed,  $\tilde X_i - E_{ij} X^j$ is a function of $(\sigma-\tau)$ and 
$ \tilde X_i + E^t_{ij}  X^j$  is  a function of $(\sigma+ \tau)$.
For the noncompact directions there are no
dual  derivatives and we have
\be  
\label{iwlt4754675389vg} 
 {\partial \over \partial\tilde x_i} = \Bigl\{ 
  {\partial \over \partial\tilde x_a}\,, 0\,\Bigr\}\,.
\ee
As a consequence, while $D_a \not= \bar D_a$ we have $D_\mu= \bar D_\mu$.

It is useful  
to introduce operators  $\square$ and   $\Delta$, both  quadratic
in the $\alpha_0$ and $\bar \alpha_0$ operators:
\be
\begin{split}
 -{\alpha'\over 2} \square  ~ &\equiv \,{1\over 2}  \alpha_0^i  \, G_{ij}  \alpha_0^j  + {1\over 2} \, \bar\alpha_0^i  \, G_{ij}  \bar\alpha_0^j 
 \,, \\[1.0ex]
 -{\alpha'\over 2}    
 \Delta ~&\equiv \, {1\over 2} \alpha_0^i  \, G_{ij}  \alpha_0^j  - 
 \,{1\over 2} \bar\alpha_0^i  \, G_{ij}  \bar\alpha_0^j  \,.
\end{split}
\ee
We note that, in general   
\be
L_0 - \bar L_0 =  N - \bar N -{\alpha'\over 2}  \, \Delta \,,
\ee
so that the level matching condition for fields with $N = \bar N$
becomes the constraint $\Delta =0$.
In terms of our derivatives, we get
\be
\begin{split}
 \square &= {1\over 2} \,D^i  \, G_{ij}  D^j  + 
{1\over 2} \, \bar D^i  \, G_{ij}  \bar D^j ~= {1\over 2} \bigl( \,D^i  D_i  + 
 \bar D^j    \bar D_j\bigr)\,, \\[0.8ex]
\Delta&=  {1\over 2}\, D^i  \,  G_{ij}  D^j \, - {1\over 2} \, \bar D^i  \,  G_{ij}  \bar D^j
\,= {1\over 2}\bigl(\,D^i  D_i  - 
 \bar D^j    \bar D_j  \bigr) \,.
\end{split}
\ee
Writing $D^2 \equiv D^iD_i$  and $\bar D^i \bar D_i = \bar D^2$ we have
\be
\label{spflkd}
\boxed{
\begin{split}
\phantom{~~\Biggl(} \square = {1\over 2} ( D^2 + \bar D^2)  \,, ~~
\Delta =  {1\over 2} ( D^2 - \bar D^2)  \,.~ 
\phantom{~~\Biggl(}
\end{split}}
\ee
An explicit computation using the expressions for $\alpha_0$
and $\bar \alpha_0$ gives
\be
\square = {1\over \alpha'} \Bigl( G^{ij}\,{\partial\over \partial x^i} \, {\partial\over \partial x^j} + 2\,(B G^{-1})_i^{~j}\, \, {\partial\over \partial \tilde x_i}\,{\partial\over \partial x^j} + (
G - BG^{-1} B)_{ij}\,{\partial\over \partial \tilde x_i}  {\partial\over \partial \tilde x_j}\Bigr)\,.
\ee
Note that the contribution  to $\square$   from  the non-compact directions is  the
expected term ${1\over \alpha'} \eta^{\mu\nu} \partial_\mu \partial_\nu$.
Recalling that $E$ is a constant, this can be rewritten as     
\be
\square = {1\over \alpha'}  
~ \partial^t ~\HH(E) ~ \partial \,,~~\hbox{with}~~\partial =   \begin{pmatrix}  {\partial \over \partial \tilde x_i} \\[1.0ex] 
{\partial \over \partial x^j} \end{pmatrix}\,.
\ee
Another short computation, together with   (\ref{iwlt4754675389vg}), 
shows that the   
operator $\Delta$ takes the form
\be
\label{deltaconstraint}
\boxed{  
\phantom{~~\Biggl(}
\Delta = -{2\over \alpha'} \sum_i {\partial \over \partial \tilde x_i}  {\partial\over \partial x^i} = 
-{2\over \alpha'}\sum_a {\partial \over \partial \tilde x_a}  {\partial\over \partial x^a} 
 \,.\phantom{~~\Biggl(}} 
\ee
Note that no background fields are 
required here. 
We can also write 
\be
\Delta = -{1\over \alpha'}  
~ \partial^t ~\eta~ \partial \,,~~\hbox{with}~~
\eta =\begin{pmatrix}0& I \\ I& 0 \end{pmatrix}\  \,.
\ee  
While $\square$ is a Laplacian for the 
 metric $\HH(E) $,
$\Delta$ is one for the $O(D,D)$ invariant metric $\eta$.

In  string field theory  
the physical state conditions $L_0+ \bar L_0-2=0$  and $L_0- \bar L_0=0$ are treated very differently.  The former arises from 
the free string field equation of motion and gives equations of the form
$\square \Phi_I=....$ for the component fields $\Phi_I(x, \ti x)$.
The latter 
 is imposed as a constraint on  
the string field,  
so that the fields with $N=\bar N$ are required to satisfy 
$$\Delta \Phi_I=0\,.$$
As usual, we include the standard $bc$ ghost system with ghost oscillators
$b_n,c_n,\bar b_n, \bar c_n$.

\medskip
\noindent
{\bf Rectangular Tori:}~Let us consider the case where 
$\hat B_{ab}=0$  
and the metric is diagonal.  
If $R_a$ denotes the 
physical radius of the circle $X^a \sim X^a + 2\pi$ we  have
\be
\hat E_{ab} = \hat G_{ab}  =  {R_a^2 \over \alpha'} \, \delta_{ab}\,, \qquad
\hat G^{ab} ={ \alpha'\over R_a^2}\, \delta^{ab}\,.  
\ee
For the derivatives we find
\be
\label{eoivvbve8i}
\begin{split}
D_a &=~{ 1\over \sqrt{\alpha'}} \,\Bigl(\,{\partial\over \partial x^a} - {R_i^2 \over \alpha'} \delta_{ab} \,{\partial \over \partial\tilde x_b}\Bigr) 
\,, ~~
\bar D_a 
= ~{ 1\over \sqrt{\alpha'}} \,  \Bigl(\,{\partial\over \partial x^a} + {R_i^2 \over \alpha'}\,\delta_{ab} {\partial \over \partial\tilde x_b}\Bigr)\,,
\\[1.0ex]
\square &= {1\over \alpha'} \Bigl(\,  \eta^{\mu\nu} \partial_\mu \partial_\nu
+ { \alpha'\over R_i^2}\,\delta^{ab} {\partial\over \partial x^a} \, {\partial\over \partial x^b}  + 
{R_i^2 \over \alpha'}\,\delta_{ab}{\partial\over \partial \tilde x_a}  {\partial\over \partial \tilde x_b}\Bigr)\,.
\end{split}
\ee
We can 
 introduce coordinates $u^a$ and $\tilde u_a$ that 
have physical lengths (repeated indices not summed)
\be
\begin{split}
u^a  &= R_a\, x^a \,, ~~\quad  u^a\sim u^a + 2\pi R_a \,,  
~~~~
\tilde u_a  
= {\alpha'\over R_a}\, \tilde x_a\,,  ~~\quad  \tilde u_a \sim  \tilde u_a + 2\pi {\alpha'\over R_a} \,.\\[1.0ex]
\end{split}
\ee
For the noncompact directions we can take  
$u^\mu = \sqrt{\alpha'} x^\mu$. We then get
\be
\begin{split}
D_a &=~{ R_a\over \sqrt{\alpha'}} \, 
\Bigl(\,{\partial\over \partial u^a} - \,
\delta_{ab}{\partial \over \partial\tilde u_b}\Bigr)\,, ~~~
\bar D_a 
= ~{ R_a\over \sqrt{\alpha'}} \,  \Bigl(\,{\partial\over \partial u^a} +  \delta_{ab} {\partial \over \partial\tilde u_b}\Bigr)\,,  \\[1.0ex]
\square &=\eta^{\mu\nu} {\partial\over \partial u^\mu} \, {\partial\over \partial u^\nu}+ \delta^{ab} {\partial\over \partial u^a} \, {\partial\over \partial u^b}  + 
\delta_{ab}\,{\partial\over \partial \tilde u_a} 
 {\partial\over \partial \tilde u_b}\,.
\end{split}
\ee

\subsection{Quadratic action from string field theory}\label{quaactfro}

The closed string field with $N = \bar N = 1$ takes the form
\be
\label{the_string_field}
\begin{split}
\ket{\Psi} & =  \int [dp] ~ \Bigl(  - {1\over 2} 
e_{ij} ( p) \,\alpha_{-1}^i \bar \alpha_{-1}^j \, c_1 \bar c_1
 + e(p)  \, c_1 c_{-1}    +  \bar e (p)  
\, \bar c_1 \bar c_{-1} \\[0.5ex]
&\hskip70pt+ i\sqrt{\alpha'\over 2} \,\bigl(
\,  f_i (p) \, c_0^+ c_1 \alpha_{-1}^i 
+ \bar f_j (p) \, c_0^+ \bar c_1 \bar \alpha_{-1}^j\Bigr)\, \ket{p} \,.
\end{split}
\ee
We have used $\int [dp]$ to denote the     integral over the continuous momenta $p_\mu$
and the sum over the discrete momenta $p_a$ and discrete winding 
$w^a$ so that, for example,  $e(p)=e(p_\mu,p_a,w^a)$.  
The string field 
has ghost number two: each term
includes two ghost oscillators acting on the ghost-number zero
state $\ket{p}$.  
In the above $c_0^\pm = \half (c_0 \pm \bar c_0)$ and we define
$b_0^\pm  =  b_0 \pm \bar b_0$, 
so that
$\{ c_0^\pm, b_0^\pm\} = 1$.
 As required, $b_0^- \ket{\Psi} =0$ because
$b_0^-\ket{p}=0$ and the ghost oscillator $c_0^-$ does not appear
in $\ket{\Psi}$.  
This   expansion of the string field features five
momentum-space component fields: $e_{ij}, e, \bar e, f,$ and $\bar f$.

We wish to construct
the quadratic action, given by
\be
\label{quad-action-def-89}
(2\kappa^2 )\,S^{(2)} =    -{2\over\alpha'}  \bra{\Psi} \, c_0^-
Q \ket{\Psi} ~ \,.
\ee
Here $Q$ is the (ghost-number one) BRST operator of the 
conformal field theory and $\bra{\Psi}$ denotes the BPZ 
conjugate of the string field $\ket{\Psi}$ in (\ref{the_string_field}). The computation of $S^{(2)}$ is straightforward\footnote{We use the inner product
$\bra{p'} \,c_{-1} \bar c_{-1} c_0^- c_0^+  c_1 \bar c_1\, \ket{p} 
= (2\pi)^{n+ 2d}\,  \delta (p-p')$.
The BRST operator is 
$
Q = -{\alpha'\over 2} \,c_0^+  \square
 + \alpha_0 \cdot \big( \alpha_{-1}  c_1 + c_{-1} \alpha_1 \bigr)
+\bar \alpha_0 \cdot \big( \bar\alpha_{-1} \bar c_1 + 
\bar c_{-1} \bar\alpha_1 \bigr)- b_0^+ ( c_{-1} c_1 + \bar c_{-1} \bar c_1 ) + \ldots$.}   and the result is
\be
\label{quad-action}
\begin{split}
(2\kappa^2)\,S^{(2)} &=    \int [dx \,d\tilde x]\,
\Bigl[\, \, {1\over 4} e_{ij} \square e^{ij} 
\,+\, 2 \bar{e} \, \square \, e
- f_i\, f^i - \bar f_i\, \bar f^i\\[1.0ex]
&\hskip55pt - \, f^i  \,\bigl(
\, \bar D^j e_{ij}   -2 \,D_i\bar e \bigr)
 + \bar f^j  \,
\bigl(\, D^i e_{ij} \, + 2\, \bar D_j e \bigr)\Bigr]\,.
\end{split}
\ee
Here $\int [dx d \ti x ] =\int  d^n x^\mu  d^d x^a  d^d\tilde x_a$ is an integral over all $n+2d$ coordinates of $\R^{n-1,1}\times T^{2d}$. 
The definitions of $\square, D,$ and $\bar D$ were 
 given 
in~\S\ref{gentorback}.  
 All indices are raised and lowered with the metric $G^{ij}$.
The gauge parameter $\ket{\Lambda}$  for the
linearised gauge transformations is  
\be
\label{gt-param}
\ket{\Lambda}  =  \int [dp] ~ \Bigl(  
 {i\over \sqrt{2\alpha'}} 
\lambda_i ( p) \,\alpha_{-1}^i  c_1
- {i\over \sqrt{2\alpha'}} 
\bar\lambda_i ( p) \,\bar\alpha_{-1}^i  \bar c_1
 + \mu(p)  \, c_0^+ \Bigr) \ket{p} \,.
\ee   
The string field $\Lambda$ has ghost number one and
is annihilated by $b_0^-$.  It encodes two vectorial gauge parameters
$\lambda_i$ and $\bar \lambda_i$ and one scalar gauge parameter $\mu$.
The consistency of the string field theory 
requires
the level-matching 
conditions (\ref{lmcondroj}). 
As a result, the fields $e_{ij},d, e,\bar e, f_i, \bar f_i$ and the gauge parameters $\lambda, \bar \lambda, \mu$ must be annihilated
by $\Delta$ (defined in (\ref{deltaconstraint})):
\be  
\Delta e_{ij}  = \Delta d=  \Delta e= \Delta \bar e = \Delta f_i
= \Delta \bar f_i = 0 \,, ~~ ~~\Delta \lambda_i =
\Delta \bar \lambda_i = \Delta \mu = 0 \,.
\ee 
The quadratic string action (\ref{quad-action-def-89}) is invariant under the gauge transformations
\be 
\label{gtfa}
\delta \ket{\Psi}=Q  \ket{\Lambda} \,.
\ee
Expanding this equation using (\ref{gt-param}) and (\ref{the_string_field}) gives
  the following gauge transformations of the component fields:
\be
\label{collgt}
\begin{split}
\delta e_{ij} &= D_i
\bar\lambda_j  +\bar D_j \lambda_i \,, \\[1.0ex]
\delta f_i &= -\half  \, \square \,\lambda_i + D_i \mu  \,,\\[1.0ex]
\delta \bar f_i  & = \phantom{-}\half  \, \square \,\bar \lambda_i
+ \bar  D_i \mu\,,\\[1.0ex]
\delta e &=  -\half D^i\lambda_i +  \mu\,, \\[1.0ex]
\delta \bar e & = \phantom{-}\half \bar D^i\bar \lambda_i +\, \mu\,.
\end{split}
\ee
We can now introduce fields $d$ and $\chi$ by
\be
d= {1\over 2}\, (e - \bar e)  \,, \qquad  \hbox{and}
\quad  \chi =  {1\over 2}\, (e+ \bar e)\,.
\ee
The gauge transformations of $d$ and $\chi$ are
\be
\begin{split}
\delta d &= -{1\over 4}  (D^i\lambda_i   + \bar D^i\bar \lambda_i )\,,\\
\delta \chi & = -{1\over 4} ( D^i\lambda_i  - \bar D^i\bar \lambda_i)+ \mu\,.
\end{split}
\ee
We can use $\mu$ to make the gauge choice 
$$\chi=0\,. $$
After this choice is made, gauge transformations with arbitrary $\lambda$
and $\bar \lambda$ require compensating $\mu$ transformations
to preserve $\chi=0$.  These
do not affect  $d$ or $e_{ij}$, as neither    
transforms under $\mu$ gauge transformations. 
It does change the gauge transformations of  $f$ and
$\bar f$, but this is of no concern  here
as these auxiliary fields  will be eliminated 
using their equations of motion. 
 Therefore, we set 
$e = d$ and $\bar e = -d$ in (\ref{quad-action}) and eliminate
the auxiliary fields $f_i$ and $\bar f_i$, using
\be
\label{f-elim-quad}
f_i = -{1\over 2} \bigl(  \bar D^j e_{ij} - 2 D_i \bar e\bigr)\,,\qquad
\bar f_j = {1\over 2}  \bigl( D^i e_{ij} + 2 \bar D_j e \bigr) \,.
\ee
The result is the following quadratic action
\be
\label{quad-action-final} 
\boxed{\phantom{\Biggl(}(2\kappa^2) \,S ^{(2)}=    \int [d x d\tilde x] \,
\Bigl[\, \, {1\over 4} e_{ij} \square e^{ij}
 + {1\over 4} (\bar D^j e_{ij})^2
+ {1\over 4} ( D^i e_{ij})^2  - 2 \, d \, D^i \bar D^j e_{ij}
\,-\, 4 \,d \, \square \, d \,\,\Bigr]\,.~~}
\ee
The gauge transformations generated by $\lambda$ are 
\be
\label{lamtrans}
\begin{split}
\delta_\lambda e_{ij}  &= ~ \bar D_j \lambda_i \,,\\[1.0ex]
\delta_\lambda d~ & = -{1\over 4} D\cdot \lambda \,,
\end{split}
\ee
and the gauge transformations generated by $\bar \lambda$ are
\be
\label{lambartrans}
\begin{split}
\delta_{\bar \lambda} e_{ij}  &= ~D_i \bar\lambda_j \,,  \\[1.0ex]
\delta_{\bar \lambda} d~ & = -{1\over 4}  \bar D \cdot \bar \lambda\,,
\end{split}
\ee
where we use a dot to indicate index contraction: $a \cdot b\equiv 
a^i b_i $.
The action is invariant under the  $\mathbb{Z}_2$ symmetry
\be
\label{z2-sym}
e_{ij} ~\to~  e_{ji}\,, \quad  D_i ~\to ~ \bar D_i\,, \quad
\bar D_i  ~\to ~ D_i \,, \quad  d ~\to ~  d~ \,,  
\ee
which,
as we will discuss later,
is related
to the invariance of the closed string theory under orientation reversal.
For our present purposes we note that 
this relates the $\delta_\lambda$
and $\delta_{\bar \lambda}$ transformations, so that 
invariance under this  $\mathbb{Z}_2$  and $\delta_\lambda$ implies invariance under
$\delta_{\bar \lambda}$.

A short computation using (\ref{spflkd}) shows that the variation 
$\delta = \delta_\lambda + \delta_{\bar\lambda}$ of the action 
(\ref{quad-action-final}) gives
\be
\label{vmorg}
(2\kappa^2)\,  
\delta  S ^{(2)}=  \int [d x d\tilde x] \, \Bigl[\, 
{1\over 2}  \, e^{ij}  \Delta  ( \bar D_j \lambda_i - D_i \bar \lambda_j) 
+ 2 d \Delta    (D\cdot \lambda - \bar D \cdot \bar \lambda) 
\Bigr]\,.
\ee
As expected, the variation  vanishes only if we use
 the constraint $\Delta =0$.   
 Note that it is sufficient for the invariance of the quadratic action that the parameters satisfy the constraints 
 $\Delta \lambda= \Delta \bar \lambda=0$. 
 We have attempted to relax the constraints by adding extra fields, but have been unable to  find a gauge invariant action without constraints. 
 
 The action (\ref{quad-action-final}) and the associated gauge transformations are completely general.
 They describe the dynamics of fluctuations about the toroidal
 background with background field $E_{ij}$.  This background
 field enters the action through the derivatives, as can be seen
 from (\ref{groihfgruu8774}).

\subsection{Comparison with conventional actions}\label{comwitcon}

We can compare our general free theory (\ref{quad-action-final})
with the one discussed in \S\ref{amultfordoub}.  
For this
we scale the coordinates
by $\sqrt{\alpha'}$ to give them dimensions of length and the derivatives (\ref{groihfgruu8774})
become
\be
\label{eoikjfkj999}
\begin{split}
D_i &= \partial_i - \tilde\partial_i -B_{ik}\tilde\partial^k  \,, \\[0.5ex]
 \bar D_i &= \partial_i + \tilde\partial_i -B_{ik}\tilde\partial^k\,,
\end{split}
\ee
where we defined
\be
\tilde \partial_i  \equiv G_{ik} \tilde \partial^k=G_{ik} {\partial\over \partial \tilde x_k}\,.
\ee  
Then  
\be
 \square = \partial^2 + \tilde\partial^2    + (B_{ij}\tilde\partial^j)^2 -2 B_{ij} \partial ^i \tilde\partial^j\,,  ~~\hbox{and}~~
  \Delta = - 2 \partial_i \tilde \partial^i \,.
  \ee 
Here $G_{ij}$ is used to raise and lower indices and $\partial ^2 = G^{ij} \partial_i \partial_j$, etc.  
For simplicity we will consider backgrounds with $B_{ij}=0$.   
The derivatives and laplacians   
above become
\be
\label{eoikjfkj99}
D_i = \partial_i - \tilde\partial_i \,, \quad 
\bar D_i = \partial_i + \tilde\partial_i\,,~~
 \square = \partial^2 + \tilde\partial^2\,,  ~~\hbox{and}~~
  \Delta = - 2 \, \partial_i \tilde \partial^i \,.   
  \ee 
We decompose the field $e_{ij}$ into its symmetric and
antisymmetric parts:
\be
e_{ij} = h_{ij}  + b_{ij} \,, \quad\hbox{with} \quad 
    h_{ij} = h_{ji} \,, ~~b_{ij} = - b_{ji} \,.
\ee
The action (\ref{quad-action-final}) then gives 
\be
\label{eorkjr}
\begin{split}
(2\kappa^2) \,S^{(2)} = \int [dx d\tilde x] ~\Bigl[ & ~~~\,{1\over 4}  \, h^{ij} \partial^2  h_{ij} 
+ {1\over 2}  (\partial^j h_{ij})^2-2\, d\, \partial^i \partial^j \, h_{ij}  - 4\, d \,  \partial^2  \,d
\\[1.0ex]
& +{1\over 4}  \, h^{ij}  \tilde \partial^2  h_{ij}
+ {1\over 2}  (\tilde\partial^j h_{ij})^2 +2\, d\, \tilde\partial^i \tilde\partial^j \, h_{ij}  - 4\, d \,  \tilde \partial^2  \,d\\[1.0ex]
& + {1\over 4}  \, b^{ij} \partial^2 b_{ij} + {1\over 2}  (\partial^j b_{ij})^2
\\[1.0ex]
& +{1\over 4}  \, b^{ij} \tilde \partial^2  b_{ij} + {1\over 2}  (\tilde\partial^j b_{ij})^2
 \\[1.0ex]
& + (\partial_k h^{ik}) (\tilde \partial^j b_{ij}) + 
(\tilde\partial^k h_{ik}) ( \partial_j b^{ij}) 
- 4 \, d\, \partial^i \tilde \partial^j  b_{ij} ~~ \Bigr]\,.
 \end{split}
\ee
To appreciate this result, we recall the standard 
action $S_{\rm{st}}$ for gravity, Kalb-Ramond,  and dilaton fields 
\be 
\label{standardaction}
(2\kappa^2)S_{\rm{st}}= \int dx  \sqrt{-{\rm g}}\, 
e^{-2\phi}  
 \Bigl[  R- {1\over 12}  H^2+ 4 (\partial \phi)^2
\Bigr]\,.
\ee
We expand to quadratic order in fluctuations using
${\rm g}_{ij} = G_{ij} + h_{ij}$,  $\phi = d + {1\over 4}G^{ij}h_{ij}$,
and ${\rm b}_{ij} = B_{ij} + b_{ij}$, with constant $G_{ij}$ and $B_{ij}$.
It follows that $H_{ijk} = \partial_i b_{jk} + \cdots$, and we find
\be
\label{norm-action}
(2\kappa^2) \,S^{(2)}_{\rm{st}} = \int dx  ~
L[\,h,b,d;\partial\,]\,,
\ee
where
\be
\begin{split}
L[\,h,b,d;\partial\,]
=  
 ~\,&{1\over 4}  \, h^{ij} \partial^2  h_{ij} 
+ {1\over 2}  (\partial^j h_{ij})^2-2\, d\, \partial^i \partial^j \, h_{ij}  - 4\, d \,  \partial^2  \, d  
\\ +& {1\over 4}  \, b^{ij} \partial^2 b_{ij} + {1\over 2}  (\partial^j b_{ij})^2\,.
 \end{split}
 \ee
Comparing with (\ref{norm-action}) we see that our action (\ref{eorkjr}) 
can be written as
\be
\label{eorkjr9}
\begin{split}
(2\kappa^2) \,S^{(2)} &= \int [dx d\tilde x]    
~\Bigl[ ~ L[\,h,b,d;\partial\,]+L[\, h,b,-d;\ti \partial\,]
 \\[1.0ex]
& \hskip50pt + (\partial_k h^{ik}) (\tilde \partial^j b_{ij}) + 
(\tilde\partial^k h_{ik}) ( \partial_j b^{ij}) 
- 4 \, d\, \partial^i \tilde \partial^j  b_{ij} ~ \Bigr]\,.
 \end{split}
\ee
While in (\ref{norm-action})  the fields depend only
on the spacetime coordinates $x^i$, here they depend on $\ti x$ also.
The  lagrangian $L$ appears twice, first
with ordinary derivatives $\pa $ and then with dual derivatives $\ti \pa$, together with $d\to -d$.
 Finally, in the last line we have unusual terms
with mixed derivatives.  They introduce 
novel quadratic couplings between
the metric and the Kalb-Ramond field!  Finally, there is 
a new coupling of the dilaton to the Kalb-Ramond field. 

We now turn to the symmetries. The linearised version of
the standard action  (\ref{norm-action})     
is invariant under linearised  diffeomorphisms:
\be
\label{diffeogt}
\begin{split}
\delta h_{ij} &= \partial_i \epsilon_j + \partial_j \epsilon_i \,, \\
\delta b_{ij} &= ~0 \,,\\
\delta d~& = - \half  \partial\cdot \epsilon\,,
\end{split}
\ee
as well  
as antisymmetric tensor gauge transformations:  
\be
\label{krgt}
\begin{split}
\delta h_{ij} &=~~0 \,, \\
\delta b_{ij} &= -\partial_i \tilde\epsilon_j + \partial_j \tilde\epsilon_i  \,,\\
\delta d~& =~0\,.
\end{split}
\ee
Note that the scalar dilaton $\phi \equiv d + {1\over 4} G^{ij} h_{ij}$ 
is  invariant under  linearised  diffeomorphisms.

The symmetries of the double field theory 
 (\ref{eorkjr})  are (\ref{lamtrans})
and 
 (\ref{lambartrans}). 
Defining
\be
\epsilon_i \equiv {1\over 2} ( \lambda_i + \bar\lambda_i) \,, \qquad
\tilde \epsilon_i \equiv {1\over 2} ( \lambda_i - \bar\lambda_i) \,, \qquad
\ee
we can rewrite  these gauge transformations in a more familiar form.
The transformations with parameter $\epsilon$ are
\be
\label{diffeo-tild}
\begin{split}
\delta h_{ij} &= \phantom{-}\partial_i \epsilon_j + \partial_j \epsilon_i \,,
  \\[1.0ex]
\delta b_{ij} &=
- (\tilde \partial_i \epsilon_j - \tilde\partial_j \epsilon_i)\,,~~\\[1.0ex]
\delta d   \,\,    & =   -\half \,\partial\cdot \epsilon   \,.
\end{split}
\ee
These give transformations of the same form as  the linearised diffeomorphisms (\ref{diffeogt})  together with  an exotic 
gauge transformation of $b_{ij}$ 
in which dual derivatives~$\ti \pa$   
act on the parameter.  
The transformations with parameter 
$\tilde\epsilon$ are   
\be
\begin{split}
\tilde\delta h_{ij} &= \phantom{-} \tilde \partial_i \tilde\epsilon_j + \tilde\partial_j \tilde\epsilon_i \,, \\[1.0ex]
\tilde\delta b_{ij} &= -(\partial_i \tilde\epsilon_j - \partial_j \tilde\epsilon_i) 
\,,~~\\[1.0ex]
\tilde\delta d   \,\,    & =  ~~ \half\,
\tilde \partial \cdot \tilde\epsilon \,.
\end{split}
\ee
Comparing with (\ref{krgt}), we see   
Kalb-Ramond gauge transformations with parameter $\ti \epsilon$ together with
gravity  transformations 
 that are  
 linearised diffeomorphisms  
 with $\pa $ replaced by  
 $\ti \pa$.
 Note that this time the scalar dilaton is  $\tilde\phi \equiv d - {1\over 4} G^{ij} h_{ij}$,
 since this is invariant  under  linearised  dual diffeomorphisms. 
 Also interesting  
is that the transformation of $d$ under these 
dual diffeomorphisms is of the same form as
  the one in (\ref{diffeo-tild}), but 
 with 
opposite sign. 
While 
the  Minkowski space theory has a 
 gauge invariant
dilaton $\phi = d + {1\over 4} h$, there is none 
in the toroidal theory.
We certainly have  $\delta \phi = 0$,
but $\tilde \delta \phi = \tilde \partial \cdot \tilde \epsilon$.  
There
is no dilaton that is invariant under both $\epsilon$ and $\ti \epsilon$ transformations.

\sectiono{Cubic action and gauge transformations} 

In this section we use  closed string field theory  
to compute  the  
cubic interactions for the string field (\ref{the_string_field})
together with the gauge transformations
with parameter (\ref{gt-param})  to linear order in the fields. 
 The computation is laborious
since there are many terms to consider but the techniques are
standard in string field theory.  

In the action  
we have kept only the terms with a total number of 
derivatives  ($D$ or $\bar D$) less than or equal to two.
In the gauge transformations
we have kept the terms linear in the fields    
and which are relevant to an  action with two derivatives.  
 This strategy was expected to
lead to an action that is exactly gauge invariant to this order,
just as  it does for string field actions  around
flat space.  The constraint $\Delta =0$ does not involve terms
with different numbers of derivatives so no complication is expected.  

The string field theory product used to define the 
interactions involves a projector.   
The string fields satisfy  the constraint $L_0-\bar L_0=0$ and
the projector imposes the constraint $L_0-\bar L_0=0$ on the product.
Such projector should lead to a projector
that imposes the constraint $\Delta =0$ in our field theory products, and thus in our interactions. We discuss this in detail in section~\ref{constandcoc}. 
As we show there, however, when the fields satisfy the $\Delta=0$ constraint
no additional projectors are needed for the cubic interactions.
 The projectors are needed in the gauge transformations, in the terms that
 involve a product of a field and a gauge parameter.  In order
 to avoid cluttering the notation we will leave them implicit.  
 As explained in section~\ref{constandcoc} the check of gauge invariance
 to this order is correctly done naively, ignoring the projectors.

The vertex operators for strings on a torus include cocycles that lead to momentum-dependent sign  factors in the
exact cubic string field theory interactions,  
and these sign factors should also appear in our cubic double field theory action. 
  These factors are not expected to affect gauge invariance to cubic order. 
We present the results of this section without cocycle-induced sign factors, 
but will discuss these further in section~\ref{constandcoc}.  

As a check of our results, we used the gauge transformations
obtained in section~\ref{simthegautra}  to independently construct the cubic term in the action by the Noether
method.  The result is exactly the same cubic action that we present here.
We have also checked that the gravitational sector of the action  
agrees with that in the standard action (\ref{standardaction}), expanded to cubic order with the help of~\cite{Goroff:1985th}, for fields independent of $\ti x$ and in a gauge in which the metric perturbation is traceless.

\subsection{Cubic terms and gauge transformations from CSFT}

The string field theory action is non-polynomial and takes the form
\begin{equation}
\label{csft_action99}
(2\kappa^2) S= -\frac{2}{\alpha'}\Bigl[  \langle \Psi|c_0^-\,
Q|\Psi\rangle + \frac{1}{3} \{ \Psi,
\Psi\,,\Psi\} +\frac{1}{3\cdot 4} \{ \Psi,
\Psi\,,\Psi, \Psi\} +\cdots\, ~\Bigr] \,.
\end{equation}
Here $\{ \Psi, \Psi, \Psi \} = \bra{\Psi} c_0^- [ \Psi, \Psi ] \rangle$
and   $\{ \Psi, \Psi, \Psi , \Psi \} = \bra{\Psi} c_0^- [ \Psi, \Psi , \Psi ] \rangle$
where $[\cdot, \cdot ]$ is the closed string product and   $[\cdot, \cdot , \cdot ]$ is a triple product  of string fields.    
The higher order terms  
require the introduction of products of all orders, 
with relations between them implied by gauge invariance.
The closed string products 
(discussed further in section~\ref{constandcoc})  
are graded commutative and 
therefore   
symmetric when
the entries are even vectors in the state space. 
The string field  $\Psi$ is even.
The  gauge transformations are
\be
\delta_\lambda \Psi = Q \lambda + [ \lambda\, ,  \Psi] + \cdots \,, 
\ee
where the dots denote terms with higher powers
of the string field $\Psi$.   
The computation of the action 
to cubic order in  
the string field (\ref{the_string_field})
gives:
\be
\label{cubic-action}
\begin{split}
(2\kappa^2)\,S &=    \int [dx d\tilde x]\,
\Bigl[\, \, {1\over 4} e_{ij} \square e^{ij} 
\,+\, 2 \bar{e} \, \square \, e
- f_i f^i - \bar f_i \bar f^i
- \, f^i \bigl(
 \bar D^j e_{ij}   -2D_i\bar e \bigr)
 + \bar f^j  \bigl(D^i e_{ij}  + 2 \bar D_j e \bigr) \\[1.2ex]
&\hskip56pt - {1\over 8} \,e_{ij}\Bigl(  -(D_k e^{kj} ) (\bar D_l e^{il})\, -(D_ke^{kl}) (\bar D_l e^{ij})  \,- 2\,(D^i e_{kl})  ( \bar D^j e^{kl} ) 
~\\[1.0ex]
&\hskip98pt 
+ 2(D^ie_{kl} )\,( \bar D^l e^{kj})\, 
+ 2  ( D^ke^{il} ) (\bar D^j e_{kl})  \Bigr) \\
&\hskip56pt + 
{1\over 2}  e_{ij} f^{i}\bar f^j 
 -{1\over 2}  f_i f^{i}\, \bar e   
+{1\over 2}  \bar f_i \bar f^{i}\,  e  
\\[1.0ex]
&\hskip56pt - {1\over 8} \,
e_{ij} \Bigl( (D^i \bar D^je) \bar e
-(D^ie)( \bar D^j\bar e)
-( \bar D^je) (D^i\bar e)
+ e\,D^i\bar D^j\bar e\Bigr)\\[1.0ex]
&\hskip56pt
-{1\over 4} 
\,f^i \Bigl( e_{ij} \bar D^j\bar e + \bar D^j (e_{ij}\bar e) \Bigr)
+{1\over 4} \, f^i \Bigl( (D_i e)\bar e  -e D_i\bar e\Bigr) 
\\[1.0ex]
&\hskip56pt
-{1\over 4}   
\,\bar f^j\Bigl(\, e_{ij} D^i e +  D^i (e_{ij}\, e)\Bigr)
+{1\over 4} \,\bar f^j\Bigl(\,(\bar D_j e)\bar e   - e\,\bar D_j\bar e\Bigr)   \Bigr]\,.
\end{split}
\ee  
All fields are assumed to satisfy the constraint $\Delta =0$.
The above action is invariant under the exchanges  
\be
e_{ij} ~ \leftrightarrow ~  e_{ji}\,, \quad  D_i ~\leftrightarrow ~ \bar D_i\,, \quad
f_i  ~\leftrightarrow ~  - \bar f_i\,, 
\quad e  ~\leftrightarrow ~  - \bar e\, .
\ee
This discrete symmetry 
implies we need only concern
ourselves with the gauge transformations generated by $\lambda$
and by $\mu$.   
Those generated by $\bar \lambda$ can be written
in terms of the $\lambda$ ones  and
the discrete transformations above.  For the $\lambda$ 
gauge transformations we find, to linear order in the fields,  
\be
\label{lambda-nonlinear}
\begin{split}
\delta_\lambda e_{ij}  &=  \bar D_j \lambda_i \,
-  \,  {1\over 4} \Bigl[\,\,
 \lambda^k D_i e_{kj}   - (D_i \lambda^k)  e_{kj} 
 +  D^k ( \lambda_ie_{kj}) + \,  (D^k \lambda_i) e_{kj}   
    - D^k( \lambda_k e_{ij})  - \,\lambda_k D^k e_{ij} ~\Bigr] 
    \\[0.9ex]
&~~~ - \, {1\over 4} \Bigl[\lambda_i \bar D_j \bar e 
 -  (\bar D_j \lambda_i)\bar e  \Bigr]  \,+ \, {1\over 2} \, \lambda_i \, \bar f_j\,, 
  \\[2.0ex]
\delta_\lambda e  &= - {1\over 2}  D^i \lambda_i  - {1\over 4} f^i \lambda_i
+ 
\,  {1\over 8}\, \bigl( e D^i \lambda_i + 2  (D^i e)\lambda_i \bigr) \,,  \\[2.0ex]
\delta_\lambda \bar e  &=~
 {1\over 16}\, \bigl( \bar e D^i \lambda_i + 2  (D^i \bar e)\lambda_i \bigr) \,.
 \end{split}
 \ee
 We have not written the gauge transformations for the auxiliary fields
 $f_i$ and $\bar f_i$ since they are quite cumbersome and will not 
 be needed.  The $\mu$ gauge transformations, to linear order
 in the fields, are
 \be
 \label{mu-nonlinear}
 \begin{split}
\delta_\mu\, e_{ij}  &= ~ 0\,,   \\[1.0ex]
\delta_\mu\, e~ &=  \mu
 -  {3\over 8}  \,\mu \, e \,, \\[1.0ex]  
\delta_\mu \,\bar e~ &=  \mu 
 +  {3\over 8}  \, \mu\, \bar  e   \,. 
\end{split}\ee
To preserve the constraint, the variation of any field must be annihilated by $\Delta$.
The field-independent terms in the variations meet this requirement as the gauge parameters are in the kernel of $\Delta$. 
 The terms  involving a product 
of a field and a gauge parameter  are not guaranteed to satisfy the constraint and a projection is needed.
 In section~\ref{constandcoc} we discuss the natural projector 
$[[ \,\cdot \,]]$ that satisfies  $\Delta [[ A ]]=0$ for arbitrary $A(x,\tilde x)$.  All terms
linear in the fields in the above gauge transformations include an
 implicit $[[ \ldots ]]$
around them.  We do not write these brackets here to avoid cluttering.

The closed string field theory
 predicts a gauge algebra that is 
 quite intricate~\cite{Zwiebach:1992ie}:  
   the bracket of two  gauge transformations is in general
a gauge transformation with field dependent structure constants
and 
the gauge algebra only
closes on-shell.  
To lowest nontrivial order
we find
\be
 [\delta_{\lambda_1} \,, \delta_{\lambda_2} ] \Psi =\delta_\Lambda  \Psi  + (\hbox{on-shell}=0\,\hbox{terms}) 
\qquad  \hbox{with}  \quad  \Lambda = [\lambda_2, \lambda_1] + \dots\,,
\ee
where the dots represent field dependent terms. 
The product of parameters $[\lambda_2, \lambda_1] $ is antisymmetric under the interchange $1 \leftrightarrow 2$ since the $\lambda$'s have ghost number one.
For  gauge parameters $\lambda_1$ and $\lambda_2$
the computation of the closed string product, keeping the lowest
number of derivatives, gives
\be
\label{ifnkjekkgjkdjdf98}
\begin{split}
\Lambda^i  =  
&\,  ~~{1\over 2} \,\Bigl[  (\lambda_2 \cdot D) \lambda_1^i 
-(\lambda_1 \cdot D) \lambda_2^i  \Bigr]\\[1.0ex]
&  +{1\over 4} \,
\Bigl[ ~ \lambda_1 \cdot D^i  \lambda_2 - \lambda_2\cdot D^i  \lambda_1  \Bigr] \\[1.0ex]
&  +{1\over 4} \,\Bigl[ \lambda_1^i (D\cdot \lambda_2)- \lambda_2^i (D\cdot \lambda_1) ~\Bigr] \, \equiv   \{ \lambda_2, \lambda_1\} ^i  \,.
\end{split}  
\ee
In the above, we introduced a bracket $\{ \cdot \,, \cdot \}$ of two
vectors, defined by the right hand side.  It is the bracket induced by the
closed string product and resembles the Lie bracket of vector fields,
but does not coincide with it.
 One can 
show that ghost number conservation implies that the commutator of two $\lambda$ transformations does not give a $\bar \lambda$   
 transformation
 nor does it give a $\mu$-transformation. In (\ref{ifnkjekkgjkdjdf98}) the projection
 brackets $[[ \dots ]]$ act on the right hand side, since any allowed gauge parameter
 must be in the kernel of $\Delta$.  
 We have verified   
  the structure
 of $\Lambda^i$ in (\ref{ifnkjekkgjkdjdf98}) by computing explicitly the leading
 inhomogeneous term in the
 commutator of two transformations on  $e_{ij}$.   
 The projectors cause no complication. 
 
 \bigskip
 \noindent
 It is of interest to see if the 
 bracket $\{\lambda_2, \lambda_1\}$   
  forms
 a Lie algebra.  The first line of (\ref{ifnkjekkgjkdjdf98}) is the Lie derivative, but
 the other two lines are exotic.  We have found that
 \be
\{ \{\lambda_2, \lambda_1\} , \lambda_3\}
+ \{\{ \lambda_3, \lambda_2\} , \lambda_1\}
+ \{\{ \lambda_1, \lambda_3\} , \lambda_2 \}  \not= 0 \,.
 \ee
We have checked that this non-vanishing result occurs even
if all products of  $\lambda$'s  are in
 the kernel of~$\Delta$.  So the failure
of the Jacobi identity is not only due to the projectors implicit in the 
bracket $\{ \,, \, \}$. 
The fact that the Jacobi identity does not hold 
is a reflection of the homotopy-Lie algebra structure of the string field theory gauge algebra.

\medskip
\noindent
\underbar{Fixing the $\mu$ gauge symmetry.}
~We noted in the quadratic theory that the $\mu$ symmetry
could be used to set $e= d$ and $\bar e = -d $ in the action.
A similar result holds at the cubic level, as we discuss now.
First note that the $\mu$ transformations in (\ref{mu-nonlinear})
give
\be
\begin{split}
\delta_\mu (e-\bar e) &=  ~0 ~ - {3\over 8} \, \mu (e+ \bar e)\,,\\
\delta_\mu (e +\bar e) &=  2\mu - {3\over 8} \mu (e- \bar e)\,.
\end{split}
\ee
As a result, the following fields 
\be
\label{defnonlineared}
\begin{split}
d&\equiv {1\over 2}\, (e - \bar e) +{3\over 64}\, (e+ \bar e)^2 \,, \\
\chi &\equiv   {1\over 2}\, (e+ \bar e)
+{3\over 32}\, (e^2- \bar e^2) \,,
\end{split}
\ee
have  transformations in which terms linear in fields vanish:
\be
\begin{split}
 \delta_\mu d&=0 \,, \\
\delta_\mu \chi &=\mu  \,.
\end{split}
\ee
We now use $\mu$  to set $\chi=0$.
Since
\be
 \chi =  {1\over 2}\, (e+ \bar e) \Bigl(1 +{3\over 16}\, (e- \bar e )\Bigr)\,,
 \ee 
the perturbative solution to $\chi=0$ is
\be e=-\bar e\,.
\ee
It then follows from (\ref{defnonlineared}) that in this gauge
\be
\label{d-nonl}
d= {1\over 2}\, (e - \bar e)\,,
\ee
and we can take
$e = d$ and $\bar e =  -d\,$ in evaluating the action and the
gauge transformations.

\medskip
\noindent
Note that $\lambda$ gauge transformations
now
require compensating $\mu$ transformations to preserve the
gauge $\chi=0$.  Indeed, it follows from
\be
(\delta_\mu + \delta_\lambda) \chi = \mu - {1\over 4} D\cdot \lambda
\,+ \, \hbox{non-linear}\,,
\ee
that we must set
$\mu = {1\over 4} D\cdot \lambda  + \ldots $
and therefore the final $\lambda$ gauge transformations 
take the form  $ \delta_\lambda +  \delta_{\mu= {1\over 4} D\cdot \lambda
+ \ldots } \,.$
Since $\delta_\mu e_{ij} =0$ and $\delta_\mu d =0$, this only affects
the auxiliary fields.  Since auxiliary fields will be eliminated, we need
not concern ourselves with these compensating gauge transformations.

\subsection{Simplifying the gauge transformations}\label{simthegautra}

We now turn  to simplifying  the $\lambda$-gauge transformations
 of $e_{ij}$ and $d$, dropping all terms of quadratic and higher order in the fields. 
The field equations for the auxiliary fields $f$ and $\bar f$ have non-linear terms, but to the order we are working it suffices to substitute for $f$ and $\bar f$ in the gauge transformation (\ref{lambda-nonlinear})  using the linearised field equations
 (\ref{f-elim-quad}). 
From (\ref{d-nonl}) we  have
\be
\delta d = {1\over 2} (\delta e-  \delta\bar e)\,.
\ee   
In the formulae for $\delta d$, 
and  $\delta e_{ij}$, we can set
 $e=d $ and $\bar e = - d$.  
  Then  (\ref{lambda-nonlinear}) gives
\be
\label{orig-gt-mu-fixed}
\begin{split}
\delta_\lambda e_{ij}  &=  \bar D_j \lambda_i \,-\,  {1\over 4} \Bigl[\,\,
 \lambda^k D_i e_{kj}   - (D_i \lambda^k)  e_{kj}  
 +   \lambda_iD^k e_{kj} + \,  2(D^k \lambda_i) e_{kj}  
  \\[0.7ex]
&~~  - (D\cdot \lambda )e_{ij}  -2 \,\lambda_k D^k e_{ij} ~\Bigr] 
+ {1\over 4} \Bigl[\lambda_i \bar D_j d 
 -  (\bar D_j \lambda_i)d  \Bigr]  \,+ \, {1\over 4} \, \lambda_i \, 
 (D^k e_{kj} + 2 \bar D_j d )\,, \\[1.3ex]
\delta_\lambda d \,\, &= {1\over 2} \Bigl[ 
- {1\over 2}  D^i \lambda_i  - {1\over 4} f^i \lambda_i
+ {1\over 8}\, ( e D^i \lambda_i + 2  (D^i e)\lambda_i )  
- {1\over 16}\, 
 ( \bar e D^i \lambda_i + 2  (D^i \bar e)\lambda_i )  \Bigr] \,.
\end{split}
\ee

Next we look for redefinitions of the fields and gauge parameters.
After some manipulation the above transformations
can be written as
\be
\begin{split}
\delta_\lambda e_{ij}  &= ~ \bar D_j \Bigl(\lambda_i + {3\over 4} \lambda_i d\Bigr) \, +  {1\over 2} \Bigl[\,
   (D_i \lambda^k)  e_{kj}  
 - \,  (D^k \lambda_i) e_{kj}    
   + \,\lambda_k D^k e_{ij} ~\Bigr] ~~~~ \\
&~~+\,   D_i \Bigl(-{1\over 4} \lambda^k e_{kj}\Bigr)  -   \delta_\lambda ( e_{ij} d)   \,  ,
 ~     \\[1.2ex]
\delta_\lambda d~  &=
- {1\over 4}  D^i \Bigl(\lambda_i  + {3\over 4} \lambda_i d\Bigr) 
+  {1\over 2}  (\lambda \cdot D) \,d\,\\[1.0ex]
&~~~
 - {1\over 4} \bar D^j \Bigl( -{1\over 4} \lambda^k e_{kj} \Bigr)
 -{1\over 32}\, \delta_\lambda (e_{ij} e^{ij})- {9\over 16}\, \delta_\lambda
\, d^2  \,.
 \phantom{\Biggl(}
 \end{split}
\ee
We redefine the gauge parameter $\lambda_i$
by taking $\lambda_i + {3\over 4} \lambda_i d \to \lambda_i$,
without
affecting the remaining terms linear in fields.  Moreover, note
that the first term on the second line in each of the above transformations
can be thought of as
linearised transformations with an effective barred
parameter  $\bar \lambda = -{1\over 4} \lambda^k e_{kj}$.
The $\delta _{\bar \lambda}$ transformation with parameter
$\bar \lambda = -{1\over 4} \lambda^k e_{kj}$ leaves the quadratic action invariant, while it changes the cubic action by terms cubic in the fields. In checking the  invariance of the quadratic plus cubic action up to terms quadratic in the fields, these 
$\delta _{\bar \lambda}$  transformations constitute a separate symmetry 
and so need not be included in the $\lambda $ transformations.
We
can therefore ignore them and we will do so.
  We then have
\be
\begin{split}
\delta_\lambda \Bigl( \,e_{ij} + e_{ij} d\Bigr)~
 &= ~ \bar D_j \lambda_i  \, +  {1\over 2} \Bigl[\,
   (D_i \lambda^k)  e_{kj}  
 - \,  (D^k \lambda_i) e_{kj}    
   + \,\lambda_k D^k e_{ij} ~\Bigr] \,,    
 ~     \\[1.5ex]
\delta_\lambda \Bigl( \, d 
 -{1\over 32}\,  e_{ij} e^{ij}- {9\over 16}\, \, d^2 \Bigr)   
 &=
- {1\over 4}  D\cdot \lambda 
+  {1\over 2}  (\lambda \cdot D) \,d\,.
 \end{split}
\ee
We redefine the fields
\be
\label{redefine-cubic}
\begin{split}
e'_{ij}  &= ~ e_{ij} +  e_{ij} d \,, \\[0.5ex]
d'   ~ & =   ~d + {1\over 32} e_{ij}e^{ij} + {9\over 16} d^2\,.
\end{split}
\ee
to give primed fields 
 that have  simple gauge transformations
\be
\label{gaugetrans99}
\begin{split}
\delta_\lambda e'_{ij}  &= ~ \bar D_j \lambda_i  \, +  {1\over 2} \Bigl[\,
   (D_i \lambda^k)  e'_{kj}  
 - \,  (D^k \lambda_i) e'_{kj}    
   + \,\lambda_k D^k e'_{ij} ~\Bigr]     \, , 
 ~     \\ 
\delta_\lambda d'  &=
- {1\over 4}  D\cdot \lambda 
+  {1\over 2}  (\lambda \cdot D) \,d'\,~\,.
 \phantom{\Biggl(}
 \end{split}
\ee
After these field redefinitions, it is convenient to 
drop the primes to simplify notation.  We 
do so in what follows.

\subsection{Simplifying the action}

We now consider the full action (\ref{cubic-action}) and 
first
fix the $\mu$ gauge symmetry by setting $e= d$ and $\bar e = -d$.
We then eliminate the auxiliary fields $f$ and $\bar f$ and,
 after a fair amount of straightforward work, we find
\be
\label{action5}
\begin{split}
(2\kappa^2)\,S &=    \int [d xd\tilde x] \,
\Bigl[\, \, {1\over 4} e_{ij} \square e^{ij}  + {1\over 4} (\bar D^j e_{ij})^2
+ {1\over 4} ( D^i e_{ij})^2  - 2 \, d \, D^i \bar D^j e_{ij}
\,-\, 4 \,d \, \square \, d~~~
\\[1.0ex]
&\hskip50pt - {1\over 8} \,e_{ij}\Bigl(   -(D_ke^{kl}) (\bar D_l e^{ij})  - 2\,(D^i e_{kl})  ( \bar D^j e^{kl} )\\[1.0ex]
&\hskip88pt + 
2(D^ie_{kl} )\,( \bar D^l e^{kj}) +2  (D^k e^{il} )( \bar D^j e_{kl} )  \Bigr)~
\\[1.0ex]
&\hskip50pt  +\,{1\over 2} d \,  \Bigl(e_{ij} \,\bar D_k \bar D^j e^{ik}  + 
e_{ij} \, D_l
D^i  e^{lj} + \, (D^ie_{ij})^2 + (\bar D^j e_{ij})^2
  \Bigr)
\\[1.0ex]
&\hskip50pt  - \,{1\over 4} 
e_{ij} \,  (D^i \bar D^jd)  d  
-{9\over 4}  e_{ij} (D^id)( \bar D^j d) 
 - {1\over 2} \,d^2 \,\square\, d\,.
  \end{split} 
\ee
This is the action expected to be invariant under the original gauge transformations
(\ref{orig-gt-mu-fixed}). Since we simplified those gauge transformations 
by the field redefinitions (\ref{redefine-cubic}) we now
perform 
these same  field redefinitions in the action.  
From (\ref{redefine-cubic}), we set
\be
\begin{split}
e_{ij}  &= ~ e'_{ij} -  e'_{ij} d' \,. \\[0.5ex]
d   ~ & =   ~d' - {1\over 32} e'_{ij}e'^{ij} - {9\over 16} d'^2\,.
\end{split}
\ee
to obtain an action in terms of 
the primed fields. 
Dropping all primes, the  
result is
\be
\label{redef-action}
\boxed{
\begin{split}
\phantom{\Biggl(}(2\kappa^2)S &=    \hskip-2pt\int [d x d\tilde x] \,
\Bigl[\, \, {1\over 4} e_{ij} \square e^{ij}  + {1\over 4} (\bar D^j e_{ij})^2
+ {1\over 4} ( D^i e_{ij})^2  - 2 \, d \, D^i \bar D^j e_{ij}
\,-\, 4 \,d \, \square \, d~~~
\\[0.6ex]
&\hskip25pt + {1\over 4} \,e_{ij}\Bigl( \,(D^i e_{kl})  ( \bar D^j e^{kl} )
 -(D^ie_{kl} )\,( \bar D^l e^{kj}) 
 - (D^k e^{il} )( \bar D^j e_{kl} )  \Bigr)~
\\[1.0ex]
&\hskip25pt +
{1\over 2} d\,  
\Bigl(  (D^ie_{ij})^2  + (\bar D^j e_{ij})^2 
+{1\over 2}(D_k e_{ij})^2 
 +{1\over 2}( \bar D_k e_{ij} )^2
 + 2e^{ij} (D_i   D^k e_{kj}  + \bar D_j  \bar D^k e_{ik} )
   \Bigr)~  \\[0.9ex]
&\hskip25pt + 
4\,e_{ij} d\, D^i \bar D^jd   
 + 4 \,d^2 \,\square\, d~\Bigr]\,.\\[2.0ex]
 \end{split} }
\ee
The discrete $\mathbb{Z}_2$ symmetry (\ref{z2-sym}) we found in the
quadratic theory is preserved here.  This  is essentially
manifest for all terms except the $e^3$ terms, where it takes
a small computation to confirm 
the symmetry.  The transformations are written
again here for convenience
\be
\label{disc-sym-final}
\boxed{
~~\mathbb{Z}_2~\hbox{transformations}:~~\phantom{\Biggl(}
e_{ij} ~\to~  e_{ji}\,, \quad  D_i ~\to ~ \bar D_i\,, \quad
\bar D_i  ~\to ~ D_i \,, \qquad  d ~\to ~  d~\,.~}
\ee
The gauge transformations are those in (\ref{gaugetrans99})
\be
\label{gaugetrans}
\boxed{\begin{split}
\phantom{\Biggl(} 
\delta_\lambda e_{ij}  &= ~ \bar D_j \lambda_i  \, +  {1\over 2} \Bigl[\,
   (D_i \lambda^k)  e_{kj}  
 - \,  (D^k \lambda_i) e_{kj}    
   + \,\lambda_k D^k e_{ij} ~\Bigr] \,,    \,~  
 ~     \\ 
\delta_\lambda d  &=
- {1\over 4}  D\cdot \lambda 
+  {1\over 2}  (\lambda \cdot D) \,d\,~.
 \phantom{\Biggl(}
 \end{split}}
\ee
The discrete symmetry (\ref{disc-sym-final}) of the action $S$ is fundamental to our analysis.  It implies that gauge transformations
with barred gauge parameters obtained from (\ref{gaugetrans})
by the discrete symmetry are 
also invariances of $S$.  The action
$S$ then has the appropriate doubled symmetry. For future reference
the barred gauge transformations are
\be
\label{gaugetrans-barred}
\begin{split}
\delta_{\bar \lambda} e_{ij}  &= ~ D_i \bar\lambda_j  \, +  {1\over 2} \Bigl[\,
   (\bar D_j \bar \lambda^k)  e_{ik}  
 - \,  (\bar D^k \bar \lambda_j) e_{ik}    
   + \,\bar \lambda_k \bar D^k e_{ij} ~\Bigr]  \,,    \,~  
 ~     \\ 
\delta_{\bar \lambda}  d  &=
- {1\over 4}  \bar D\cdot \bar \lambda 
+  {1\over 2}  (\bar\lambda \cdot \bar D) \,d\,~.
 \phantom{\Biggl(}
 \end{split}
\ee
In all of the above gauge transformations, 
there is an implict 
projection $[[ \, \cdot \, ]]$  
to the kernel of $\Delta$  for the terms linear in the fields.

As a check of the action $S$ we used the Noether method to
construct a cubic term to be added to the quadratic action for which the action is invariant under (\ref{gaugetrans}), 
up to terms cubic or higher in the fields.  The result was precisely the action $S$ 
given above.  
We note that the cubic action  can be rewritten in a suggestive way (up to quartic terms)  as
\be
\label{invar-action}
(2\kappa^2)\, S     
=    \int [dx d\tilde x]\,  \,    
e^{-2d}\,
\Bigl[\, 
-{1\over 4}K
  - 2 \, e_{ij}  D^i \bar D^j d
+2 (Dd)^2+2 (\bar Dd)^2\Bigr] \,.   
\ee
Here $K=K_2+K_3$, where
\be  
K_2= (D^i e_{ij})^2
+    ( \bar D^j e_{ij})^2 
+  \frac 1 2  ( D^ke_{ij})^2 +  
\frac 1 2  (\bar D^ke_{ij})^2 
 +2 e^{ij}(D_iD^ke_{kj}
+\bar D_j\bar D^ke_{ik} ) \,,   
\ee
coincides, up to total derivatives, with the quadratic Lagrangian for $e_{ij}$  and
\be  K_3
=
- \,e_{ij}\Bigl(   \,(D^i e_{kl})  ( \bar D^j e^{kl} )
- (D^ie_{kl} )\,( \bar D^l e^{kj}) -  (D^k e^{il} )( \bar D^j e_{kl} )  \Bigr) \, ,
\ee
coincides with the cubic Lagrangian for $e_{ij}$.
This suggests that $K$ may give the leading terms in the expansion of some  curvature.

We can now reconsider the algebra of gauge transformations
discussed around 
equation~(\ref{ifnkjekkgjkdjdf98}). Our field redefinitions
cause the mixing of the unbarred and barred transformations, so some of the
simplicity is lost.  Nevertheless the answers are still reasonably compact.
The commutation of two gauge transformations with parameters 
$(\lambda_1, \bar \lambda_1)$ and $(\lambda_2, \bar \lambda_2)$ is
a gauge transformation with parameters $(\Lambda, \bar\Lambda)$ that to leading
order are field independent:
\be
\label{iwltfttaovm99}
\begin{split}
\phantom{\Biggl[}~~\Lambda^i   =
&\,  ~~{1\over 2} \,\Bigl[ ~ (\lambda_2 \cdot D + \bar \lambda_2 \cdot \bar D) \, \lambda_1^i 
- (\lambda_1 \cdot D + \bar \lambda_1 \cdot \bar D) \lambda_2^i  \Bigr] 
\\[1.0ex] &
+{1\over 4} \,
\Bigl[ ~ \lambda_1 \cdot D^i  \lambda_2 - \lambda_2\cdot D^i  \lambda_1  \Bigr] - {1\over 4} \,
\Bigl[ ~ \bar \lambda_1 \cdot D^i  \bar \lambda_2 - \bar\lambda_2\cdot D^i  \bar \lambda_1  \Bigr] \,,
\\[1.3ex]
\bar \Lambda^i = & 
~~{1\over 2} \,\Bigl[~   (\lambda_2 \cdot D + \bar \lambda_2 \cdot \bar D)\, \bar\lambda_1^i 
-\, (\lambda_1 \cdot D + \bar \lambda_1 \cdot \bar D)\, \bar \lambda_2^i  \Bigr] 
\\[1.0ex] &
- {1\over 4} \,\Bigl[ ~ \lambda_1 \cdot \bar D^i  \lambda_2 - \lambda_2\cdot \bar D^i  \lambda_1  \Bigr]
+ {1\over 4} \,\Bigl[ ~ \bar\lambda_1 \cdot \bar D^i  \bar \lambda_2 - \bar \lambda_2\cdot \bar D^i  \bar \lambda_1  \Bigr]~.
\end{split}
\ee
The constraint $\Delta=0$ on the parameters is used in
calculating the algebra. 
The same caveats discussed earlier apply here.   The 
commutator of gauge
transformations  to all orders in the fields
is expected to 
 include field dependent structure constants as well as
terms that vanish on-shell.  
There is 
an implicit projection $[[\,\cdot ]]$ in the above right-hand
sides so that  $(\Lambda, \bar\Lambda)$ are in the kernel of $\Delta$.
Finally,
the brackets $[\,\cdot , \cdot]$ implicit above do not satisfy the Jacobi identity.

\subsection{Conventional field theory limits}\label{fieldlimits}

In this section we examine the gauge transformations
in the limits where there is dependence on  either just $x$ or just  $\tilde x$ coordinates
and show that we recover the expected results.
Interestingly, these two limits require 
two different sets of field
redefinitions and these break the discrete $\mathbb{Z}_2$
symmetry of the theory.

We wish to compare our results with the gauge transformations of the conventional (undoubled) theory 
of a metric $\Gg_{ij}(x^k)$,   
Kalb-Ramond field  $\bb_{ij}(x^k)$, and a dilaton $\phi (x^k)$.  
Under diffeomorphisms with parameter $\xi^i$ and 
antisymmetric gauge transformations with parameter
$\alpha_i$, the first two  
 fields transform as
\be   
\label{diftran}
\begin{split}
\delta \Gg_{ij}&= {\cal L}_\xi \Gg_{ij}, \\[0.5ex]
\delta \bb_{ij}&= {\cal L}_\xi \bb_{ij} +\partial _i \alpha _j-\partial _j \alpha _i\,.
\end{split}
\ee
For the dilaton we have    
\be
\label{difdil}
\delta \phi
=  \,\xi^i \partial _i \phi \,. \hskip30pt
\ee
Here $ {\cal L}_\xi$ is the Lie derivative with respect to $\xi^i$. For
any rank two tensor $r_{ij}$,  the Lie derivative with respect to $\xi^i$ 
 takes the form
\be
{\cal L}_\xi\, r_{ij}  =  (\partial_i \xi^k) \, r_{kj} + (\partial_j \xi^k) \, r_{ik}
+\xi^k \partial_k r_{ij}\,.
\ee
The above form of the standard diffeomorphisms is all we need to
compare with our results.  It is interesting, however, to 
write the transformations more geometrically. 
We first note that (\ref{diftran}) can be written as
\be
\label{diftran99}
\begin{split}
\delta \Gg_{ij}&=  \nabla_i \xi_j +  \nabla_j \xi_i\, ,\\[0.5ex]
\delta \bb_{ij}&= H_{ijk}\xi ^k +\partial _i \ti \xi _j-\partial _j \ti \xi _i\,,
\end{split}
\ee
where $ \nabla$ is the covariant derivative with Levi-Civita connection $\Gamma$,
$H$ is the field strength 
\be
H_{ijk}=\partial _i \bb_{jk}   + \partial_j \bb_{ki} +
\partial _k \bb_{ij} 
\,, \ee
and we have defined
\be
\xi_i \equiv \Gg_{ij}\xi^j,
\qquad
\ti \xi_i \equiv \alpha_i -\bb_{ij}\xi^j\,.
\ee
Introducing the field 
\be
 {\cal E}_{ij}=\Gg_{ij}+\bb_{ij}\,,
 \ee
 the transformations can  be written as  transformations of ${\cal E}$: 
\be
\label{geomtransf}
\begin{split}
\delta {\cal E}_{ij}&= 
 \nabla_i \xi_j +  \nabla_j \xi_i +H_{ijk}\xi ^k+\partial _i \ti \xi _j-\partial _j \ti \xi _i \\[2.0ex]
\to \qquad \delta {\cal E}_{ij}&=  \hat\nabla_i \xi_j + \hat \nabla_j \xi_i +\partial _i \ti \xi _j-\partial _j \ti \xi _i\,,
\end{split}
\ee
where
$\hat \nabla$ is the derivative  for the connection with torsion
\be
\hat \Gamma ^k_{ij}= \Gamma ^k_{ij}-   \Gg^{kl}H_{ijl}
=\frac 1 2 \, \Gg^{kl}\,(
\partial _i {\cal E}_{lj}
+\partial _j {\cal E}_{il}
-\partial _l {\cal E}_{ij})\,.
\ee
The transformation (\ref{geomtransf}) encodes nicely the full
gauge structure of the fields.
For the dilaton transformation (\ref{difdil})  
it is convenient to define a field $\ddd$ by
\be
e^{-2\ddd}\equiv \sqrt{-\Gg}\, e^{-2\phi} \,.   
\ee
Since $\sqrt{-\Gg}$ is a density we find that $e^{-2\ddd}$ is also a density:
\be
\label{dens}
\delta e^{-2\ddd}= \partial _i (e^{-2\ddd}\xi^i)\,.
\ee

\medskip
Returning to the task at hand, we split the fields into constant background fields $G,B$ plus fluctuations
\be
\begin{split}  
\Gg_{ij}&=G_{ij}+h_{ij}\, , \\ 
\bb_{ij}&=B_{ij}+b_{ij}\, . 
\end{split}
\ee
The transformations (\ref{diftran}) then imply transformations
for the fluctuations.  A short computation shows that they can be written as
\be
\begin{split}
\delta h_{ij}&=\partial _i \epsilon _j+\partial _j \epsilon _i+
  {\cal L}_\xi h_{ij}\, , \\[0.5ex]
\delta b_{ij}&= \partial _i\ti  \epsilon _j-\partial _j \ti \epsilon _i+{\cal L}_\xi b_{ij} \,,
\end{split}
\ee
where
\be
 \epsilon _i= G_{ij} \xi ^j, \qquad
\ti \epsilon_i = \alpha_i -B_{ij}\xi^j\,.
 \ee
 Defining as usual the field $\fluc e_{ij}$ that puts together the two types
of fluctuations, 
 \be
 {\fluc e}_{ij}=h_{ij}+b_{ij}\,,
 \ee
we readily find that it  transforms as
\be
\label{gaugetransa}
\begin{split}
\delta  \fluc e_{ij}&=(\partial _i \epsilon _j+\partial _j \epsilon _i)+   (\partial _i\ti  \epsilon _j-\partial _j \ti \epsilon _i)+{\cal L}_ \epsilon  \fluc e_{ij}\\[1.0ex]
&=(\partial _i \epsilon _j+\partial _j \epsilon _i)+   (\partial _i\ti  \epsilon _j-\partial _j \ti \epsilon _i)
+((\partial _i \epsilon^k) \fluc e_{kj}+ (\partial _j \epsilon^k) \fluc e_{ik} 
+\epsilon^k \partial _k  \fluc e_{ij} ) \,,
    \end{split}
\ee 
where  indices are raised and lowered 
using $G_{ij}$.  This is our final form for the conventional gauge transformations, to be compared with the result arising from the cubic
theory we have constructed.

Our analysis requires both unbarred and barred
gauge parameters, so we put together the gauge transformations
(\ref{gaugetrans}) and (\ref{gaugetrans-barred}) to obtain the transformations  
\be
\begin{split}
\delta e_{ij}  &= ~ \bar D_j \lambda_i  \, +  {1\over 2} \Bigl[\,
   (D_i \lambda^k)  e_{kj}  
 - \,  (D^k \lambda_i) e_{kj}    
   + \,\lambda_k D^k e_{ij} ~\Bigr] \\
   & ~+  D_i \bar\lambda_j  \, +  {1\over 2} \Bigl[\,
   (\bar D_j \bar \lambda^k)  e_{ik}  
 - \,  (\bar D^k \bar \lambda_j) e_{ik}    
   + \,\bar \lambda_k \bar D^k e_{ij} ~\Bigr] \, , 
   \end{split}
   \ee
   as well as
   \be
\label{gaugetrans-combo}
\delta  d  =
- {1\over 4}  (D\cdot \lambda +\bar D\cdot \bar \lambda )
+  {1\over 2}  (\lambda \cdot D+\bar\lambda \cdot \bar D) \,d\,~. \hskip40pt
\ee
We can rearrange the former in the  suggestive form  
\be
\label{mastereqns}
\begin{split}
\delta e_{ij}  &= ~ \bar D_j \lambda_i  \,+  D_i \bar\lambda_j  +  
   (D_i \lambda^k)  e_{kj}   +
   (\bar D_j \bar\lambda^k)  e_{ik}    
     + {1\over 2}\,(\lambda_k D^k +  \bar \lambda_k \bar D^k  )e_{ij} ~~
      \\[0.5ex]
   & ~ \,  
-    {1\over 2} \,\bigl(
   D_i \lambda^k  
 +   D^k \lambda_i\bigr) e_{kj}    -{1\over 2} \,\bigl(
   \bar D_j \bar\lambda^k  
 + \,  \bar D^k \bar \lambda_j  \bigr) e_{ik}  \,.
   \end{split}
   \ee
The first line, as we will see, contains 
terms that combine naturally to form 
Lie derivatives.
The above 
transformations are expected to receive  corrections  of quadratic and  higher order in $e_{ij}$, while those for $\fluc e_{ij}$ above are exact.
The fields  $e_{ij}$,  $\fluc e_{ij}$ are related by non-linear field redefinitions
 $\fluc e_{ij}=e_{ij}+O(e^2)$ 
 \cite{Michishita:2006dr} and 
 next we shall seek such redefinitions to bring the
 transformations of $e_{ij}$ to the same form as those for $\fluc e_{ij}$.
 We then undertake a similar analysis for the T-dual system, and find a different field redefinition is needed.
 
\medskip
We now examine the above gauge transformations in two limits.
The first is that when fields have no $\tilde x^i$ dependence.  
The second is that when fields have no $x^i$ dependence.
It is convenient in both cases to use linear combinations
of the  gauge parameters:
\be
\epsilon_i \equiv {1\over 2} ( \lambda_i + \bar\lambda_i) \,, \qquad
\tilde \epsilon_i \equiv {1\over 2} ( \lambda_i - \bar\lambda_i)  \,.   
\ee
We now consider the two possible limits.

\subsubsection{Fields with no $\tilde x$ dependence.}\label{nox}
In this case we can set $\tilde \partial$
equal to zero in the derivatives (\ref{groihfgruu8774}).
It follows then that  $D = \bar D = \partial$, absorbing $\sqrt{\alpha'}$ into
the definition of the coordinates.  All indices are raised
or lowered with $G^{ij}$ and~$G_{ij}$. 
The transformations with parameter $\epsilon$ are obtained from 
(\ref{mastereqns}) setting $\lambda_i = \bar \lambda_i = \epsilon_i$:
\be
\begin{split}
\delta_\epsilon e_{ij}  &= ~ \partial_j \epsilon_i  \,+  \partial_i \epsilon_j  +  
   (\partial_i \epsilon^k)  e_{kj}   +
   (\partial_j \epsilon^k)  e_{ik}    
     + \epsilon_k \partial^k e_{ij} 
      - {1\over 2} (\delta_\epsilon {e_i}^{\,k}) e_{kj}   
    -{1\over 2} (\delta_\epsilon e^k_{~j}) e_{ik}  \\[0.5ex]
   &= ~ \partial_j \epsilon_i  \,+  \partial_i \epsilon_j  +  
   (\partial_i \epsilon^k)  e_{kj}   +
   (\partial_j \epsilon^k)  e_{ik}    
     + \epsilon_k \partial^k e_{ij} 
   -    {1\over 2} \delta_\epsilon ({e_i}^{\,k} e_{kj})  \,. 
     \end{split}
   \ee
We therefore have
\be
\label{ntegdf}
\delta_\epsilon \Bigl( e_{ij}  +  {1\over 2} {e_i}^{\,k} e_{kj}  \Bigr)
= ~ \partial_j \epsilon_i  \,+  \partial_i \epsilon_j  +  
   (\partial_i \epsilon^k)  e_{kj}   +
   (\partial_j \epsilon^k)  e_{ik}    
     + \epsilon_k \partial^k e_{ij} \,.
 \ee
It follows that the field
 \be
 \label{eplisdf}
 e_{ij} ^+\equiv e_{ij}  +  {1\over 2} {e_i}^{\,k} e_{kj} \,,
 \ee
transforms as
\be
\label{gaugetransb}
\begin{split}
\delta_\epsilon   
\, e_{ij}^+&=(\partial _i \epsilon _j+\partial _j \epsilon _i) +{\cal L}_ \epsilon e_{ij}  ^+ \,, \end{split}
\ee 
up to terms of order $(e^+_{ij})^2$.

\smallskip
\noindent
The $\tilde \epsilon$-gauge transformations are obtained 
 from 
(\ref{mastereqns}) setting $\lambda_i = -\bar \lambda_i = \tilde\epsilon_i$:
\be   
\tilde\delta_{\tilde \epsilon}\, e_{ij}  = ~ \partial_j \tilde \epsilon_i  \,-  \partial_i \tilde \epsilon_j  
-    {1\over 2} (\delta_{ \tilde \epsilon} {e_i}^{\,k}) e_{kj}    -{1\over 2} (\delta_{ \tilde \epsilon} e^k_{~j}) e_{ik}  \,,
   \ee
so that
\be
\label{ntbf}
\tilde \delta _{ \tilde \epsilon}\,e_{ij}^+
= ~ \partial_j \tilde \epsilon_i  \,-  \partial_i \tilde \epsilon_j \,,
\ee
\noindent
up to terms of order $(e^+_{ij})^2$.
The transformations for 
$ e_{ij} ^+$ are precisely the standard gauge transformations
(\ref{gaugetransa}), up to higher order terms.  This is what we
wanted to show.

Note that the full field (background plus fluctuation) with natural gauge transformations~is
\be
\label{fullfield}
 {\cal E}_{ij}\equiv E_{ij} +e_{ij}  ^++\hbox{cubic terms} =G_{ij} +B_{ij}+
 e_{ij}  +  {1\over 2} {e_i}^{\,k} e_{kj}  +\hbox{cubic terms}\,,    
 \ee
 so that $\fluc e_{ij} =e_{ij}  + 
  {1\over 2} {e_i}^{\,k} e_{kj}$, up to cubic terms.   
 We now show that ${\cal E}$    
 has the expected gauge 
 transformations.  
Indeed, for  
$\tilde\epsilon$ transformations 
(up to terms of quadratic in fields)   we have  
\be
\tilde\delta_{\tilde \epsilon}\, {\cal E}_{ij} = ~ \partial_j \tilde \epsilon_i  \,-  \partial_i \tilde \epsilon_j \,.
\ee
The $\epsilon$ transformations are a little more intricate.
We first compute the Lie derivative of ${\cal E}$:   
\be
\begin{split}  
{\cal L}_\epsilon {\cal E}_{ij}  &=   (\partial_i \epsilon^k) \, {\cal E}_{kj} + (\partial_j \epsilon^k) \, {\cal E}_{ik}
+\epsilon^k \partial_k {\cal E}_{ij}\\
&=   (\partial_i \epsilon^k) \, E_{kj} + (\partial_j \epsilon^k) \, E_{ik}
+{\cal L}_\epsilon e_{ij}^+\\
&= \partial_i \epsilon_j +  \partial_j \epsilon_i 
- \partial_i ( \, B_{jk}\epsilon^k) + \partial_j (B_{ik}\epsilon^k)
+{\cal L}_\epsilon e_{ij}^+\\
&= \delta_\epsilon {\cal E}_{ij}  + \tilde\delta_{B\epsilon}  {\cal E}_{ij}\,,
\end{split}
\ee
where we used (\ref{gaugetransb}), noted that 
$\delta_\epsilon {\cal E}_{ij} = \delta_\epsilon e^+_{ij}$,
 and recognised the presence of a $\tilde\delta$
transformation with parameter $\tilde \epsilon_i = B_{ij} \epsilon^j$.
We thus have a symmetry $\bar \delta_{\epsilon}$ for which the transformation
of ${\cal E}$ is through the Lie derivative 
(up to terms quadratic in fields):   
\be
\bar \delta_\epsilon {\cal E}_{ij} \equiv (  \delta_\epsilon  + \tilde\delta_{B\epsilon} ) {\cal E}_{ij}=  {\cal L}_\epsilon {\cal E}_{ij} \,.
\ee

The gauge transformation of $d$ is obtained
from (\ref{gaugetrans-combo}).  
For the $\delta_\epsilon$ transformations ($\lambda = \bar\lambda = \epsilon)$
we find, up to terms quadratic in fields,
\be
\delta_\epsilon 
 d=  - \frac 1 2 \,\partial \cdot \epsilon  + \epsilon \cdot \partial d \,.
\ee
This can be rewritten as
\be
\label{densc}
\delta_\epsilon e^{-2d}= \partial _i (e^{-2d}\epsilon ^i)\,,
\ee
and agrees with 
(\ref{dens}) if $\ddd $ is the same as $d$, up to 
terms cubic in the fields.  A short calculation shows that 
$\tilde \delta_{\tilde \epsilon} d = 0$, as would be expected.

\subsubsection{Fields with no $x$ dependence.} \label{ssccx}

The  configuration dual 
to the one considered in~\S\ref{nox}
has fields  independent of $x^a$.
 In order to avoid the  complication of indices running
over  non-compact directions and  toroidal directions we will 
consider 
the case in which
there is no $x$ dependence at all;
neither on the non-compact $x^\mu$ nor on the toroidal $x^a$.  
We will simplify further  by only considering the transformations with 
parameters $\lambda _a$ and $\bar\lambda_a$   and the components $e_{ab}$ of $e_{ij}$.
 
For  fields that do not depend on $x$, the derivative $\partial_i$ vanishes and the derivatives
(\ref{groihfgruu8774}), absorbing $\sqrt{\alpha'}$ into
the definition of the coordinates, take the form
\be
\label{r9jlg}
D_a = - \, \hat E_{ac} \, \tilde\partial^c \,, \qquad
\bar D_a = ~ \, \hat E_{ca} \, \tilde\partial^c  \,.
\ee
 We expect from T-duality 
that  the theory based on $\tilde x$ 
coordinates sees the dual background $\hat E'_{ab} = \hat G'_{ab}+ \hat B'_{ab} 
= \hat E^{-1}_{ab}$. 
As we will see later, the dual metric $\hat G'$ is related to 
the original metric $\hat G$ by
\be
\label{metrixsdlkf}
{\hat G}'^{-1} = \hat E\, \hat G^{-1} \, \hat E^t  = \hat E^t\, \hat G^{-1} \, \hat E\,.
\ee 
Note that ${\hat G}'^{-1}$ 
naturally has lower indices, just like $\hat E$ does.
For example, from the above we see that
$({\hat G}'^{-1})_{ab}  = \, \hat E_{\,ac}  \hat G^{cd}  \, \hat E_{bd} \,$.
Thus we use ${\hat G}'^{-1}$ to {\em lower} indices of primed objects:
\be
A'_a \equiv  ({\hat G}'^{-1})_{ab} A'^b \,.
\ee
The fields 
and gauge parameters appropriate
here are 
field redefinitions of the original fields and gauge parameters whose forms are suggested by T-duality transformations. 
We
introduce the fluctuation field
\be
\label{newfdlflkd}
{e'}^{ab} \equiv  - (\hat E^{-1} )^{ac} \,  e_{cd}  \, (\hat E^{-1} )^{db} \,.
\ee
Note that this field has 
upper  indices.   In order to compute
the gauge transformations we use 
\be
\label{newfdlflkd3}
\delta {e'}^{ab} =  - (\hat E^{-1} )^{ac} \,  \delta e_{cd}  \, (\hat E^{-1} )^{db} \,.
\ee
For the gauge parameters we introduce new primed ones through the
relations
\be
\label{dflkdrotijkf}
\lambda_a =  - \, \hat E_{\,ab} \, \lambda'^b\,, \qquad 
\bar\lambda_a =  ~ \, \hat E_{\,ba} \, \bar \lambda'^b\,.
\ee
All other gauge parameters will be taken to vanish.

Consider first the dilaton transformations (\ref{gaugetrans-combo}), where
  indices, of course,  are   contracted with the original metric $\hat G ^{-1}$.
 We write this out explicitly   
 \be
 \begin{split}
 \delta\, d  &= -{1\over 4}  \hat G^{ab}  \bigl(  D_a \lambda_b + \bar D_a \bar 
 \lambda_b\bigr)  + {1\over 2}  \hat G^{ab}  \bigl(  \lambda_a D_b + \bar\lambda_a \bar D_b\bigr) \, d \\[0.5ex]
 &=-{1\over 4}  \hat G^{ab}  \bigl(  \hat E_{ac} \tilde \partial^c\, \hat E_{bd} \lambda'^d   + \hat E_{ca} \tilde \partial^c   \hat E_{db} \bar 
 \lambda'^d\bigr)  + {1\over 2}  \hat G^{ab}  \bigl( \hat E_{ac}  \lambda'^c 
 \hat E_{bd} \tilde \partial^d  + \hat E_{ca} \bar\lambda'^c  \hat E_{db} \tilde
 \partial^d \bigr)\,d\,,\\[0.5ex]
&=  -{1\over 4}  ({\hat G}'^{-1})_{cd}  \bigl(   \tilde \partial^c\, \lambda'^d   
+  \tilde \partial^c    \bar 
 \lambda'^d\bigr)  + {1\over 2}  ({\hat G}'^{-1})_{cd} \bigl(  \lambda'^c 
 \tilde \partial^d  +  \bar\lambda'^c   \tilde
 \partial^d \bigr)\,d\,,
 \end{split}
 \ee 
 where we made use of (\ref{r9jlg}) and (\ref{dflkdrotijkf}) to obtain the second
 line and  (\ref{metrixsdlkf}) to obtain the last line.    This can now be rewritten as
\be
\label{gaugetrans-combop}
\delta   d  =
- {1\over 4} \, \ti \partial\cdot [\lambda '+ \bar \lambda ' ] +  {1\over 2}  \, [\lambda ' +\bar\lambda']\cdot \ti \partial d\,~, 
\ee
 where
  indices are contracted with $\hat G'^{-1}$.
Taking 
 $\bar \lambda '=  \lambda '= \epsilon'$,
 this gives
 \be
\label{gaugetrans-combops}
\delta   d  =
- {1\over 2}  \,\ti \partial\cdot\epsilon' +   \epsilon' \cdot \ti \partial d\,~,
\ee  
which can be rewritten as
 \be
\label{denscs}
\delta e^{-2d}= \ti \partial \cdot(e^{-2d}\epsilon ') \,,
\ee
and is of the same form as (\ref{dens}).   Gauge transformations $\tilde \delta$ of the dilaton
with $\lambda = - \bar \lambda'$ 
 vanish on account of (\ref{gaugetrans-combop}).
 
Let us now turn to the gauge transformations
 (\ref{mastereqns}) of $e_{ij}$ where, again,
all indices are raised with $G^{ij}$.  We rewrite this  result with
lower-indexed fields and derivatives and explicit $G^{-1}$ factors:
\be
\label{mastereqns998}
\begin{split}
\delta e_{ab}  &\equiv   ~ \bar D_b \lambda_a  \,+  D_a \bar\lambda_b  +  
  \hat G^{cd} (D_a \lambda_d)  e_{cb}   +
 \hat G^{cd} (\bar D_b \bar\lambda_d)  e_{ac}    
     + {1\over 2}\, \hat G^{cd} 
     (\lambda_c D_d +  \bar \lambda_c \bar D_d  )e_{ab} ~~
      \\[0.5ex]
   & ~ \,  
-    {1\over 2} \,\hat G^{cd}\bigl(
   D_a \lambda_d  
 +   D_d \lambda_a\bigr) e_{cb} 
   -{1\over 2} \,\hat G^{cd}\bigl(
   \bar D_b \bar\lambda_d  
 + \,  \bar D_d \bar \lambda_b  \bigr) e_{ac}  \,.
   \end{split}
   \ee
Our task now is to manipulate the right hand side above. We replace $D$ and 
$\bar D$ by the explicit forms in (\ref{r9jlg}), write the gauge parameters in terms of the primed gauge parameters in (\ref{dflkdrotijkf}), simplify using (\ref{metrixsdlkf}), and finally evaluate (\ref{newfdlflkd3}).   This takes some effort, but the
result is relatively simple:
\be
\label{mastereqns993}
\begin{split}
\delta e'^{ab}  &=
 ~ \tilde\partial^b \lambda'^a \,
 +\tilde\partial^a   \bar\lambda'^b ~+  
 \,(\tilde\partial^a  \lambda'_c)   e'^{cb}  +
( \tilde \partial^b    \bar\lambda'_c) 
     e'^{ac}      ~
     + {1\over 2}\, 
     (   \lambda'_c   \tilde\partial^c  
      +  \bar \lambda'_c   \tilde\partial^c )e'^{ab} ~~
      \\[0.5ex]
   &\hskip5pt  ~ \,  
-    {1\over 2} \,({\hat G}'^{-1})_{cd}\Bigl[ 
  \bigl( \tilde\partial^a \lambda'^c  +  \tilde\partial^c \lambda'^a  \bigr)  \,  e'^{db}   + e'^{ac}  \bigl(
   \tilde\partial^b \bar\lambda'^d  +  \tilde\partial^p \bar\lambda'^d   \bigr)  \Bigr]  \,.
   \end{split}
   \ee
We first take the case when  
 $\lambda ' = \bar\lambda ' = \epsilon '$.  
We find
\be
\label{mastereqns990}
\delta e'^{ab}  =
 ~ \tilde\partial^b \epsilon'^a \,
 +\tilde\partial^a   \epsilon'^b ~+  
 \,(\tilde\partial^a  \epsilon'_c)   e'^{cb}  +
( \tilde \partial^b    \epsilon'_c) 
     e'^{ac}      ~ +   \epsilon'_c   \tilde\partial^c  \, e'^{ab} ~  
-    {1\over 2} \, \delta \bigl( 
 e'^{ac}  \, ({\hat G}'^{-1})_{cd} e'^{db}   \bigr)\,,  
   \ee
which gives
\be
\label{mastereqns989} 
\delta \Bigl( e'^{ab} +    {1\over 2} \, 
 e'^{ac}  \, ({\hat G}'^{-1})_{cd} e'^{db}   \Bigr) \ =
 ~ \tilde\partial^b \epsilon'^a \,
 +\tilde\partial^a   \epsilon'^b ~+  
 \,(\tilde\partial^a  \epsilon'_c)   e'^{cb}  +
( \tilde \partial^b    \epsilon'_c) 
     e'^{ac}      ~ +   \epsilon'_c   \tilde\partial^c  \, e'^{ab} ~   \,.
   \ee
Note that the tilde derivatives naturally have the index up.  The parameters
$\epsilon'$
naturally have the index down, just like the coordinates, so that an infinitesimal  diffeomorphism takes the form
 $ \tilde x'_a= \tilde x_a + \epsilon'_a $.

  For the case 
 $\bar \lambda '= - \lambda '= - \tilde\epsilon'$,
 equation
 (\ref{mastereqns993}) gives
\be
\label{mastereqns993x5}  
\begin{split}
\tilde\delta e'^{ab}  &=
 ~ \tilde\partial^b \tilde\epsilon'^a \,
 -\tilde\partial^a   \tilde\epsilon'^b  
 \,  -  {1\over 2}  \delta \Bigl( 
  e'^{ac}   \,({\hat G}'^{-1})_{cd}  e'^{db}    \Bigr)  \,,     
   \end{split}
   \ee
so that we now have
\be
\label{mastereqns993x6}  
\begin{split}
\tilde\delta \Bigl( e'^{ab} +    {1\over 2} \, 
 e'^{ac}  \, ({\hat G}'^{-1})_{cd} e'^{db}    \Bigr)  &=
 ~ \tilde\partial^b \tilde\epsilon'^a \,
 -\tilde\partial^a   \tilde\epsilon'^b  
 \,  \,.
   \end{split}
   \ee
 We define
\be
\label{sadsadaak}
\bar e^{ab} \equiv  e'^{ab}  +  {1\over 2}  
  e'^{ac}   \,({\hat G}'^{-1})_{cd}  e'^{db}  \,,
\ee
so that, to this order, our gauge transformations  take the form
\be
\label{mastereqns988xx}
\begin{split} 
\tilde\delta \bar e^{ab}   &=
 ~ \tilde\partial^b \tilde\epsilon'^a \,
 -\tilde\partial^a   \tilde\epsilon'^b \,, \\[1.0ex]
\delta  \bar e^{ab}     &=
 ~ \tilde\partial^b \epsilon'^a\,
 +\tilde\partial^a   \epsilon'^b ~+  
 \,(\tilde\partial^a  \epsilon'_c)  \bar e^{cb}  +
( \tilde \partial^b    \epsilon'_c) 
    \bar e^{ac}      ~ +   \epsilon'_c   \tilde\partial^c  \, \bar e^{ab} \,.
  \end{split}
   \ee
These are the expected transformations.

\medskip
To give a clearer interpretation we now introduce a field that incorporates
the background and the fluctuation.  If we denote by  $\hat E^{ab}$ the inverse background field $\hat E^{ab} = \{ \hat E^{-1}\}$
we now define
\be
\label{834gn37u}
 {\cal E}^{ab} \equiv  \hat E^{ab} + \bar e^{ab} +O({e'}^3) \,.
\ee
 We now show that this has the expected gauge transformations, up to terms of order ${e'}^2$.
We clearly have
\be
\label{sfgwgwrrg}
\tilde\delta {\cal E}^{ab}  = \tilde\partial^b \tilde\epsilon'^a \,
 -\tilde\partial^a   \tilde\epsilon'^b\,.
\ee

Next, we aim to write $\delta  \bar e^{ab}  =\delta \EE^{ab}  $ in terms of a Lie derivative. The Lie derivative 
of ${\cal E}^{ab}$ follows from the  tensorial transformation 
\be
{\cal E}'^{ab} (\tilde x') =   {\partial \tilde x_c\over \partial \tilde x'_a}\,
     {\partial \tilde x_d\over \partial \tilde x'_b} \,\,  {\cal E}^{cd} (\tilde x)\,.
\ee
The result is 
\be{\cal L}_{\epsilon '} {\cal E}^{ab} = 
 \,(\tilde\partial^a  \epsilon'_c)   {\cal E}^{cb}  +
( \tilde \partial^b    \epsilon'_c) 
     {\cal E}^{ac}      ~ +   \epsilon'_c   \tilde\partial^c  \, {\cal E}^{ab}\,.
\ee
Using (\ref{834gn37u}) this gives:
\be
{\cal L}_{\epsilon '} {\cal E}^{ab} = 
 \,(\tilde\partial^a  \epsilon'_c)   \hat {E}^{cb}  +
( \tilde \partial^b    \epsilon'_c)    \hat{E}^{ac}      
    + \,(\tilde\partial^a  \epsilon'_c)  \bar e^{cb}  +
( \tilde \partial^b    \epsilon'_c) 
     \bar e^{ac}      ~ +   \epsilon'_c   \tilde\partial^c  \,\bar  e^{ab}\,.
\ee
Noting that $\hat E^{ab} = {\hat G}'^{ab} + {\hat B}'^{ab}$, 
that $\hat G'$ raises indices, 
and using (\ref{mastereqns988xx}),
we find
\be
\delta {\cal E}^{ab}=\delta \bar e^{ab} =
{\cal L}_{\epsilon '} {\cal E}^{ab} 
-  \,\tilde\partial^a ( \epsilon'_c   \hat{B}^{cb})  +
 \tilde \partial^b   ( \epsilon'_c     \hat {B}^{ca} )     
 \,.
\ee
Since the last two terms give a symmetry of the form (\ref{sfgwgwrrg}),
the theory
contains a gauge symmetry
\be
\delta {\cal E}^{ab} 
 = {\cal L}_{\epsilon'} {\cal E}^{ab} \,. \ee

We conclude by comparing the field redefinition used  here to
that used in the absence of $\tilde x$ dependence, namely
 $e_{ij} \to e_{ij}^+$ in (\ref{eplisdf}). 
 For this purpose it is convenient to re-express the 
 present field redefinition (\ref{sadsadaak})  
\be
\label{sadsadaakas}
e'^{ab} \to  e'^{ab}  +  {1\over 2}  
  e'^{ac}   \,({\hat G}'^{-1})_{cd}  e'^{db}  \,.
\ee
in terms of the lower-indexed field $e_{ab}$.  For this we 
use (\ref{newfdlflkd}), which gives
$e_{cd} = -\hat E_{ce} \,e'^{ef} \,\hat E_{fd}$.
Thus multiplying  (\ref{sadsadaakas}) from the left 
and from the right by $\hat E$, we find
\be
\label{sadsadaakas99}
e_{ab} ~\to  ~e_{ab}  -  {1\over 2}  
  e_{ac}   \,(\hat E^{-1} {\hat G}'^{-1} \hat E^{-1})^{cd}  e_{db}  \,. 
\ee
Using (\ref{metrixsdlkf})  we see that
 \be
\label{newfdminis}
e_{ab} \to e_{ab}^- \equiv e_{ab}
- \frac 1 2 e_{ac} (\hat G^{-1}\, \hat E^t\hat E^{-1})^{cd}\, e_{db}\,.
\ee
The above shows that the field redefinition $e'^{ab}\to \bar e^{ab}$ is equivalent to $e_{ab}\to e^-_{ab}$.  If  $\hat B_{ab}=0$ we find $e_{ab}^-
= e_{ab}
- \frac 1 2 e_a^{\,\,\,c} \,e_{cb}\,,$  an expression that differs
by a crucial sign from  $e_{ij}^+= e_{ij} + \frac 1 2 e_i^{\,\,k}\, e_{kj}$.

Note that the field redefinition $ e\to e^+$ needed to bring the $\epsilon$ transformations
to the form of $x$-diffeomorphisms  (in the $\ti x$-independent case) differs
from the field redefinition $e\to e^-$  needed to bring the $\ti \epsilon$ transformations
to the form of $\ti x$-diffeomorphisms  (in the $ x$-independent case). While our symmetries contain both $x$-diffeomorphisms and $\ti x$-diffeomorphisms in certain limits, it is not clear how, or even if, they fit together to form diffeomorphisms of the doubled torus.

\sectiono{T-duality of the action}\label{Tdualityofaction}

We have written a field theory action (\ref{redef-action}) that represents
the dynamics of certain fluctuations ($e_{ij}$ and $d$)
about the background $E_{ij}$.  T-duality states that the 
closed string physics
around backgrounds $E$ and $E'$ related by an $O(d,d,\mathbb{Z})$
transformation are identical.  In the string field theory context this
was proven in~\cite{Kugo:1992md} by showing that the string field
theories formulated around $E$ and $E'$ are equivalent.  In fact
these theories are related by a homogeneous field redefinition.  This 
field redefinition does not mix fields at different mass levels; on a given field
it shuffles momenta and winding, as well as the various polarizations.
For this reason, it is to be expected that  
our construction, which only keeps the $N = \bar N = 1$ fields, should have a T-duality symmetry.    In this section we prove that $T$-duality
is a property of the action we have constructed. 
In string field theory, there are cocycle-induced sign factors in the T-duality transformations
\cite{Kugo:1992md}.\footnote{See  \cite{Hellerman:2006tx} for a review of 
the role of cocycles in T-duality in the first-quantized formalism.}
Our cubic action does not include the 
momentum dependent sign factors that arise from cocycles and so
our T-duality transformations do not include such factors either. 
As we discuss in Section~\ref{constandcoc}, such sign factors may be needed in some circumstances and could  affect the  duality transformations.

We also  establish that the action  
 is invariant under the background
change $B_{ij} \to - B_{ij}$.  This discrete symmetry is not part of
the group of $O(d,d,\mathbb{Z})$ symmetries,   but plays an important role in the theory.  
We conclude by discussing  a natural generalization
of  the Buscher rules that may describe T-duality transformations of toroidal 
backgrounds that fail to have $U(1)$ isometries due to explicit 
dependence on both coordinates and dual coordinates of the tori.
Again, this discussion is modulo cocycle-induced sign factors.

\subsection{Duality transformations}\label{dualfeirjnf}

We begin by reviewing a few properties of 
duality transformations.  The group elements  $g\in O(D,D; \mathbb{Z})$ are 
$2D\times 2D$ matrices of integers
that leave the metric $\eta$ invariant:
\be
\label{condsooo}
g^t \eta g = \eta \,, \qquad\eta= \begin{pmatrix}0& I \\ I& 0 \end{pmatrix}\,.
\ee
One readily sees that $\det g = \pm 1$.

Our indices $i$ run over $D=n+d$ values,  so that the coordinates $x^i$ split into   $n$ non-compact directions  $x^\mu$ and $d$ compact ones $x^a$.
If $n=0$ and all dimensions are compact, then 
the doubled torus has  $2D$   periodic coordinates $x^i, \ti x_i$ transforming in the fundamental representation of $O(D,D;\Z)$.
If there are $n$ non-compact directions, 
we shall be interested in the $O(d,d;\Z)$ 
subgroup of $O(D,D;\Z)$   that preserves $x^\mu$
and which acts only on the $2d$ periodic coordinates $x^a,\ti x_a$.
It is this $O(d,d;\Z)$ subgroup that is a symmetry of the string theory, but it will be useful to represent its action in terms of the $2D\times 2D$ matrix $g$.

As in \S\ref{gentorback},    
we write
$E = G + B$, with 
$D\times D$ matrices
$E =  \{ E_{ij} \}$,  $G=  \{ G_{ij} \}$,  and $B = \{ B_{ij} \}$. We also
use  $G^{-1} =  \{ G^{ij}\}$. 
If we write the $2D\times 2D$ matrix
\be
g= \begin{pmatrix}a& b \\ c& d \end{pmatrix}\,,
\ee
 the group action on the background is
\be
\label{eeprime}
E ' =  g(E) =   ( a E + b ) (c E + d)^{-1} \,.
\ee
We emphasize that we restrict ourselves to matrices $g$ in the $O(d,d;\Z)$ subgroup of $O(D,D;\Z)$.   
This means that explicitly we have the $D\times D$ matrices:
\be
a = \begin{pmatrix} \hat a & 0 \\ 0 & 1\end{pmatrix}\,, 
~~ b =  \begin{pmatrix} \hat b & 0 \\ 0 & 0\end{pmatrix}\,,
~~ c =  \begin{pmatrix} \hat c & 0 \\ 0 & 0\end{pmatrix}\,,
~~ d =  \begin{pmatrix} \hat d & 0 \\ 0 & 1\end{pmatrix}\,, 
\ee
where $\hat a, \hat b, \hat c$, and $\hat d$ are $d\times d$ matrices such
that 
\be 
\hat g= \begin{pmatrix}\hat a& \hat b \\ \hat c& \hat d \end{pmatrix} 
\in \, O (d, d, \mathbb{Z})\,.
\ee
(We use hats for $d\times d $ matrices.)
It is straightforward to verify that if $\hat g \in O(d, d, \mathbb{Z})$ 
then $g \in O(D, D, \mathbb{Z})$.
The background $E$ is a matrix of the form
\be
E = \begin{pmatrix}  \hat E & 0 \\ 0 &  \eta \end{pmatrix}\,, ~~\hbox{with}
~~ \hat E = \hat G + \hat B = 
 [\hat E_{ab}] ~~\hbox{and} ~~ \eta = [\eta_{\mu\nu}]   
\, .\ee
It follows from the transformation in (\ref{eeprime}) that
\be
\label{eorifkljgeriu}
E' =  \begin{pmatrix} \hat E' & 0 \\ 0 & \eta \end{pmatrix} \,, ~~
\hbox{with} ~~  \hat E' =  (\hat a \hat E + \hat b ) (\hat c \hat E + \hat d)^{-1} \,.
\ee
This is the expected transformation of the background metric: the
background $\hat E$ in the torus is transformed by an element
of $O(d,d,\mathbb{Z})$ while the Minkowski background is left unchanged.

It is a familiar result that the   
non-linear  transformation (\ref{eeprime}) of $E$ 
 becomes the linear transformation of 
the $2D\times 2D$ matrix $\HH$ defined in (\ref{genmet}): 
\be
\label{linHtransf}
 \HH(E')  = g\,\HH(E)g^t \, .    
\ee

It is useful to introduce the $D\times D$  matrices 
$M$ (written as $M_i{}^j$) and $\bar M$ (written as $\bar M_i{}^j$)
defined by the relations
\be
\label{mbarm}
\begin{split}
M&\equiv  d^t - \,E \, c^t\, = \begin{pmatrix} {\hat d}^t - \hat E\, \hat c^t
& 0 \\ 0 & 1 \end{pmatrix} \,,\\[1.0ex]
\bar M  &\equiv  d^t  + E^t  c^t  =  \begin{pmatrix} \hat d^t + \hat E^t\, \hat c^t
& 0 \\ 0 & 1 \end{pmatrix}\,.
\end{split}
\ee
The matrices $M$ and $\bar M$ control the transformation
of the metric $G$ 
obtained from  (\ref{eeprime}) 
by splitting $E'$ into symmetric and antisymmetric parts,  $E' = G' + B'$.
Indeed, equation (4.10) in~\cite{Kugo:1992md} gives
\be  
\label{oeihrluge}
\begin{split}  
(\hat d + \hat c \hat E)^t \,\, {\hat G}'  \,\,(\hat d + \hat c \hat E)  & =  
\hat G \,,\\[0.3ex]
(\hat d - \hat c \hat E^t)^t \, {\hat G}'  \,(\hat d  -\hat c \hat E^t)  & = \hat  G\,.
\end{split}
\ee
These relations, together with (\ref{mbarm})  quickly lead to
\be
\label{fsetmat4678599}
\begin{split}
G^{-1} &= (\bar M^t )^{-1}\,  G^{\prime -1}  \,\bar M^{-1}\,,\\[1.0ex]
G^{-1} &= (M^t)^{-1}  \, G^{\prime -1}\, M^{-1} \,.    
\end{split}
\ee
Two more identities from~\cite{Kugo:1992md} (eqns.\,(4.19)) are
useful to us:
\be
\label{jnbswvgn}
\begin{split}
\hat b^t - \hat E \hat a^t &= -  (\hat d^t - \hat E \hat c^t)   \hat E'\,,\\[1.0ex]
\hat b^t + \hat E^t \hat a^t &= ~( \hat d^t + \hat E^t \hat c^t)  \hat E'^t\,.
\end{split}  
\ee
In terms of the $D\times D$ matrices the above relations give,
\be
\label{jnbvgn}
\begin{split}
b^t - E a^t &= -  M  E'\,,\\[1.0ex]
b^t + E^t a^t &= ~\bar M\, E'^t \,.   
\end{split}
\ee
Finally, a perturbation of the background $E+\delta E$ transforms to $E'+\delta E'$ where
\be
\delta E'=M^{-1} \delta E (\bar M^t)^{-1}\,,
\ee
so that
\be
\label{sdfjlsd}
\begin{split}
\delta E_{ij}  &= M_{i}{}^k  \, \bar M_{j}{}^l  \, \delta E'_{kl} ~\,.
\end{split}
\ee

\subsection{Duality invariance}

We will begin with the
action (\ref{redef-action}) written around a background $E$ and
with  fields $(e_{ij},d)$ collectively
denoted by $\Psi$.  Setting  $2\kappa^2 =1$ we have 
\be
\label{initaction}
S(E, \Psi) = \int dx^\mu\, d\X\, \, \mathcal{L}\, \Bigl[D_k , \bar D_l , G^{-1}\,;\, 
e_{ij}(x^\mu,\X),  d(x^\mu,\X) \Bigr]\,.
\ee 
Here $\X$ is  a $2d$-column vector of coordinates
\be
\X \equiv  \begin{pmatrix} \tilde x_a \\ x^a  \end{pmatrix}\,,
\ee
and $\int d\X \equiv  \int d\tilde x dx$.   
  The action
  (\ref{initaction}) 
  is constructed from lower-index derivatives $D_i,\bar D_j$ and the lower-indexed $e_{ij}$ fields (together with $d$) with all index 
 contractions using the  
metric $G^{-1}$.
The action depends on the background $E$ 
through $G^{-1}$ and the derivatives $D,\bar D$ (see (\ref{groihfgruu8774})).  

We will establish equivalence 
between the theory on  the 
background $E$ and the 
  theory formulated on a  background 
\be
E' = g(E) \, \quad \hbox{with} \quad  g = \begin{pmatrix}a& b \\ c& d \end{pmatrix}\,,
\ee 
where  $g$
is in the  $O(d,d;\mathbb{Z})$ subgroup of $O(D,D;\Z)$,
 as explained in \S\ref{dualfeirjnf}.

It is notationally convenient to introduce extra coordinates
$\ti x_\mu$, so that we have $2D$ coordinates $X$ with
\be  
X \equiv  \begin{pmatrix} \tilde x_i \\ x^i  \end{pmatrix}\,, ~~
\hbox{where} \quad  \tilde x_i = \begin{pmatrix} \tilde x_a \\ \ti x_\mu  \end{pmatrix}\,, ~~ x^i = \begin{pmatrix} x^a \\ x^\mu
\end{pmatrix}\,.
\ee
We will only consider fields that are independent of the extra coordinates 
$\ti x_\mu$, so that these coordinates play no role.
With the help of these coordinates the action (\ref{initaction}) can be written
as 
\be
\label{initactionxxs}
S(E, \Psi) = \int dX\, \, \mathcal{L}\, \Bigl[D_k , \bar D_l , G^{-1}\,;\, 
e_{ij}(X),  d(X) \Bigr]\,.
\ee 
Here
\be
\int dX \equiv  \int dx^\mu dx^a  d\ti x_a \,,
\ee
with no integration over the trivial coordinates $\ti x_\mu$.
Our argument will apply to any action that is of the form (\ref{initactionxxs})
and with  indices contracted in the way we describe below.

There is a natural action of $O(D,D)$ on the $2D$ 
coordinates $X$ but, as before,  
we only consider the $O(D,D;\Z)$ transformations in the $O(d,d;\Z)$ subgroup that  preserves $x^\mu$ and $\ti x_\mu$ and respects the periodicities of the $x^a,\ti x_a$.
Such a  
 transformation takes $X$ to $X'$ where
\be
\label{vmpss}
X' = \begin{pmatrix} \tilde x' \\ x'  \end{pmatrix}=
  g X  =   \begin{pmatrix}a& b \\ c& d \end{pmatrix}
 \begin{pmatrix} \tilde x \\ x  \end{pmatrix}
 =  \begin{pmatrix} a\tilde x + b x \\ c\tilde x + d x  \end{pmatrix}\,.
\ee
Then  $O(d,d;\Z)$ transformations   act as diffeomorphisms of the doubled torus
$T^{2d}$, the subgroup of the large diffeomorphisms $GL(2d;\Z)$ preserving $\eta$.
Our ansatz for the transformation of $e,d$ in the general case follows from the 
transformations found in~\cite{Kugo:1992md}:
\be
\label{clhsb88vgn}
\begin{split}
e_{ij} (X) &= M_{i}{}^k  \, \bar M_{j}{}^l  \, e'_{kl} (X')\,,\\[0.5ex]
d(X) &=  d'(X')\,.
\end{split}
\ee
Using this to write the fields $e,d$ in terms of $e',d'$ in 
in (\ref{initactionxxs}) 
gives
\be
\label{initaction2}
S(E, \Psi (\Psi')) = \int   dX'\, \, \mathcal{L}\, \Bigl[D_i , \bar D_j , G^{-1}\,;\, 
M_{i}{}^k  \, \bar M_{j}{}^l  \, e'_{kl} (X'),  d'(X') \Bigr]\,,
\ee 
where we have used
\be
\int dX \equiv  \int dx^\mu   dx^a d \ti x_a  = \int dx^\mu  d(x')^a d (\ti x')_a
 =\int dX'\,,
 \ee
since the Jacobian of the transformation is unity.

The transformation  (\ref{vmpss}) of $X$  
implies  that the (lower-indexed) 
 derivatives acting on the new fields can be rewritten in terms
of primed derivatives based on $E'$ as follows
 \be
\label{sghsabvgn99}
\begin{split}
D &= ~~ M D' \,, \\[0.5ex]
\bar D &=   ~~\bar M  D'\, ,
\end{split}
\ee
as we will
show below.
Then 
the action becomes
\be
\label{initaction3}
S(E, \Psi (\Psi')) = \int   dX'\, \, \mathcal{L}\, \Bigl[M D'  , \bar M  D' , G^{-1}\,;\, 
M_{i}{}^k  \, \bar M_{j}{}^l  \, e'_{kl} (X'),  d'(X') \Bigr]\,.
\ee 
If we can show that this is equal to
\be
\label{initaction33}
  S(E', \Psi') = \int  dX'\, \, \mathcal{L}\, \Bigl[ D', \bar D',\,   G^{\prime -1} \,;\, 
 e'_{ij} (X'),  d'(X') \Bigr]  \,,
\ee 
then we
will have 
\be S(E', \Psi') =S(E, \Psi (\Psi')) \,,
\ee
establishing the   desired physical equivalence.  

To show this, we 
need to keep track of which indices transform with an $M$ and which with an  
 $\bar M$.
For  this argument, we introduce a notation in which lower indices $i$ transform with an $M$ and lower
 indices $\bar i$ transform with a $\bar M$, while upper indices transform with the inverses of these matrices. Then (\ref{sghsabvgn99}) implies that the derivatives are $D_i, \bar D_{\bar j}$ while 
 (\ref{clhsb88vgn}) implies that $e_{i\bar j}$ has a  first index which is unbarred and a second which is barred.
 The two forms for the transformation of the metric in (\ref{fsetmat4678599}) imply that we can write $G^{-1}$ 
 with with two unbarred indices 
 as $G^{ij}$ or with two barred ones
as  $G^{\bar i\hskip1pt \bar j}$.
 For any action in which all unbarred indices are contracted amongst
 themselves using $G^{ij}$ and all the barred indices are contracted amongst themselves using 
  $G^{\bar i \hskip 1pt\bar j}$, 
  equation  
  (\ref{fsetmat4678599}) implies that all factors of $M$ and $\bar M$
  will cancel. 
    This gives the equality of (\ref{initaction3}) and
  (\ref{initaction33}), as required. 
The index contractions in the cubic action (\ref{redef-action}) 
indeed obey this rule. 
We see from the string field (\ref{the_string_field}) that the first
index in $e_{ij}$ is tied to an unbarred oscillator while the second is
tied to a barred oscillator.  It is clear 
from the commutation relations  (\ref{erocgkjer}) that
contractions always relate two un-barred or two barred operators, but
cannot ever mix them. The same is true for the derivatives $D$ and 
$\bar D$ that arise from unbarred and barred zero modes,
as shown in (\ref{derorign}). It follows that the action derived
from the string field theory obeys the 
stated contraction rules, and so must be T-dual in this way.

\medskip
To complete the above proof we must derive (\ref{sghsabvgn99}). 
Consider the action of derivatives with respect to  $x$ and $\tilde x$
on functions of $X'$.  As a preliminary, 
short calculations using (\ref{vmpss}) give
\be
\begin{split}
{\partial \over \partial x}  F( X') &=   \Bigl(  b^t {\partial \over \partial \tilde x'}
+ d^t {\partial \over \partial x'} \Bigr)  F( X')\,,\\[1.0ex]
{\partial \over \partial \tilde x}  F( X') &=   \Bigl(  a^t {\partial \over \partial \tilde x'}+ c^t {\partial \over \partial x'} \Bigr)  F( X')\,.
\end{split}
\ee
We then have for $D_i$ 
\be
\begin{split}
D  F(X')  &= {1\over \sqrt{\alpha'}} \, \Bigl( {\partial \over \partial x} 
- E\,  {\partial \over \partial \tilde x} \Bigr)  F(X') \\
&= {1\over \sqrt{\alpha'}} \, \Bigl( b^t {\partial \over \partial \tilde x'}
+ d^t {\partial \over \partial x'}
- E \Bigl[ a^t {\partial \over \partial \tilde x'}+ c^t {\partial \over \partial x'} \Bigr] \Bigr) F(X') \\
&= {1\over \sqrt{\alpha'}} \,\Bigl(
 (d^t - E c^t) {\partial \over \partial x'}
+  ( b^t - E a^t)  {\partial \over \partial \tilde x'}
 \Bigr) F(X')\,.
\end{split}
\ee
Making use of (\ref{jnbvgn})
\be
D  F(X')  =  M~{1\over \sqrt{\alpha'}} \,\Bigl(
 {\partial \over \partial x'}
-E'  {\partial \over \partial \tilde x'}
 \Bigr) F(X') = M~D' F(X')\,,
\ee
as we wanted to show. 
We repeat for the   derivative $\bar D_i$:
\be
\begin{split}
\bar D  F(X')  &= {1\over \sqrt{\alpha'}} \, \Bigl( {\partial \over \partial x} 
+ E^t\,  {\partial \over \partial \tilde x} \Bigr)  F(X') \\
&= {1\over \sqrt{\alpha'}} \, \Bigl( b^t {\partial \over \partial \tilde x'}
+ d^t {\partial \over \partial x'}
+ E^t \Bigl[ a^t {\partial \over \partial \tilde x'}+ c^t {\partial \over \partial x'} \Bigr] \Bigr) F(X') \\
&= {1\over \sqrt{\alpha'}} \,\Bigl(
 (d^t + E^t c^t) {\partial \over \partial x'}
+  ( b^t + E^t a^t)  {\partial \over \partial \tilde x'}
 \Bigr) F(X')\,.
\end{split}
\ee
Making use of (\ref{jnbvgn})
\be
\bar D  F(X')  =  \bar M~{1\over \sqrt{\alpha'}} \,\Bigl(
 {\partial \over \partial x'}
+E'^t   
 {\partial \over \partial \tilde x'}
 \Bigr) F(X') = \bar M ~\bar D' F(X')\,,
\ee
as we wanted to show.  This completes our proof of (\ref{sghsabvgn99}),
and therefore our proof of $T$-duality. 

\subsection{Inversion}

We now give some explicit formulae
 relevant to  the $\mathbb{Z}_2$ duality transformation
that simultaneously
exchanges all tori coordinates $x^a$ and $\tilde x_a$.
This duality  transforms the toroidal background
with
\be 
\label{sdfaszs}
\hat g= \begin{pmatrix}\hat a& \hat b \\ \hat c& \hat d \end{pmatrix} 
= \begin{pmatrix} 0& 1 \\ 1 & 0 \end{pmatrix}  \in \, O (d, d, \mathbb{Z})\,.
\ee
Explicitly, $ \hat b_{ab}=\delta _{ab}$ and  $ \hat c^{ab}=\delta ^{ab}$, introducing metrics that can naturally raise and lower indices in what follows.
Using (\ref{eorifkljgeriu}) we find that the toroidal part of the background
is transformed to:
\be
\hat E'  = 
 \hat E^{-1}\, , ~~\qquad~~
 E' = \begin{pmatrix}\hat E^{-1} &0\\0 & \eta   \end{pmatrix}\,.
\ee
We also have from (\ref{oeihrluge})
\be  
\label{oeihrlug9e}
\hat E^t \,\, {\hat G}'  \hat E  =\hat E \, {\hat G}'  \hat E^t   = \hat G\,.  \ee
Taking inverses and solving for ${\hat G}'^{-1}$ we find
\be
\label{metrixsdlkf99}
{\hat G}'^{-1} = \hat E\, \hat G^{-1} \, \hat E^t  = \hat E^t\, \hat G^{-1} \, \hat E\,.
\ee 
This 
gives equation (\ref{metrixsdlkf}) which was  used to investigate the
gauge transformations of the theory in which fields depend only on $\tilde x$.
We also have from  (\ref{mbarm})
 \be
\label{mbarmn}
M  = \begin{pmatrix}  - \hat E\, 
& 0 \\ 0 & 1 \end{pmatrix}\,, ~~
\bar M  =  \begin{pmatrix}  \hat E^t
& 0 \\ 0 & 1 \end{pmatrix}\,,
\ee
so that
\be
M_a{}^b= -\hat E_{ac}\delta ^{cb}, \qquad
\bar M_a{}^b= \hat E_{ca}\delta ^{cb} \,,
\ee
using $ \hat c^{ab}=\delta ^{ab}$.
The transformation for the field $e$ was given in (\ref{clhsb88vgn})
and takes
the form
$e_{ab} (X) = M_{a}{}^c  \, \bar M_{b}{}^d  \, e'_{cd} (X')$,  
since the matrices
$M$ and $\bar M$ are block diagonal.  We then find
\be
e_{ab} (X) 
= - \hat E_{ac} \, {e'}^{cd}(X')\,
\hat E_{db} \, ,
\ee
where
$ {e'}^{cd}= e'_{ab} \delta ^{ac}\delta ^{bd}$, giving
an  $e'$ with upper indices, which was the natural convention used in \S\ref{ssccx}.
If we solve for $e'$ we immediately obtain (\ref{newfdlflkd}). 
The above results justify the starting point of the analysis in
\S\ref{ssccx}.

\subsection{The discrete symmetry $B\to - B$}

It is well known  that the background
change $B_{ij} \to -B_{ij}$ in a toroidally compactified 
theory is a symmetry of the closed string theory.
Since $B_{ij}$ couples electrically to the string, this symmetry 
is a consequence of the orientation invariance of the theory.
We now show that the discrete symmetry discovered in the 
action guarantees the invariance of the physics under $B_{ij} \to -B_{ij}$.

We begin with our  action 
\be
S(E, \Psi) = \int dX\, \, \mathcal{L}\, \Bigl[D , \bar D , G^{-1}\,;\, e_{ij}(X), 
d(X) \Bigr]\,.
\ee 
The replacement $B \to -B$ makes $E \to  G - B$, which means
\be
E \to  E^t\,.
\ee
This does not affect the metric $G$,  
but the action formulated
with background $E^t$ has the derivatives
changed. Given (\ref{groihfgruu8774}), we have 
\be
S(E^t, \Psi) = \int 
dX\, \, \mathcal{L}\, \Bigl[{ 1\over \sqrt{\alpha'}} \,\Bigl(\,{\partial\over \partial x} - E^t \,{\partial \over \partial\tilde x}\Bigr)
 ,~   { 1\over \sqrt{\alpha'}} \,  \Bigl(\,{\partial\over \partial x} + {E}\, {\partial \over \partial\tilde x}\Bigr), ~G^{-1}\,;\, e_{ij}(X), 
d(X) \Bigr]\,.
\ee 
We now redefine the fields  as
\be
\begin{split}
e_{ij} (X) &=  e'_{ji} ( X' )\,,\\[0.5ex]
d(X) &=  d' ( X' )\,,
\end{split}
\ee
with 
\be
X' =  \begin{pmatrix} \tilde x' \\ x'  \end{pmatrix} = 
 \begin{pmatrix} -\tilde x \,\\ ~x  \end{pmatrix}\,.
\ee
The effect of this change on the derivatives is to reverse the sign 
of the terms carrying a tilde coordinate, so by now
\be
S(E^t , \Psi(\Psi')) =~ \int 
dX'\, \, \mathcal{L}\, \Bigl[ \bar D' , \,
 D'  , G^{-1}\,;\, e'_{ji}(X'),\,  d'(X') \Bigr]\,,
\ee
where we also used  $dX = dX'$.
The above action has exactly the replacements associated with the discrete
symmetry (\ref{disc-sym-final}) that  leaves the action invariant, so 
\be
S(E^t , \Psi (\Psi')) =~ \int 
dX'\, \, \mathcal{L}\, \Bigl[D', \,
\bar D' , G^{-1}\,;\, e'_{ij}(X') ,
 d'(X') \Bigr]  =  S(E, \Psi')\,.
\ee
This shows the physical equivalence of the actions
formulated around $E$ and around $E^t$.

\subsection{Field redefinitions, Buscher rules, and generalised T-duality}
\label{frbusgenT}

If the fields $e_{ij}$ and $d$ depend on the spacetime coordinates $x^i=(x^\mu,x^a)$ but are independent of the dual coordinates $\ti x_a$,  then
there is a conventional
low-energy effective theory.
The effective field theory for these fields
obtained using string field theory  must be
equivalent to the standard  
string low-energy effective field theory
 (\ref{standardaction}) with higher-derivative $\alpha ' $ corrections.
 The standard theory is written in terms of  the total field $\mathcal{E}_{ij}$ 
which defines    $\Gg_{ij}$ and $\bb_{ij}$ fields   that have the standard diffeomorphism and 
anti-symmetric tensor gauge transformations  (\ref{diftran99}). 
The map from string field theory to   the standard effective  field theory
has been studied in~\cite{Michishita:2006dr}, but has not been found explicitly.
We have shown in (\ref{fullfield})  
 that 
\be
\label{fullfielda}
 {\cal E}_{ij}= \Gg_{ij} + \bb_{ij} =  E_{ij} +
 e'_{ij}  +  {1\over 2} {e'_i}^{\,k} e'_{kj}  + \hbox{cubic corrections} \,.   \ee
 Here
 $e'_ {ij}$ is the field   used for the double field theory
 and is related to 
 the string field theory variable $e_{ij}$  arising in (\ref{the_string_field}) by (\ref{redefine-cubic}), so that
$e'_{ij}  = ~ e_{ij} +  e_{ij} d$,
making it clear that the dilaton $d$   mixes in.
In the following, we will use only $e'_{ij}$ and drop the primes.

The full non-linear relation will include $\alpha '$ corrections involving  
 derivatives of $e_{ij}$ and $d$ and string loop corrections. 
It is also subject to field redefinition ambiguities~\cite{Michishita:2006dr}. 
To zeroth    order in $\alpha '$, however, the relation should contain no derivatives, 
 on dimensional grounds and 
 because  it is used to match two two-derivative actions. Then at zeroth order in $\alpha '$ and at 
 string tree level there must be some algebraic function $f(e,d)$, so that
\be
\label{fullfieldb}
 {\cal E}_{ij}\equiv E_{ij} +
  f_{ij}(e,d) \, , \qquad f_{ij}(e,d) =e_{ij}  +  {1\over 2} {e_i}^{\,k} e_{kj}  
 + \hbox{cubic corrections}\,, 
 \ee 
 with $ {\cal E}_{ij}$ transforming as in (\ref{geomtransf}).
 Moreover, this  relation should apply both for the compactified and uncompactified theory.
 The field $ {\cal E}_{ij}$    combines the background and fluctuations geometrically.

 When the fields are independent of the torus coordinates 
 $x^a$ as well as $\ti x_a$,
the $U(1)^d$ torus action is an isometry leaving the fields invariant
and T-duality acts through the Buscher rules \cite{buscher, GivRoc}. 
The full metric $\Gg_{ij}$ and Kalb-Ramond field $\bb_{ij}$ depend on $x^\mu$ and are independent of $x^a,\ti x_a $ and so transform according to the extension \cite{GivRoc} of the Buscher rules 
for the torus:  
\be
\label{caleeprime}
{\cal E} ' =  g({\cal E}) =   ( a {\cal E} + b ) (c {\cal E} + d)^{-1} \,.
\ee
These transformations are expected to receive $\alpha '$ corrections and possibly string loop corrections, but to zeroth order in $\alpha ' $ and string tree level, they are the complete non-linear transformations.
This can now be compared with the T-duality transformations of $e_{ij},d$ found above.
As the coordinates $x^\mu$ do not transform, 
the dilaton is
invariant and (\ref{clhsb88vgn})   
gives   
\be
\label{clhsb88vsdgn}
\begin{split}
e_{ij} (x) &= M_{i}{}^k  \, \bar M_{j}{}^l \, e'_{kl} (x)\,, \\[0.5ex]
d(x) \,&= \, d'(x)\,.
\end{split}
\ee
For infinitesimal $e_{ij}$,  we have   
$ {\cal E}_{ij}\ = E_{ij} +
 e_{ij}  +  O(e^2)$ and using (\ref{sdfjlsd})   
we see that the expansion   
of (\ref{caleeprime}) gives a linear transformation of $e_{ij}$ which is precisely (\ref{clhsb88vsdgn})  plus quadratic corrections.   
The requirement that the relation (\ref{fullfieldb}) should map the linear transformation
(\ref{clhsb88vsdgn}) to the fractional linear transformation places stringent constraints on the function $f_{ij}$,
as discussed in~\cite{Michishita:2006dr}.  
A simple explicit function that is compatible with these two forms of the T-duality transformations was found in~\cite{Michishita:2006dr}, but requiring such compatibility does not fix the function uniquely.

Let us now return to the case in which the fields depend on the torus coordinates $x^a$ as well as $x^\mu$ (but not $\ti x_a$), so that  
massive Kaluza-Klein modes with momenta on the torus exist. The dependence on $x^a$ means that the $U(1)^d$ torus action does not preserve the fields and so the usual  Buscher rules do not apply.
Nonetheless, our linear transformations (\ref{clhsb88vgn}) 
 for $e,d$ 
 still apply, and  
 the full fields with geometric gauge transformations are 
 still   given by (\ref{fullfieldb}).
The function $f_{ij}$ in  (\ref{fullfieldb})
still converts  linear duality transform transformations
into  fractional linear transformations.
As a result, we learn that the linear  transformation of $e_{ij},d$ implies 
the non-linear transformations of 
${\cal E},d$ given by
\be
\label{newbuscher}
\begin{split}
{\cal E} ' (X')&=  g({\cal E}(X)) =   ( a {\cal E}(X) + b ) (c {\cal E}(X) + d)^{-1} \,,\\[0.3ex]
d'(X')&=~d(X)\,.
\end{split}
\ee
Here, the argument $X$
refers to $(x^\mu, x^a, \ti x_a=0)$
and $X' = g X$.  
For inversion in all $d$ circles,
$X'$ is given by $(x'^\mu, x'^a , \ti x_a')= (x^\mu, 0,  x^a )$
so that the T-dual of a configuration with dependence   on $(x^\mu, x^a)$ is one with dependence on $(x^\mu, \ti x_a)$, 
as   expected.
Conversely,  the T-dual of a configuration with dependence   on $(x^\mu, \ti x_a)$  is one with dependence on $(x^\mu, x^a)$.

We now turn to the general case in which the fields $e_{ij}$  and $d$ depend on 
$\ti x_a$ as well as
$x^\mu,x^a$. Then the T-duality transformations of $e_{ij},d$ are still given by (\ref{clhsb88vgn}).
In this case it is not so clear how we should define the total field ${\cal E}(X)$
as we   no longer have a conventional field theory description to guide us.
Moreover,  as we saw in \S\ref{ssccx},  
 different field redefinitions are useful in different contexts.
A natural definition, however, is to take
${\cal E}_{ij}\equiv E_{ij} +
 e_{ij}  +  f_{ij}(e,d)$ with the {\it same} algebraic function $f$ that arose above in the map from string field theory 
 to the effective field theory,  
 so that we recover the results above in the case in which 
there is no dependence on $\ti x_a$.  
If we do so,  
the fact that $f_{ij}$ maps our linear T-duality transformations to fractional linear ones implies that the transformation of this 
${\cal E}_{ij}$ is again  
given by  
(\ref{newbuscher}) but now with general dependence 
on the coordinates $(x^\mu, x^a, \ti x_a)$.
This is a simple and manifestly  $O(d,d;\Z)$ compatible 
candidate for the generalisation of the Buscher rules to the case with general dependence on $(x^\mu, x^a, \ti x_a)$.

The fact that the double field theory action satisfies
\be
S(E,e,d)=S(E',e',d') \,,
\ee
implies that the transformation of both the background 
$E$ and the fields $e,d$ is a symmetry of the action.
If the action can be rewritten in terms of the total field ${\cal E}$ so that
$S(E,e,d)=S({\cal E},d)$, then it will be manifestly independent of the split into a background field $E$ and a fluctuation $e$ and will be invariant under  the T-duality transformations
  (\ref{newbuscher})
\be
S({\cal E},d)=S({\cal E}',d') \,.
\ee

\medskip
The matrix $\HH(E)$ defined in (\ref{genmet}) for the background field $E$
has a natural  
generalisation for the total field $\EE = \Gg + \bb$.
We define the $2D\times 2D$ matrix $\HH({\cal E})$ by 
\be
\label{genmetpl}
 \HH({\cal E}) = \begin{pmatrix}  ~\Gg - \bb \Gg ^{-1}
\bb ~  &  \bb \Gg ^{-1}  \\[2.0ex]  -\Gg ^{-1} \bb  &  \Gg ^{-1} \end{pmatrix}\,.
\ee
It follows from (\ref{linHtransf}) that the background transformation
 (\ref{newbuscher}) 
 induces a simple linear transformation for $ \HH(\EE)$
 \be
 \label{dfghssdgf9}
  \HH(\EE '(X')) = g\, \HH(\EE (X))\, g^t \,.
\ee
From (\ref{dfghssdgf9}), the
inverse $\GG({\cal E})  \equiv  (\HH({\cal E}))^{-1}$
transforms as
\be
\label{dfghssdgf}
\GG ({\cal E}) =  \, g^t \,   \GG ({\cal E}' (X')) \, g \,.  
\ee
This
can be written suggestively  using  $X' = g X$: 
 \be
  \label{dfghssdgf2}
  \GG (X)
   =\left( \frac {\partial X' }  {\partial X} \right)^t  \GG' (X')\left( \frac {\partial X' }  {\partial X} \right) \, .
\ee
where
$\GG'(X')=\GG(\EE'(X') ) $  and $\GG(X)=\GG(\EE (X)) $.
This shows that $\GG$ behaves as a covariant tensor under $O(D,D)$ transformations.  Indeed,  $\GG$ defines a duality invariant 
line element
 \be
 \label{dualinvmetric}
ds^2
 = dX ^t\, \GG(  \EE(X))\, dX\,.
 \ee
The metric   $\GG$  and its relation to the generalised metric in generalised
geometry is discussed in~\S\ref{coandopenque}.

\sectiono{Constraint, cocycles, and null subspaces}\label{constandcoc}

In this section we discuss some of the subtle issues that
arise in our construction.  We have referred to these 
at various points in the earlier sections.  We begin
with  an examination
of the constraint that requires fields and gauge parameters to be
in  ker$(\Delta)$, namely, the kernel of the second-order differential operator 
$\Delta$.  We
define the natural linear projection $[[\, \cdot \,]]$ that takes an arbitrary 
double field to this kernel.  We then turn to a discussion of
cocycles and sign factors.  It is possible that the nonlinear completion
of the theory will involve these sign factors.  Finally, we conclude
with a discussion of null spaces that 
arise from the restriction to
double fields that have no winding in some suitable T-dual frame,
resulting in a conventional field theory for that non-winding sector.

\subsection{The constraint and projectors}\label{theconandpro}

The constraint $L_0- \bar L_0=0$ is applied to all fields and gauge parameters.
The product of two fields satisfying the constraint will not  satisfy it in general, and for this reason
 the string product $[ \cdot, \cdot]$  
includes an explicit projection onto states that satisfy
$L_0- \bar L_0=0$.
It also has
 an insertion of $b_0^-$  that ensures
that the string product is annihilated by $b_0^-$. Schematically, we have
\be
\label{stringsvertex}
[\Psi_1 , \Psi_2 ]  \equiv   \int {d\theta\over 2\pi} e^{ i \theta (L_0 - \bar L_0) }  b_0^- [ \Psi_1 , \Psi_2]'  = \delta_{L_0 - \bar L_0, 0} ~b_0^- [ \Psi_1 , \Psi_2]' \,,
\ee
where the primed bracket $[\cdot \, , \, \cdot]'$ inserts the states in
the three-punctured sphere that defines the vertex. 
 The
$b_0^-$ insertion implies that the string product has an intrinsic ghost number
of minus one:  $\hbox{gh} ([A, B]) = \hbox{gh} (A) + \hbox{gh}(B) -1$.  
This inclusion of the projection in  the string product in covariant closed string field
theory leads to the  failure of   a Jacobi identity  and this then requires further higher order interactions resulting in a  
non-polynomial theory.  Concretely, one finds~\cite{Zwiebach:1992ie}
\be
\label{csft-fh}
\begin{split}
0 =&\,  Q [B_1, B_2, B_3] +  [QB_1, B_2, B_3] +(-1)^{B_1} [ B_1, QB_2,
B_3] + (-1)^{B_1+
B_2} \, [B_1, B_2, QB_3] \\[1ex] &\hskip-5pt + (-)^{B_1} [B_1, [B_2,
B_3]] + (-1)^{B_2 (1+ B_1)} \, [B_2, [B_1, B_3]] + (-1)^{B_3 (1+B_1+ B_2)} \, [B_3,
[B_1, B_2]] \,.
\end{split}
\ee
If the string product $[\cdot \,, \cdot\,, \cdot ]$ satisfied a Jacobi-like identity, the terms on the
second line would add up to zero.  Since they do not, one requires
an elementary triple product  represented by $[\cdot \,, \cdot\,, \cdot ]$
and used to define a quartic elementary interaction.  This triple product
(as well as all higher ones) must also include a projection to states
that satisfy $L_0- \bar L_0 =0$.
The failure of $Q$ to be a  derivation
 of this product  is  equal to the 
violation of the Jacobi identity.  The above relation
is part of the defining relations of the $L_\infty$
 homotopy Lie-algebra~\cite{Zwiebach:1992ie,Lada:1992wc}.

\medskip
Consider now the states with  $N= \bar N = 1$.
Projection down to the  
 physical space with $\Delta =0$  is most easily discussed in momentum space.  Consider
a field
$\phi$
 (with $N= \bar N = 1$)  with definite 
 momenta and winding numbers
$(w^a, p_a)  =  ( m^{a}\,, n_a)$ with  $ a = 1 , 2, \ldots , d\,.$
Then 
\be
\Delta \phi = 0 \quad \leftrightarrow \quad \sum_a n_a  m^{a} 
\equiv  n \,m = 0\,.
\ee
We 
combine the winding  $m^a$ 
and the momentum   $n_a$   
of $\phi$ into a $2d$-column vector $v$: 
\be
\label{vvectors}
 v = \begin{pmatrix}  m \\ n \end{pmatrix}  \, \in \mathbb{Z}^{2d} \,,
\ee
and  define the  inner product  
with respect to the $O(d,d)$ invariant metric $\hat \eta$
\be
\label{circprod}
v \circ v'\equiv   v^T \, \hat \eta \,  v'  =   \bigl( m , n ) 
\begin{pmatrix} 0& 1 \\ 1& 0 \end{pmatrix} 
\begin{pmatrix}   m' \\ n'  \end{pmatrix}  = m  n'  +  n m'  \,.
\ee
Since $v \circ v = 2 n_a m^a$, 
the $\Delta=0$ constraint on the vector takes the form
\be
\Delta \phi= 0  \quad \leftrightarrow \quad  v \circ v= 0 \,.
\ee
In words, the vector $v$ is null with respect to $\eta$. 
A field with definite momentum and winding must satisfy this condition
to be allowed.
 A general superposition of such allowed fields is
also allowed, since $\Delta$ is a linear operator.
 If we have fields $\phi$ and $\phi'$  with
  null momenta $v$ and $v'$, the product $\phi \phi'$ has momentum $v+v'$ which is not null in general.
 The  product will only satisfy the constraint if the momenta are orthogonal:
\be
\Delta (\phi \phi') = 0  ~~\leftrightarrow~~ 
(v + v')\circ (v+ v')=0  \quad \leftrightarrow \quad  v \circ 
v' = 0 \,.
\ee
The  constraint is not satisfied by products  
since $\Delta$ is a second-order differential operator.  
We enforce the constraint as follows.  
Given a general field $A(x^\mu, x^a , \tilde x_a)$,
a Fourier series for the compact dimensions yields 
\be
A(x^\mu, x^a , \tilde x_a) =  \sum_{v\in \mathbb{Z}^{2d}}  \hat A (x^\mu, v) \, e^{i v^T  \mathbb{X} }  \, = \, \sum_{v\in \mathbb{Z}^{2d}}  \hat A (x^\mu, v) \, e^{i m^a \tilde x_a + i n_a x^a }  \,. 
\ee
Since $\Delta = -{2\over\alpha'} \partial_a \tilde \partial^a$ we find
\be
\Delta A= {1\over \alpha'} \sum_{v\in \mathbb{Z}^{2d}} 
v\circ v\,  \, \hat A (x^\mu, v) \, 
 \, e^{i v^T  \mathbb{X} }  \, \,. 
\ee
A canonical projection of a general $A$   into a field $[[ A ]]$ that   satisfies the $\Delta=0$ constraint is defined by
\be
[[A]] \equiv  \sum_{v\in \mathbb{Z}^{2d}} \,\delta_{v\circ v, 0}\,
 \hat A (x^\mu, v) \, e^{i v^T  \mathbb{X} }  \,  \,. 
\ee
The role of the Kronecker delta is to retain only the Fourier 
components of the field whose momenta are null.  It is now clear
that \be
\Delta [[A]] =0\, .
\ee    It is also clear from the definition that with constants $\alpha$ and
$\beta$ and functions $A$ and $B$ we have
\be
[[\alpha A+ \beta B]] = \alpha\, [[A]] + \beta \,[[ B]] \,.
\ee
   The operation $[[\, \cdot \,]]$ is 
a linear map from the space of functions on the doubled torus to
the kernel of~$\Delta$.  It is a projector because 
applying it twice has
the same effect as applying it once. The operation $[[ \,\cdot \,]]$ implements the
$\Delta=0$ constraint in the same way that the Kronecker
delta in (\ref{stringsvertex}) 
implements  the level matching 
constraint for a general string field. 
For  constrained  fields $A(x,\ti x)$ and $B(x, \tilde x)$, the product
$[[  A(x,\ti x)  B(x,\ti x)]]$ projects onto  those Fourier modes
$\hat A(x,v) \hat B(x,v')$ with $v,v'$ both null and orthogonal, $v \circ 
v' = 0 $.

The closed string product includes the projector $\delta_{L_0-\bar L_0, 0}$ because 
the product of two allowed states 
should give an allowed state. 
We must 
therefore use the projection $[[ \,\cdot \,]]$ in the gauge transformations
 (\ref{gaugetrans}) to ensure that the gauge variations are allowed
 variations of the fields.  This means that, properly
written, the gauge transformations are 
\be
\label{gaugetrans99f}
\begin{split}
\phantom{\Biggl(} 
\delta_\lambda e_{ij}  &= ~ \bar D_j \lambda_i  \, +  {1\over 2} \, \
\Bigl[\Bigl[\,
   (D_i \lambda^k)  e_{kj}  
 - \,  (D^k \lambda_i) e_{kj}    
   + \,\lambda_k D^k e_{ij} ~\Bigr]\Bigr] \,,    \,~  
 ~     \\
\delta_\lambda d  &=
- {1\over 4}  D\cdot \lambda 
+  {1\over 2} \,\bigl[\bigl[ (\lambda \cdot D) \,d\, \bigr] \bigr]\,~.
 \end{split}
\ee
Happily, there is no need to  use the projection $[[ \,\cdot \,]]$
 in the cubic action.  The action is  correct as
written 
in equation (\ref{redef-action}).
This is not difficult to explain.  Let $A$ be a field
that satisfies $\Delta A =0$ and $B$ be a field that does not.
Then, we claim  that   
\be
\label{nop}
  \int \, A \,[[\, B\, ]] =\int  \, A  \, B \,,
\ee
where the integral is over $x^\mu, x^a$, and $\tilde x_a$.
It follows from the above    
 that the projection of $B$ to the kernel of $\Delta$ is not needed.
This is clear in momentum space.
The integration implies that any Fourier mode of  $B$ with
momentum $v$ 
can only couple to a Fourier mode of $A$ with momentum $(-v)$.
Then any Fourier mode of  $B$ with
momentum $v$ that is
not allowed cannot contribute since $(-v)$ is also not allowed and thus
cannot be found in $A$, as $A$ satisfies the constraint. 
Next consider  
the cubic term in the action. Since this arises from the string field theory term 
$\langle \Psi, [ \Psi \,, \Psi ]\rangle$
the structure we obtain must be a sum of terms of the form 
$\int \phi_1 \, [[\phi_2 \phi_3]]$.  Since $\phi_1$ satisfies the constraint, equation (\ref{nop})
shows that the projector is not needed for the product $\phi_2 \phi_3$. 
Therefore, we do not need to include additional projectors
in the quadratic and cubic terms in the action.
Similar remarks  apply to the check of gauge invariance of the action.  
The above gauge
transformations induces terms of the form $\int \phi_1 \,[[\lambda \phi_2]]$.  Again, the projector  is not needed
to the order to which we are working,
 and we can proceed naively. 

This convenient simplification may disappear for terms in the action
quartic in fields.   The terms that arise from the elementary quartic
interaction $\langle \Psi, [ \Psi \,, \Psi\,, \Psi ]\rangle$ of the closed
string field theory action would have 
the form $\int  \phi_1 [[ \phi_2 \phi_3 \phi_4]] $. Again,
because of (\ref{nop})
the projector is not needed and this term
equals $\int  \phi_1 \phi_2\phi_3 \phi_4 $.  On the other hand, terms that arise from integrating out other fields will have a projector  
of the form
$\int  \phi_1  \phi_2\, [[ \phi_3 \phi_4]]$.  This projector
cannot be eliminated.  It is clear that given four fields, the projector
can be inserted in three inequivalent ways -- the number of ways in
which the fields can be partitioned into groups of two.  It seems tempting
to believe that terms with these three inequivalent positions of the projector 
may be related to  terms with no projector through 
identities  in the spirit of (\ref{csft-fh}).

\subsection{Cocycles}

We now address another important issue.  Closed string vertex operators
 in toroidal backgrounds have cocycles -- operators that are included to 
 ensure standard commutation properties~\cite{Frenkel,Gross:1985rr}. 
 If ${\cal V}^0_{v_\alpha}$ denotes the 
 naive vertex operator for a state with
 momenta and winding specified by $v_\alpha$ one finds
 \be
\label{ftu99vgnb}
{\cal V}^0_{v_2} (z_2, \bar z_2)
~{\cal V}^0_{v_1} (z_1, \bar z_1)~=  ~e^{i\pi \,v_1 \circ v_2 }~{\cal V}^0_{v_1} (z_1, \bar z_1)
~{\cal V}^0_{v_2} (z_2, \bar z_2)\,.
\ee
The phase factor can be equal to minus one, in which case we have
the unpleasant fact that vertex operators for bosons anticommute.
A cocycle operator is included multiplicatively to define vertex operators
${\cal V}_{v_\alpha}$ that always commute.  
These cocycles affect the signs
of correlation functions.  As a result, there are extra
signs that are introduced in the three-string vertex~\cite{Hata:1986mz,Maeno:1989uc,Kugo:1992md}.  Up to field redefinitions, the sign factor 
that affects the amplitude  $\langle {\cal V}_{v_{\alpha}} {\cal V}_{v_{\beta}}
{\cal V}_{v_\gamma}\rangle$ is
\be
\label{fullcoc}
\epsilon_{\alpha\beta\gamma} =  e^{i\pi (n^{\alpha} m^{\alpha}  
+ n^{\gamma} m^{\beta} )}  \,.
\ee
Of course $v_{\alpha} + v_{\beta} + v_{\gamma} =0$.  The sign factor can be shown
to be cyclic invariant.  Under exchange
of $\alpha$ and $\beta$ labels, for example, this sign factor changes as follows:
\be
\epsilon_{\beta\alpha\gamma} = \epsilon_{\alpha\beta\gamma}\,
e^{i\pi v_\alpha\circ v_\beta}\, \,.
\ee
The sign factor (\ref{fullcoc}) is nontrivial:  it cannot be removed by redefinitions
of the states corresponding to the vertex operators.  For our case of interest
the situation is somewhat simpler. We have $n^{\alpha} m^{\alpha}=0$ because
all states satisfy the $\Delta =0$ constraint. As a result,
(\ref{fullcoc}) becomes 
\be
\label{smallcoc}
\epsilon_{\alpha\beta\gamma} =  e^{i\pi\,  n^{\gamma} m^{\beta} }  \,.
\ee
Despite appearances to the contrary this sign factor is fully symmetric under
exchange of labels. 
This can be understood as follows.
First recall the simple fact that given three null vectors that add up to zero, 
the vectors are mutually orthogonal.  This shows that 
$v_\alpha\circ v_\beta=v_\alpha\circ v_\gamma=
v_\beta\circ v_\gamma =0$
and the sign factor  associated with exchanges vanishes.  
A symmetric  
sign factor could be trivial, but we have not been able to
show that (\ref{smallcoc}) is trivial. 
Note, however, that  any three (constrained) states coupled by 
the three string vertex have momenta which are orthogonal and 
therefore the associated cocycle-free 
vertex operators commute (see (\ref{ftu99vgnb})).

The cubic action we have written did not include cocycle-induced sign
factors. If present, such signs also appear in the gauge transformations
and in the duality transformations. 
It is known that string field theory gauge invariance
to  ${O}(\Lambda \Psi^2)$ 
holds with or without such
sign factors and this may explain why our construction succeeded so
far without any sign factors. 
It is to next order that the sign factors are claimed to be needed
for gauge invariance~\cite{Hata:1986mz}. 
The cocycle-induced sign factors are non-trivial 
and required for the full  
string field theory, but  their role may be different 
for the  double field theory we are focussing on. 
We hope to return to this question in the future.

\subsection{Spaces large and small}

For fields with arbitrary dependence on the coordinates $\X$ of $T^{2d}$, the momenta can be 
arbitrary $v$'s in the
{\em full} momentum lattice $\mathbb{Z}^{2d}$ introduced in \S\ref{theconandpro}.
The constraint 
\be
\label{thispaper}
 v\circ v=0\,,  
\ee  
restricts us to the null 
subspace of $\mathbb{Z}^{2d}$, which we refer to as the  {\em large} space. 
The equation $v\circ v=0$ defines a light-cone in $\R^{2d}$ with metric $\hat \eta$, and the large
 space consists of the  
 points on this light cone with integer coordinates.
For general string states with $N\ne \bar N$, this light-cone is replaced by the hyperboloid
\be
\label{sfttoday}
\half \, v\circ v  =  N - \bar N \,.  
\ee 

A $2d$ dimensional space with metric of signature $(d,d)$ can have totally null $d$-dimensional subspaces  (called totally isotropic subspaces in the mathematics literature) in which the indefinite metric restricts to zero, so that all tangent vectors to the subspace  are null and mutually orthogonal.
We shall be interested in totally null subspaces $T^d\subset T^{2d}$.
Writing the metric as $ds^2= 2dx^ad\ti x_a$, we see that
the $d$-torus with coordinates $x^a$ and the dual torus with coordinates $\ti x_a$ are both totally null, and any $T^d$ obtained from these by acting with $O(d,d;\Z)$ will also be totally null.
If we let the coordinates of a null subspace be $y^a$ and those of the complement be $\ti y_a$,
then
  the metric $\hat \eta $ is $ds^2= 2dy^ad\ti y_a$
and 
\be
\Delta   = 
-{2\over \alpha'}\sum_a {\partial \over \partial \tilde y_a}  {\partial\over \partial y^a} \,.   
\ee
For fields that are independent of $\ti y$, the constraint
$\Delta   = 0$ is automatically satisfied.
 Moreover, all products of fields satisfy
the constraint $\Delta=0$ and no projection $[[ \, \cdot \, ]]$ is necessary.
Nor are cocycles needed, 
as all momenta $v$ for such fields are null and mutually orthogonal, so 
  all vertex operators are mutually
local. 
Then the restriction of the full double field theory to fields dependent only
 on the coordinates $y^a$ of such a null subspace (together with $x^\mu$) should give a conventional local
field theory without cocycles or projectors.
For the $T^d$ with coordinates $x^a$, this should be the conventional field theory with action
(\ref{standardaction}) (after field redefinitions, and compactified on $T^d$),
while for other choices it should be a dual theory
related to this by an  $O(d,d;\Z)$ transformation.
However, these theories can be written in a duality covariant way, by taking the double field theory and 
restricting the momentum space
to a {\em small} space where all vectors $v$ are not only null, but 
also mutually orthogonal.
With this restriction, the double field theory has no cocycles, constraints or projectors.
It would be very interesting to obtain the full nonlinear version
of our action under this simplifying assumption. 
The result may be related to the work of Siegel~\cite{Siegel:1993th}
who constructed 
a realization of T-duality in the massless sector under
the assumption that all momenta are orthogonal.

\sectiono{Comments and open questions}\label{coandopenque}

A striking feature  of string field theory on a torus is  that general solutions  involve
fields on the doubled torus instead of conventional spacetime fields. As a result, the theory is very different from that suggested by conventional effective field theories that, like supergravity limits of superstrings, 
miss  key stringy features.   
The theory on a torus is a case which is  nontrivial enough to be
  interesting  yet  is simple enough to be tractable. 
One of our goals here has been to seek a subsector of this theory that is almost as 
simple as a conventional field theory but which is rich enough to include much of the magic of string theory.

We have begun the construction of an intriguing 
double field theory of  massless    fields 
$h_{ij},b_{ij},d$ depending on both $x$ and $\tilde x$.
We have used string field theory to find the action to cubic order and showed that its variation under gauge transformations, found to linear order in the fields, vanishes to the requisite order.  
By including both winding and momenta we do not have a regime
where all excitations have parametrically small energy and the theory
may not arise as a simple decoupling limit of string theory. 
If we view our construction as an effective
 field theory for a natural
set of excitations (some of which may have large energy), the
string field theory suggests that an action and gauge transformations
should exist to all orders in the field, although the explicit calculation of these becomes much harder at higher orders.
The unusual features of string field theory include the explicit projectors
to the kernel of $L_0 - \bar L_0$,  cocycle-induced sign factors in the vertices, and the homotopy Lie algebra structure of the string products.
 These are all expected to  
  play a role in the double field theory, although they have been largely avoided at the cubic level.
Of course, the $\Delta =0$ constraint on fields and gauge parameters,
which arises from the $L_0 - \bar L_0=0$  constraint in string field theory, 
has played a central role. It was absolutely crucial, even for linearised
gauge invariance.

 It has long been
known that the $L_0 - \bar L_0 =0$ constraint is fundamental and 
all attempts to formulate closed string field theory without imposing 
this off-shell condition  on the fields and parameters
have so far failed.  Such a formulation 
could exist, but  a very significant conceptual 
advance may be needed to find it.  In our  massless theory the level-matching constraint became $\Delta =0$.
  Our attempts to relax this constraint
failed, but we hope that understanding the constraint in the simpler setting
of the double field theory may shed light on the constraint in the full string field theory.

\medskip 
It is natural to speculate on the full non-linear form of the theory. 
We noted that the free theory includes gauge parameters that 
suffice to describe ``double-diffeomorphisms" or 
linearised
diffeomorphisms
of the doubled space $\R^{n-1,1}\times T^{2d}$.  The nonlinear extension
shows that the symmetry of the theory 
appears to be considerably more intricate.  
In addition to 
linearised
diffeomorphisms, the gauge parameters 
 generate doubled gauge transformations of the antisymmetric tensor field
so that there is an interesting mixture of the two symmetries.
  Second, there is  the projection of the gauge parameters to the kernel of 
$\Delta$.  The full symmetry 
has an algebra that appears to be different from that of
 diffeomorphisms on the doubled space $\R^{n-1,1}\times T^{2d}$, 
 but does include   
 the diffeomorphisms of various undoubled subspaces
$\R^{n-1,1}\times T^{d}$ obtained by keeping only the $x^a$ coordinates, or 
keeping only
a  set of coordinates obtained from the $x^a$ by T-duality.
As we have noted at various points, the full symmetry of the theory
may turn out to be that of a homotopy-Lie algebra, or some related structure.
 It would be interesting
to see what field theory structures arise to define the higher products inherent
in such algebra.  In a homotopy Lie algebra we have field dependent structure
constants and a gauge algebra that only closes on-shell.  These features
are coherently organised and described by the products.  
While diffeomorphisms
define a conventional Lie algebra, the larger symmetry of our theory 
most likely does~not.  Perhaps the most important open question
related to the action is that of cocycles.  The construction of the quartic
terms in the action will have to face this issue, as well as the possibility
that explicit projectors to the kernel of $\Delta$ will be needed.

The string field theory treats the background $E$ and the fluctuation $e$ rather differently, but 
gives a treatment to all orders in an arbitrary fluctuation $e$.
In section 4.5, we introduced a total field $\EE(X)$ combining both background and fluctuation, showing that it had the right geometric gauge transformations when independent of $\ti x$. 
The  Buscher transformation of  $\EE$ was extended  to the case with dependence on both 
$x$ and $\ti x$,  providing a  generalisation of  T-duality
   of the kind proposed in \cite{Dabholkar:2005ve}. 
Rewriting   the double field theory in terms of $\EE$ 
would give a version of the theory independent of the split 
 into background and fluctuation, and thus with some degree
 of background independence.  
Although arbitrary geometries would be allowed,
our formulation would remain very much tied to the topology $\R^{n-1,1}\times T^d$; other topologies would have different zero-mode structures.

\medskip 
Some of the structures in  our work 
also  arise in generalised geometry, but with important differences.
Generalised geometry  \cite{Hitchin,Gualtieri} treats structures on a $D$ dimensional manifold $M$ on which there is a natural action of the group 
$O(D,D)$.
This typically
involves doubling the tangent space of a   manifold $M$ (replacing the tangent bundle $T$ with $T\oplus T^*$). Tensor indices then run over  twice the usual range, but there is dependence only  on the $D$ coordinates  of $M$. If $M$ is equipped with a metric and B-field  ${\cal E}_{ij}= \Gg_{ij} + \bb_{ij}$, these can be usefully combined into the $2D\times 2D$ matrix $ \HH({\cal E})$ 
given by (\ref{genmetpl}).
The inverse matrix $\GG ({\cal E})= \HH^{-1}$
is the {\it generalised metric} \cite{Gualtieri}.  It is a $2D\times 2D$ matrix but  depends only on the $D$ coordinates of $M$.
 Generalised geometry 
is then the study of conventional geometry with a
metric and  $B$-field on $M$, packaged in a useful way.

In our work, by contrast, we  restrict to 
$D$-dimensional 
manifolds $M=\R^{n-1,1}\times T^d$ 
 and find
that string theory leads us to $\R^{n-1,1}\times T^{2d}$, with
 a doubling of the torus coordinates but no doubling of the range of tensor indices.
 We found it notationally useful to double the coordinates of $\R^{n-1,1}$ also, to give a space 
 $M_{\rm{doubled}}$ with dimension $2D$. 
 Our fields depend non-trivially on the doubled torus coordinates $(x^a,\ti x_a)$ and on the Minkowski coordinates $x^\mu$, but do not depend on the extra dual
 Minkowski coordinates $\ti x_\mu$.
 It follows that we can use the double field theory fluctuations to define
 the field  ${\cal E} =E_{ij} + e_{ij} +\ldots$ in (\ref{fullfielda}) that depends
 nontrivially on  $(x^\mu, x^a, \ti x_a)$.
We then 
define an  $\HH({\cal E})$ 
 by (\ref{genmetpl}) and its inverse $\GG({\cal E})$.  
If ${\cal E}$ depends only on  the coordinates $(x^\mu, x^a)$ of $M$, then $\GG ({\cal E})$  
is a generalised metric on $M$, but here we 
generalise  
to allow dependence on $\ti x_a$ also. 
Since
 $\GG({\cal E})$  is a $2D\times 2D$ matrix function on 
 the  
  $2D$ dimensional space $M_{\rm{doubled}}$ 
  that depends on 
  $(x^\mu, x^a, \ti x_a)$, it is a candidate for  
 a conventional metric on $M_{\rm{doubled}}$.
 We have seen in (\ref{dualinvmetric}) that the line element 
 $ds^2 = dX ^t\, \GG(  \EE(X))\, dX  $
   is   invariant under T-duality transformations, which act as large 
   diffeomorphisms of $T^{2d}$.  
   The metric $\GG$ is constrained,
   because $\HH$ is:  $ \eta \HH = \HH^{-1} \eta$, and is further restricted by the requirement that $\Delta$ annihilate $e_{ij}$ and $d$. 
  Then $\GG$ is a natural and interesting object that could play an important role 
  in the formulation of double field theory.

 \medskip  
 Our work has been 
  concrete and explicit. 
 It has long been known that the toroidal coordinates in closed
 string theory should be doubled due to the presence of winding modes and we have 
given a precise sense to this, 
 showing that the dual coordinates enter on an equal footing with the spacetime coordinates and that fields depend on both spacetime and dual coordinates.
 We have seen that double
 field theory exists   as a free theory and 
  when we include 
 the lowest-order interactions.  A number of key features have been identified precisely.
 The symmetry structure is 
 novel and remains
 to be fully understood
  and the full nonlinear theory remains to be found.  
We have seen that  doubled fields can be used to define a kind of geometry on the doubled 
space that reduces to conventional spacetime geometry on the original torus or to a dual geometry on the dual torus.
This geometry is fully {\it dynamical}  -- it 
 depends on {\it all} of the coordinates of the doubled space, it
 evolves according to field equations and is subject to constraints.
 This leads to the 
 conclusion that the full doubled geometry is physical: the dual dimensions should not be viewed as an auxiliary structure or a gauge artifact. 
It is therefore reasonable to expect that doubled geometry
will feature prominently in the eventual understanding
of the nature of space and the role of geometry in string theory.

\vspace{0.6cm}

{\bf \large Acknowledgments:}  
We would like to thank the KITP in Santa Barbara for hospitality
during the 2009  {\it Fundamental Aspects of Superstring Theory} program
and D.~Gross for his questions and comments.  We are grateful to N.~Moeller,  whose
computer programs helped sort out signs of ghost correlators needed to
compute the action and gauge transformations.
We are happy to acknowledge
helpful conversations with M.~Green, W. Siegel, W.~Taylor, and
 A. Tseytlin.
The work of 
B.Z. is supported in part by the U.S.
DOe grant De-FC02-94eR40818.


\baselineskip 15pt




\begin{thebibliography}{99}


\small


\bibitem{Giveon:1994fu}
  A.~Giveon, M.~Porrati and E.~Rabinovici,
  ``Target space duality in string theory,''
  Phys.\ Rept.\  {\bf 244} (1994)~77
  [arXiv:hep-th/9401139].


\bibitem{Kugo:1992md}
  T.~Kugo and B.~Zwiebach,
  ``Target space duality as a symmetry of string field theory,''
  Prog.\ Theor.\ Phys.\  {\bf 87}, 801 (1992)
  [arXiv:hep-th/9201040].
  

\bibitem{Zwiebach:1992ie}
  B.~Zwiebach,
  ``Closed string field theory: Quantum action and the B-V master equation,''
  Nucl.\ Phys.\  B {\bf 390}, 33 (1993)
  [arXiv:hep-th/9206084].
  
\bibitem{Hata:1986mz}
  H.~Hata, K.~Itoh, T.~Kugo, H.~Kunitomo and K.~Ogawa,
  ``Gauge String Field Theory For Torus Compactified Closed String,''
  Prog.\ Theor.\ Phys.\  {\bf 77}, 443 (1987).

\bibitem{Maeno:1989uc}
  M.~Maeno and H.~Takano,
  ``Derivation of the cocycle factor 
  of vertex in closed bosonic string field theory on torus,"
    Prog.\ Theor.\ Phys.\  {\bf 82}, 829 (1989).
  
\bibitem{Hull:2004in}
  C.~M.~Hull,
  ``A geometry for non-geometric string backgrounds,''
JHEP {\bf 0510} (2005) 065,  arXiv:hep-th/0406102.


  
\bibitem{Tseytlin:1990nb}
A.~A.~Tseytlin,
``Duality Symmetric Formulation Of String World Sheet Dynamics,''
Phys.\ Lett.\ B {\bf 242}, 163 (1990); 
``Duality Symmetric Closed String Theory And Interacting Chiral Scalars,''
Nucl.\ Phys.\ B {\bf 350}, 395 (1991).

\bibitem{Siegel:1993th}
  W.~Siegel,
  ``Superspace duality in low-energy superstrings,''
  Phys.\ Rev.\  D {\bf 48}, 2826 (1993)
  [arXiv:hep-th/9305073];
  ``Two vierbein formalism for string inspired axionic gravity,''
  Phys.\ Rev.\  D {\bf 47}, 5453 (1993)
  [arXiv:hep-th/9302036].
  
    

\bibitem{VanRaamsdonk:2003gj}
Ê M.~Van Raamsdonk,
Ê ``Blending local symmetries with matrix nonlocality in D-brane effective
Ê actions,''
Ê JHEP {\bf 0309}, 026 (2003)
Ê [arXiv:hep-th/0305145].
Ê 


\bibitem{Duff:1990}
M.~J.~Duff,
``Duality Rotations In String Theory,''
Nucl.\ Phys.  {B \bf 335},  
610 (1990).


\bibitem{Hull:1988dp}
  C.~M.~Hull,
  ``Covariant Quantization Of Chiral Bosons And Anomaly Cancellation,''
  Phys.\ Lett.\ B {\bf 206} (1988) 234.

\bibitem{Hull:1988si}
  C.~M.~Hull,
  ``Chiral Conformal Field Theory And Asymmetric String Compactification,''
  Phys.\ Lett.\ B {\bf 212}, 437 (1988).


\bibitem{Maharana:1992my}
J.~Maharana and J.~H.~Schwarz,
``Noncompact symmetries in string theory,''
Nucl.\ Phys.\ B {\bf 390}, 3 (1993)
[arXiv:hep-th/9207016].




\bibitem{Hull:2006va}
  C.~M.~Hull,
  ``Doubled geometry and T-folds,''
  JHEP {\bf 0707} (2007) 080
  [arXiv:hep-th/0605149].

\bibitem{Berman:2007yf}
  D.~S.~Berman and D.~C.~Thompson,
  ``Duality Symmetric Strings, Dilatons and O(d,d) Effective Actions,''
  Phys.\ Lett.\  B {\bf 662}, 279 (2008)
  [arXiv:0712.1121 [hep-th]].
  
  
\bibitem{Berman:2007xn}
  D.~S.~Berman, N.~B.~Copland and D.~C.~Thompson,
  ``Background Field Equations for the Duality Symmetric String,''
  Nucl.\ Phys.\  B {\bf 791}, 175 (2008)
  [arXiv:0708.2267 [hep-th]].


\bibitem{HackettJones:2006bp}
  E.~Hackett-Jones and G.~Moutsopoulos,
  ``Quantum mechanics of the doubled torus,''
  JHEP {\bf 0610}, 062 (2006)
  [arXiv:hep-th/0605114].


\bibitem{buscher} 
  T.~H.~Buscher,
  ``A Symmetry of the String Background Field Equations,''
  Phys.\ Lett.\  B {\bf 194}, 59 (1987); 
  ``Path Integral Derivation of Quantum Duality in Nonlinear Sigma Models,''
  Phys.\ Lett.\  B {\bf 201}, 466 (1988).


\bibitem{GivRoc}
  A.~Giveon and M.~Rocek,
  ``Generalized duality in curved string backgrounds,''
  Nucl.\ Phys.\  B {\bf 380}, 128 (1992)
  [arXiv:hep-th/9112070].
 


 

\bibitem{Gregory:1997te}
  R.~Gregory, J.~A.~Harvey and G.~W.~Moore,
  ``Unwinding strings and T-duality of Kaluza-Klein and H-monopoles,''
  Adv.\ Theor.\ Math.\ Phys.\  {\bf 1} (1997) 283
  [arXiv:hep-th/9708086].


\bibitem{Tong:2002rq}
  D.~Tong,
  ``NS5-branes, T-duality and   world-sheet instantons,''
  JHEP {\bf 0207}, 013 (2002)
  [arXiv:hep-th/0204186].


\bibitem{Harvey:2005ab}
  J.~A.~Harvey and S.~Jensen,
  ``Worldsheet instanton corrections to the Kaluza-Klein monopole,''
  JHEP {\bf 0510} (2005) 028
  [arXiv:hep-th/0507204].

\bibitem{Okuyama:2005gx}
  K.~Okuyama,
  ``Linear sigma models of H and KK monopoles,''
  JHEP {\bf 0508} (2005) 089
  [arXiv:hep-th/0508097].



\bibitem{Witten:2009xu}
  E.~Witten,
  ``Branes, Instantons, And Taub-NUT Spaces,''
  arXiv:0902.0948 [hep-th].



\bibitem{Dabholkar:2005ve}
  A.~Dabholkar and C.~Hull,
  ``Generalised T-duality and non-geometric backgrounds,''
  arXiv:hep-th/0512005.
  
\bibitem{Hull:2009sg}
  C.~M.~Hull and R.~A.~Reid-Edwards,
  ``Non-geometric backgrounds, doubled geometry and generalised T-duality,''
  arXiv:0902.4032 [hep-th].
  

\bibitem{Frenkel}
  I.~B.~Frenkel and V.~G.~Kac,
  ``Basic Representations of Affine Lie Algebras and Dual Resonance Models,''
  Invent.\ Math.\  {\bf 62}, 23 (1980); 
  P.~Goddard and D.I.~Olive, ``Algebras, Lattices and Strings",
  in Vertex Operators in Mathematics and Physics.   Publications of the
  Mathematical Sciences Research Institute, Berkeley, No.3 (Springer-Verlag, 1984) 51-96.

\bibitem{Gross:1985rr}
  D.~J.~Gross, J.~A.~Harvey, E.~J.~Martinec and R.~Rohm,
  ``Heterotic String Theory. 2. The Interacting Heterotic String,''
  Nucl.\ Phys.\  B {\bf 267}, 75 (1986); 
   Nucl.\ Phys.\  B {\bf 256}, 253 (1985).



\bibitem{Ghoshal:1991pu}
  D.~Ghoshal and A.~Sen,
  ``Gauge and general coordinate invariance in nonpolynomial closed string theory,''
  Nucl.\ Phys.\  B {\bf 380}, 103 (1992)
  [arXiv:hep-th/9110038].
 
\bibitem{Michishita:2006dr}
  Y.~Michishita,
  ``Field redefinitions, T-duality and solutions in closed string field
  theories,''
  JHEP {\bf 0609}, 001 (2006)
  [arXiv:hep-th/0602251].

\bibitem{Lada:1992wc}
  T.~Lada and J.~Stasheff,
  ``Introduction to SH Lie algebras for physicists,''
  Int.\ J.\ Theor.\ Phys.\  {\bf 32}, 1087 (1993)
  [arXiv:hep-th/9209099].






\bibitem{Hellerman:2006tx}
  S.~Hellerman and J.~Walcher,
  ``Worldsheet CFTs for flat monodrofolds,''
  arXiv:hep-th/0604191.
  

\bibitem{Goroff:1985th}
  M.~H.~Goroff and A.~Sagnotti,
  ``The Ultraviolet Behavior Of Einstein Gravity,''
  Nucl.\ Phys.  B{\bf 266}, 709 (1986).





\bibitem{Alvarez:1996vt}
  E.~Alvarez and Y.~Kubyshin,
  ``Is the string coupling constant invariant under T-duality?,''
  Nucl.\ Phys.\ Proc.\ Suppl.\  {\bf 57} (1997) 44
  [arXiv:hep-th/9610032].


%
  



%
\bibitem{Hitchin}
N.~Hitchin,
``Generalized Calabi-Yau manifolds,''
 Q. J. Math.  {\bf 54}  (2003), no. 3, 281--308,
arXiv:math.DG/0209099.
%


\bibitem{Gualtieri}
M.~Gualtieri,
``Generalized complex geometry,"
PhD Thesis (2004).
arXiv:math/0401221v1 [math.DG]


\end{thebibliography}
\end{document}